\documentclass[preprints,article,accept,pdftex,moreauthors]{Definitions/mdpi} 
\usepackage{amsmath}
\firstpage{1} 
\pubvolume{1}
\issuenum{1}
\articlenumber{0}
\pubyear{2025}
\copyrightyear{2025}
\datereceived{ } 
\daterevised{ }
\dateaccepted{ } 
\datepublished{ } 
\hreflink{https://doi.org/}

\pdfoutput=1

\Title{Detecting Unusual Trading Patterns on Cryptocurrency Exchanges by Means of Complexity Measures}

\TitleCitation{Wash trading}

\Author{Jakub Zwydak $^{1}$ \orcidD{},, Marcin Wątorek $^{1}$\orcidE{}, Jarosław Kwapień $^{2}$ \orcidC{}, Stanisław Drożdż $^{1,2,}$*\orcidA{}, 
}

\AuthorNames{Jakub Zwydak, Marcin Wtorek, Jarosław Kwapień, Stanisław Drożdż}

\address{%
$^{1}$ \quad Faculty of Computer Science and Mathematics, Cracow University of Technology, ul. Warszawska 24, 31-155 Kraków, Poland; \\
$^{2}$ \quad Complex Systems Theory Department, Institute of Nuclear Physics, Polish Academy of Sciences, ul. Radzikowskiego 152, 31-342 Kraków, Poland;}

\corres{Correspondence: marcin.watorek@pk.edu.pl; stanislaw.drozdz@ifj.edu.pl;}

\abstract{
Artificial transaction generation remains an important source of potential market 
manipulation on cryptocurrency exchanges, as it may distort reported liquidity and reduce market transparency. This study proposes a diagnostic framework for detecting unusual trading patterns based on complexity and statistical-structure measures derived from high-frequency trade-level data. The analysis considers log-returns, trading volume, and transaction counts, using tail distributions, autocorrelation functions, multifractal characteristics, approximate entropy, and detrended cross-correlations. The methodology is applied to BTC, ETH, and XRP traded on Binance, Bitget, KuCoin, and Kraken over the period from April 1 to June 30, 2025. The results reveal a pronounced anomaly on Bitget for BTC and ETH after mid-May 2025. The number of transactions increases sharply, but there is no proportional increase in traded volume or return fluctuations. This regime is characterised by numerous low-volume trades, weaker autocorrelations, reduced multifractal organisation, higher short-pattern irregularity, and weaker cross-correlations involving the transaction-count series. These features are consistent with a noise-like component in trading activity and may indicate artificially increased transaction counts, although they do not provide direct proof of wash trading. The findings show that complexity-based indicators can be useful for detecting exchange-specific trading anomalies that remain hidden in price-based measures.
}

\keyword{Wash trading; trading patterns; trading anomalies; cryptocurrency market; centralised exchanges; high-frequency trading; complexity measures} 

\begin{document}

\section{Introduction}

Cryptocurrency markets have become an important and rapidly evolving segment of the global financial system~\cite{WatorekM-2021b}. Their continuous operation, global accessibility, high level of automation, heterogeneous and still-evolving regulatory frameworks, and fragmented exchange structure distinguish them from traditional financial markets. At the same time, these features make cryptocurrency markets particularly vulnerable to market manipulation, liquidity misreporting, and other forms of artificial trading activity~\cite{Chen2022,Amiram2025}. Among such practices, wash trading is especially relevant. It refers to transactions that create the appearance of trading activity without a genuine transfer of economic risk, thereby inflating reported volume, misleading liquidity indicators, and reducing market transparency~\cite{Pennec2021,Victor2021WashTrading,Cong2022CryptoWashTrading}.

This problem is particularly important for centralised cryptocurrency exchanges, where reported activity is used by market participants and data aggregators to assess exchange reliability. Artificially inflated activity may improve an exchange's apparent position in market rankings, attract additional order flow, and create a misleading impression of market depth~\cite{Amiram2025}. It may also weaken the informational content of public market data by distorting the relation between trading activity, volume, and price formation. Therefore, detecting unusual trading patterns is important not only for identifying potential manipulation but also for assessing market quality.

In the case of centralised exchanges, detecting wash trading and related forms of artificial activity is particularly difficult. Publicly available trade-level data usually include prices, volumes, and timestamps but do not provide information on trader identities, account ownership, or the direction of beneficial control. As a result, individual transactions cannot usually be classified directly as artificial. Instead, detection must rely on indirect evidence based on statistical regularities, market microstructure relations, and deviations from patterns typically observed in genuine trading activity. Previous studies have shown that artificial or suspicious trading can leave measurable traces in transaction size distributions, rounding patterns, volume dynamics, and other behavioural characteristics of trade 
data~\cite{Cong2022CryptoWashTrading,DiFrancescoMaesa2024,sila2025wash}.

Wash-trading detection may be more direct in decentralised exchanges~\cite{Watorek2024} and NFT markets~\cite{SzydloP-2024a,WatorekM-2024a}, where transaction records are publicly available on the blockchain. In such environments, the flow of tokens between addresses can be traced, enabling the identification of repeated transfers between related wallets, circular trading patterns, self-trading structures, or abnormal ownership changes~\cite{Gan2023,DeCollibusFM-2024a}. This has enabled the development of blockchain-based forensic tools based on address networks, token-flow graphs, and transaction histories~\cite{Victor2021WashTrading,Chen2024,gan2024,Niu2024,tovsic2025}. However, this approach cannot be directly applied to centralised exchanges, where trades are matched internally, and public data do not reveal the identities of counterparties.

Financial markets are complex systems composed of heterogeneous agents interacting across multiple time scales~\cite{KwapienJ-2012a}. Their collective activity gives rise to robust statistical regularities, such as heavy-tailed return distributions, volatility clustering, long-range dependence in activity measures, and non-trivial relations between trading volume, transaction intensity, and price fluctuations~\cite{MantegnaStanley1995,GopikrishnanP-1998a,GopikrishnanP-1999a,Cont2001,BouchaudJP-2010a,Safari2026}. Cryptocurrency markets exhibit most of these stylised facts as well~\cite{BegusicS-2018a,DrozdzS-2018a,DrozdzS-2019a,DrozdzS-2020a,JamesN-2021b,Abdullaev2026}, including multifractal properties of price, volume, and transaction count fluctuations~\cite{TakaishiT-2018a,daSilvaFilho-2018a,MensiW-2019a,StavroyiannisS-2019a,Stosic2019,takaishi2020market,han2020long,KwapienJ-2022a,KakainakaS-2022b,Ali2024,CHOI2026} and multiscale cross-correlations~\cite{WatorekM-2021b,watorekfutnet2022,DrozdzS-2023a}. These regularities provide a natural reference point for identifying market states that deviate from the structure expected under normal trading conditions. In this study, the benchmark for 'usual' market behaviour is therefore defined empirically rather than by a single universal parametric process, since transaction intensity depends strongly on liquidity, asset type, exchange microstructure, and intraday market conditions. An unusual trading pattern is understood as a persistent departure from these regularities.

From this perspective, artificial trading activity can be interpreted as a disturbance of the natural complexity of market dynamics. If a large number of trades is generated mechanically, for example, through excessive order fragmentation, coordinated automated strategies, or artificial liquidity-provision mechanisms, then the number of transactions may become decoupled from genuine trading volume and price dynamics. Such a process may produce many small trades without a proportional increase in traded volume or return fluctuations. It may also alter autocorrelation structure, weaken multiscale cross-correlations, reduce multifractal organisation, or change the regularity of the transaction-count series. Therefore, complexity-based measures can be used to identify persistent deviations from the stylised statistical organisation of market activity that are not captured by standard price-based indicators.

This paper investigates unusual trading patterns on centralised cryptocurrency exchanges using complexity and statistical structure measures derived from high-frequency transaction data. The analysis is performed for BTC, ETH, and XRP traded on Binance, Bitget, Kraken, and KuCoin over the period from April 1 to June 30, 2025. Tick-by-tick transactions are aggregated into 1-min time series of log-returns $R_{\Delta t=1\mathrm{min}}(t)$, trading volume $V_{\Delta t=1\mathrm{min}}(t)$, and the number of transactions $N_{\Delta t=1\mathrm{min}}(t)$. The aim is not to identify individual wash trades, which would require account-level or order-book information, but to determine whether the statistical organisation of market activity is consistent with regular trading behaviour across exchanges and assets.

The methodological framework combines several complementary approaches. First, the distribution properties of absolute log-returns, trading volume, and the number of transactions are analysed to identify exchange-specific deviations in tail behaviour. Second, autocorrelation functions are used to characterise temporal persistence in volatility, volume, and transaction-count dynamics. Third, multifractal analysis is applied to examine the scaling organisation of fluctuations across time scales. Fourth, detrended cross-correlation analysis is used to quantify scale-dependent dependencies between $|R|$, $V$, and $N$, following the idea that cross-correlations in non-stationary financial time series may depend strongly on the observation scale~\cite{PodobnikB-2008a,ZebendeGF-2011a,KwapienJ-2015a}. Finally, the detrended cross-correlations and approximate entropy (ApEn)~\cite{Pincus1991} are calculated in rolling windows to identify time-localised regime changes and changes in the regularity of the analysed time series.

The main contribution of this study is to demonstrate that complexity-based characteristics can reveal exchange-specific anomalies that are not visible from price dynamics alone. The proposed framework evaluates whether transaction counts, traded volume, and return fluctuations preserve the statistical relationships expected under regular market conditions. In this way, it complements standard price-based diagnostics by targeting distortions in the internal organisation of market activity. This motivates the use of econophysics and complexity-based methods as diagnostic tools for assessing market quality and liquidity reliability in digital asset exchanges.

\section{Methods}

\subsection{Methods based on detrending}

The identification of unusual trading patterns requires tools that can separate genuine temporal structure from local non-stationarities. This is particularly important for cryptocurrency data, where trading activity is highly intermittent ~\cite{MakarovI-2020a,WatorekM-2021b,Alexander2022,Petukhina2021}, periods of low and high activity alternate over time~\cite{Aleti2021,Watorek2023,Wang2020,Hansen2024}, and trading volume may fluctuate strongly~\cite{DrozdzS-2023a,Dyhrberg2018}. In such conditions, standard correlation or fluctuation measures may be affected by trends, local bursts of activity, or regime changes. Therefore, detrending-based methods are used here to characterise both the scaling properties of individual time series and the cross-correlations between market variables.

The methodological framework used here is based on multifractal detrended fluctuation analysis (MFDFA) and multifractal detrended cross-correlation analysis (MFCCA). MFDFA~\cite{KantelhardtJ-2002a} is an extension of detrended fluctuation analysis (DFA~\cite{PengCK-1994a}), while MFCCA~\cite{ZhouWX-2008a,OswiecimkaP-2014a} generalizes detrended cross-correlation analysis (DCCA~\cite{PodobnikB-2008a}) to different fluctuation amplitudes. These methods are particularly useful for financial and cryptocurrency time series, where non-stationarity, heavy tails, volatility clustering, and multiscale correlations are commonly observed~\cite{JiangZQ-2019a,KwapienJ-2023a,Drozdz2025futnet}.

In the present context, detrending-based methods serve two complementary purposes. First, they allow for the characterisation of the scaling organisation of individual time series, such as log-returns, trading volume, and the number of transactions. Second, they provide a scale-dependent measure of cross-correlation between pairs of variables: between absolute log-returns vs volume, absolute log-returns vs the number of transactions, and volume vs the number of transactions. 

Let one consider two time series $U=\{u(i)\}_{i=1}^T$ and $V=\{v(i)\}_{i=1}^T$ of length $T \gg 1$ that are sampled at the same time instants $i$. First, both time series are integrated to form their profiles $\tilde{U}$ and $\tilde{V}$
\begin{equation}
\tilde{u}(i)= \sum_{j=1}^i u(j), \quad \tilde{v}(i)= \sum_{j=1}^i v(j),
\label{eq::mfdcca.profiles}
\end{equation}
respectively. For a given scale $s$, the profiles are divided into segments of length $s$. To avoid discarding data at the end of the series, the segmentation is performed twice, starting from the beginning and from the end of each profile. This gives $2M_s$ segments in total.

In each segment, local trends are removed independently. A polynomial trend $P_{\nu}^{(l)}$ of order $l$ is fitted to each profile and subtracted from it. In this study, a polynomial of degree $l=2$ was used, which is a common choice for financial time series~\cite{OswiecimkaP-2013a,JiangZQ-2019a}. The detrended residuals are defined as
\begin{eqnarray}
\nonumber
x(\nu s + k) = \tilde{u}(\nu s + k) - P_{\textrm{X},\nu}^{(l)}(k),\\
y(\nu s + k) = \tilde{v}(\nu s + k) - P_{\textrm{Y},\nu}^{(l)}(k),
\label{eq::mfdcca.detrending}
\end{eqnarray}
where $k=1,...,s$ and $\nu=0,...,2M_s-1$. For each segment, the local covariance and the local variances are then calculated:
\begin{eqnarray}
\nonumber
f_{\textrm{XY}}^2(s,\nu) = {1 \over s} \sum_{k=1}^s x(\nu s + k) y(\nu s + k),\\
f_{\textrm{XX}}^2(s,\nu) = {1 \over s} \sum_{k=1}^s x^2(\nu s + k),\label{eq::mfdcca.covariances}\\
f_{\textrm{YY}}^2(s,\nu) = {1 \over s} \sum_{k=1}^s y^2(\nu s + k).
\nonumber
\end{eqnarray}
The quantities $f_{\textrm{XX}}^2(s,\nu)$ and $f_{\textrm{YY}}^2(s,\nu)$ describe the fluctuation amplitudes of the individual detrended profiles, while $f_{\textrm{XY}}^2(s,\nu)$ measures their local covariance at scale $s$.

The segment-wise quantities are next averaged over all segments and raised to a real power $q$. This leads to the bivariate fluctuation function $F_{\textrm{XY}}^q(s)$ and the univariate fluctuation functions $F_{\textrm{XX}}^q(s)$ and $F_{\textrm{YY}}^q(s)$:
\begin{eqnarray}
\nonumber
F_{\textrm{XY}}^q(s) =  {1 \over 2 M_s} \sum_{\nu=0}^{2 M_s - 1} {\rm sign} \left[f_{\textrm{XY}}^2(s,\nu)\right] |f_{\textrm{XY}}^2(s,\nu)|^{q/2} 
\\
F_{\textrm{XX}}^q(s) = {1 \over 2 M_s} \sum_{\nu=0}^{2 M_s - 1} \left[f_{\textrm{XX}}^2(s,\nu)\right]^{q/2}
\label{eq::mfdcca.fluctuation-functions}\\
\nonumber
F_{\textrm{YY}}^q(s) = {1 \over 2 M_s} \sum_{\nu=0}^{2 M_s - 1} \left[f_{\textrm{YY}}^2(s,\nu)\right]^{q/2}
\end{eqnarray}
The sign function in the definition of $F_{\textrm{XY}}^q(s)$ preserves the sign of local covariances and ensures that the bivariate fluctuation function remains real for different values of $q$~\cite{OswiecimkaP-2014a}. The parameter $q$ controls the relative contribution of fluctuations of different amplitudes. Negative values of $q$ give more weight to small fluctuations, while positive values of $q$ emphasise large fluctuations.

The procedure is repeated for a range of scales $s$. For fractal time series, the univariate 
fluctuation functions follow power-law scaling:
\begin{eqnarray}
[F_{\textrm{XX}}^q(s)]^{1/q}=F_{\textrm{XX}}(q,s) \sim s^{h_{\textrm{X}}(q)}, \quad [F_{\textrm{YY}}^q(s)]^{1/q}=F_{\textrm{YY}}(q,s) \sim s^{h_{\textrm{Y}}(q)}.
\label{eq::mfdcca.univariate.scaling}
\end{eqnarray}
The functions $h_{\textrm{X}}(q)$ and $h_{\textrm{Y}}(q)$ are generalised Hurst exponents. A constant dependence of $h(q)$ on $q$ corresponds to monofractal behaviour. In contrast, a 
clear dependence of $h(q)$ on $q$ indicates multifractality. The case $q=2$ corresponds to the standard Hurst exponent.

When $h(q)$ depends on $q$, the corresponding singularity spectrum $f(\alpha)$ can be obtained through the Legendre transform~\cite{HalseyTC-1986a,KantelhardtJ-2002a}:
\begin{equation}
\alpha = h(q) + q {dh(q) \over dq}, \quad f(\alpha) = q(\alpha - h(q)) + 1,
\label{eq::singularity spectra}
\end{equation}
where $\alpha$ is the singularity, or H\"older, exponent and $f(\alpha)$ is the multifractal spectrum. The width of this spectrum reflects the range of scaling exponents present in the time series, while its asymmetry indicates whether small or large fluctuations contribute more strongly to the multifractal structure~\cite{DrozdzS-2015a}.

The same fluctuation functions can be used to define the $q$-dependent detrended cross-correlation coefficient $\rho(q,s)$~\cite{KwapienJ-2015a}:
\begin{equation}
\rho(q,s) = {F_{\textrm{XY}}^q(s) \over \sqrt{ F_{\textrm{XX}}^q(s) F_{\textrm{YY}}^q(s) }}.
\label{eq::rhoq}
\end{equation}
This coefficient is a generalisation of the detrended cross-correlation coefficient 
$\rho_{\rm DCCA}$~\cite{ZebendeGF-2011a}. It quantifies the strength of detrended cross-correlations between two time series as a function of the time scale $s$ and the fluctuation order $q$.

The parameter $q$ plays the same filtering role as in the fluctuation functions. For $q<2$, the 
coefficient is more sensitive to smaller fluctuations, whereas for $q>2$, larger fluctuations dominate. In this study, the main analysis is based on $\rho(q=2,s)$, which corresponds to equally weighted fluctuations.

Importantly, $\rho_q(s)$ does not require the fluctuation functions to exhibit a clear power-law dependence on $s$. Therefore, it can also be applied to time series that do not display well-developed fractal or multifractal scaling. This makes it particularly suitable for detecting unusual trading patterns, because anomalous activity may itself destroy or distort scaling properties.

\subsection{Approximate and Sample Entropy}
\label{Entr}
Approximate entropy (ApEn) is a statistic used to quantify the regularity and unpredictability of a time series~\cite{Pincus1991}. In practical terms, it measures how often similar patterns of length $\mathrm{dim}$ remain similar when one additional point is included. Therefore, a higher value of ApEn indicates that local patterns are less repeatable, the dynamics are more irregular, and short-term evolution is less predictable. Conversely, a lower ApEn value corresponds to more regular, repetitive, and locally predictable dynamics.

A time series $X=\{X(1),X(2),\ldots,X(L)\}$ is embedded in $m \geqslant 1$ dimensions by constructing vectors of length $m$ with delay $\tau \geqslant 1$:
\begin{equation}
\mathbf{X}^{(m,\tau)}_i =
\left[
X(i), X(i+\tau), \ldots, X(i+(m-1)\tau)
\right],
\label{eq::apen.vectors}
\end{equation}
where $i=1,2,\ldots,L_{m,\tau}$ with $L_{m,\tau}=L-(m-1)\tau$.

The similarity between two reconstructed vectors is evaluated using the maximum norm:
\begin{equation}
d\left(\mathbf{X}^{(m,\tau)}_i,\mathbf{X}^{(m,\tau)}_j\right)=\max_{0\leq k \leq m-1}\left|X(i+k\tau)-X(j+k\tau)\right|.
\label{eq::apen.distance}
\end{equation}
Two vectors are treated as similar if their distance does not exceed the tolerance radius $r$.

For each vector $\mathbf{X}^{(m,\tau)}_i$, the relative number of similar vectors is defined as
\begin{equation}
C_i^{(m,\tau)}(r)
=
\frac{1}{L_{m,\tau}} \sum_{j=1}^{L_{m,\tau}}
\Theta\left(
r-d\left(\mathbf{X}^{(m,\tau)}_i,\mathbf{X}^{(m,\tau)}_j\right)
\right),
\label{eq::apen.ci}
\end{equation}
where $\Theta(\cdot)$ is the Heaviside function.

The average logarithmic probability of finding similar patterns of length $m$ is then given by
\begin{equation}
\Phi^{(m,\tau)}(r)=\frac{1}{L_{m,\tau}}
\sum_{i=1}^{L_{m,\tau}}
\ln C_i^{(m,\tau)}(r).
\label{eq::apen.phi}
\end{equation}
Approximate entropy is defined as the difference between the corresponding quantities for embedding 
dimensions $m$ and $m+1$:
\begin{equation}
\mathrm{ApEn}(m,\tau,r)=\Phi^{(m,\tau)}(r)-\Phi^{(m+1,\tau)}(r).
\label{eq::apen.definition}
\end{equation}

In the main calculations, the embedding dimension and delay were fixed at $m=2$ and $\tau=1$, respectively, while the similarity radius was set to $r=0.2\sigma_X$, where $\sigma_X$ denotes the standard deviation of the analysed time series $X$. To assess the robustness of the observed patterns, additional calculations were also performed for $m=3,4,5$, with $\tau=1$ and $r=0.2\sigma_X$ kept fixed. Since $\mathrm{ApEn}(\cdot)$ depends on the choice of $m$, $\tau$, and $r$, comparisons between different parameter settings should focus mainly on qualitative changes over time rather than on absolute values.

To address the known finite-sample bias of ApEn caused by self-matching, Sample Entropy (SampEn) was additionally calculated~\cite{Richman2000}. In contrast to ApEn, self-matches are excluded in SampEn, and the negative logarithm of the conditional probability that two similar patterns of length $m$ remain similar when their length is increased to $m+1$ is estimated.

For the embedded vectors $\mathbf{X}^{(m,\tau)}_i$, the number of matched vector pairs of length $m$ was defined as
\begin{equation}
B_m(r)=\#\left\{(i,j): i\neq j,\ 
d\left(\mathbf{X}^{(m,\tau)}_i,\mathbf{X}^{(m,\tau)}_j\right)\le r
\right\}.
\label{eq:sampen_Bm}
\end{equation}

Similarly, the number of matched vector pairs of length $m+1$ was defined as
\begin{equation}
A_m(r)=\#\left\{(i,j): i\neq j,\ 
d\left(\mathbf{X}^{(m+1,\tau)}_i,\mathbf{X}^{(m+1,\tau)}_j\right)\le r
\right\}.
\label{eq:sampen_Am}
\end{equation}

Sample entropy was then defined as
\begin{equation}
\mathrm{SampEn}(m,\tau,r)
=
-\ln\left(\frac{A_m(r)}{B_m(r)}\right).
\label{eq:sampen}
\end{equation}

If either $A_m(r)$ or $B_m(r)$ was equal to zero, the corresponding value was treated as undefined and was not interpreted. This issue is particularly important for large $m$, small $r$, and short rolling windows, where the number of matched patterns may become too small for stable estimation.

\section{Data specification}
\label{Dataspec}

The empirical analysis conducted in this study is based on transaction-level, tick-by-tick data for three major cryptocurrencies: Bitcoin (BTC), Ethereum (ETH), and XRP. The instruments are quoted against USDT, except for Kraken, where USD-quoted pairs are used. The sample covers the period from April 1, 2025, to June 30, 2025. The data were obtained from publicly available archives provided by four major cryptocurrency exchanges: 
Binance~\cite{Binance}, Bitget~\cite{Bitget}, KuCoin~\cite{Kucoin}, and Kraken~\cite{Kraken}. These platforms were selected from among widely recognised centralised exchanges~\cite{CoinGeckoRank} in order to represent trading venues with different levels of liquidity, trading intensity, and market structure. Binance is one of the largest cryptocurrency exchanges in terms of trading volume and is characterised by high market liquidity across a broad range of currency pairs. Bitget is a major centralised exchange known for its derivatives market and copy-trading services. Kraken is one of the oldest cryptocurrency exchanges and is often associated with high security standards and a stronger regulatory orientation. KuCoin is a large retail-oriented exchange with a broad selection of listed assets.

Tab.~\ref{tab::stats} presents descriptive statistics for the analysed trading pairs. The statistics include the average trading volume per time interval $\langle V_{\Delta t} \rangle$ and the average transaction size $\langle V \rangle$, both expressed in units of the base asset, the average number of transactions per time interval $\langle N_{\Delta t} \rangle$, and the fraction of time intervals without transactions $\%0N_{\Delta t}$ for $\Delta t=1$ min. Clear differences between exchanges are visible. Binance is the most liquid exchange in the sample, with the highest average transaction count and average trading volume. Moreover, there are no 1 min intervals without transactions on this exchange during the period under study ($\%0N_{\Delta t}=0$). Interestingly, Binance does not exhibit the lowest average transaction size. The smallest value of $\langle V \rangle$ occurs on Bitget for BTC and ETH, and on KuCoin for XRP.

The highest average transaction sizes are observed on Kraken and are several times larger than those recorded on the other exchanges considered. This indicates that, on average, larger transactions are executed on Kraken. As a result, Kraken ranks second only to Binance in average trading volume, though the average number of transactions is substantially lower. In terms of $\langle N_{\Delta t} \rangle$, Kraken ranks third for BTC, fourth for ETH, and second for XRP.

Bitget exhibits a different trading pattern. Small transactions dominate on this exchange, making Bitget second only to Binance in the number of BTC and ETH transactions, while ranking third in average trading volume. The exchange with the least frequent trading activity is KuCoin, as indicated by the highest fraction of 1 min intervals without transactions ($\%0N_{\Delta t}$). At the same time, transactions on KuCoin are characterised by a relatively small average size, which results in the lowest average trading volume among the analysed exchanges.

\begin{table}[H]
\caption{Descriptive statistics of the considered trading pairs: average volume
$\langle V_{\Delta t} \rangle$ and average transaction size $\langle V \rangle$,
both expressed in units of the base asset, average number of transactions
$\langle N_{\Delta t} \rangle$, and fraction of intervals without transactions
$\%0N_{\Delta t}$ for $\Delta t=1$ min.}
\label{tab::stats}

\centering
\small
\setlength{\tabcolsep}{3pt}

\begin{tabularx}{\textwidth}{l*{12}{>{\centering\arraybackslash}X}}
\toprule
& \multicolumn{4}{c}{\textbf{BTC}}
& \multicolumn{4}{c}{\textbf{ETH}}
& \multicolumn{4}{c}{\textbf{XRP}} \\
\cmidrule(lr){2-5}
\cmidrule(lr){6-9}
\cmidrule(lr){10-13}

\textbf{Exchange}
& \boldmath{$\langle V_{\Delta t} \rangle$}
& \boldmath{$\langle N_{\Delta t} \rangle$}
& \boldmath{$\langle V \rangle$}
& \boldmath{$\%0N_{\Delta t}$}
& \boldmath{$\langle V_{\Delta t} \rangle$}
& \boldmath{$\langle N_{\Delta t} \rangle$}
& \boldmath{$\langle V \rangle$}
& \boldmath{$\%0N_{\Delta t}$}
& \boldmath{$\langle V_{\Delta t} \rangle$}
& \boldmath{$\langle N_{\Delta t} \rangle$}
& \boldmath{$\langle V \rangle$}
& \boldmath{$\%0N_{\Delta t}$} \\
\midrule

Binance
& 3.12 & 410.32 & 0.008 & 0
& 314.7 & 569.8 & 0.55 & 0
& 27752 & 105.4 & 263.42 & 0 \\

Bitget
& 0.34 & 113.9 & 0.003 & 0.09
& 8.73 & 114.36 & 0.08 & 0.05
& 1229.7 & 3.9 & 315.14 & 0.09 \\

Kraken
& 1.01 & 22.6 & 0.045 & 0
& 16.7 & 10.9 & 1.53 & 0.04
& 13755.9 & 13.76 & 999.99 & 0.02 \\

KuCoin
& 0.06 & 6.44 & 0.009 & 0.23
& 4.17 & 31.22 & 0.13 & 0.08
& 660.3 & 3.57 & 184.75 & 0.45 \\

\bottomrule
\end{tabularx}
\end{table}

For further analysis, the tick-by-tick transaction data were aggregated into 1-min intervals. This procedure resulted in time series of prices $P_{\Delta t=1\mathrm{min}}(t)$, trading volume $V_{\Delta t=1\mathrm{min}}(t)$, and the number of transactions $N_{\Delta t=1\mathrm{min}}(t)$. Logarithmic returns were calculated from prices as $R(t_i) = \ln P(t_{i+1}) - \ln P(t_i)$, where $i = 1, \ldots, T-1$.

\begin{figure}[ht!]
\centering
\includegraphics[width=0.49\textwidth]{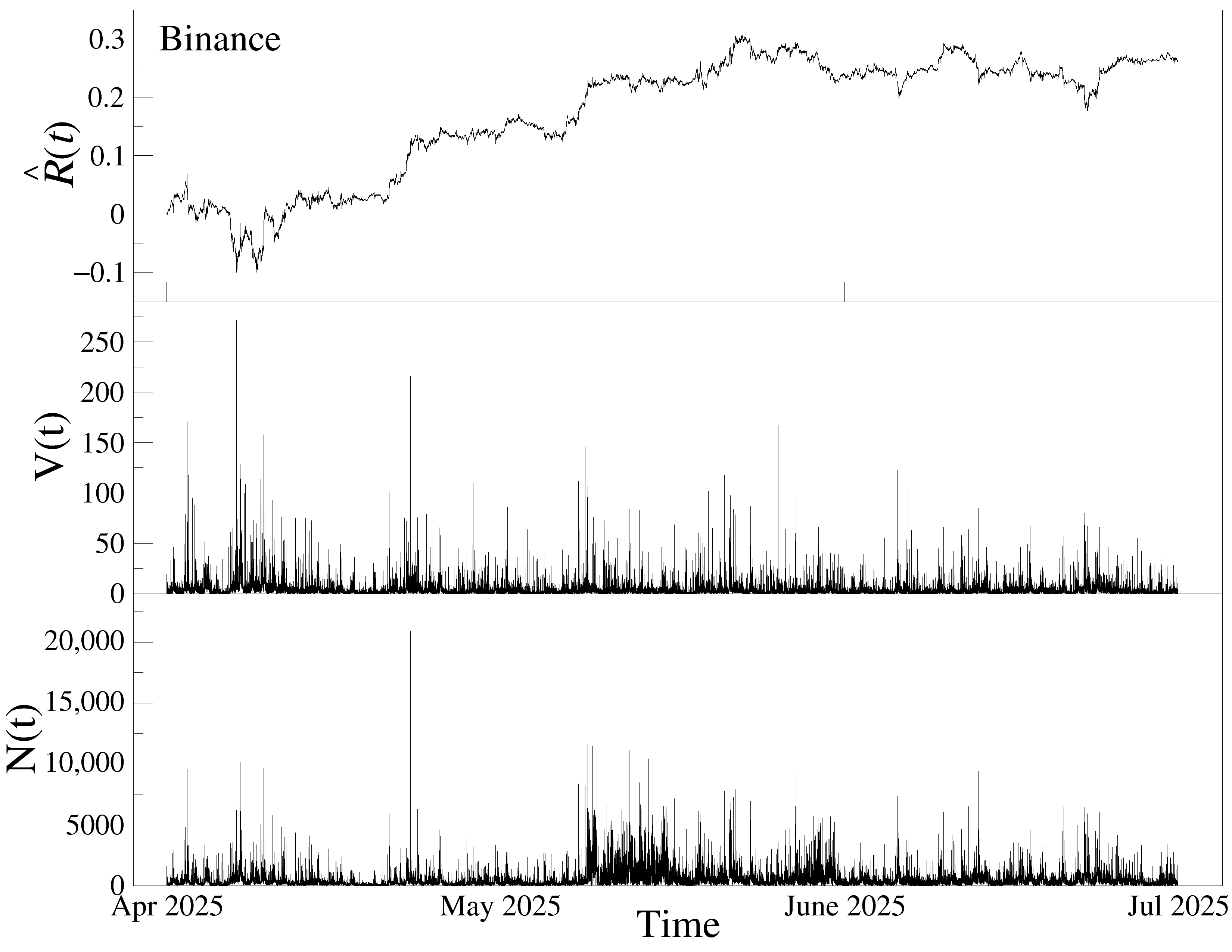}
\includegraphics[width=0.49\textwidth]{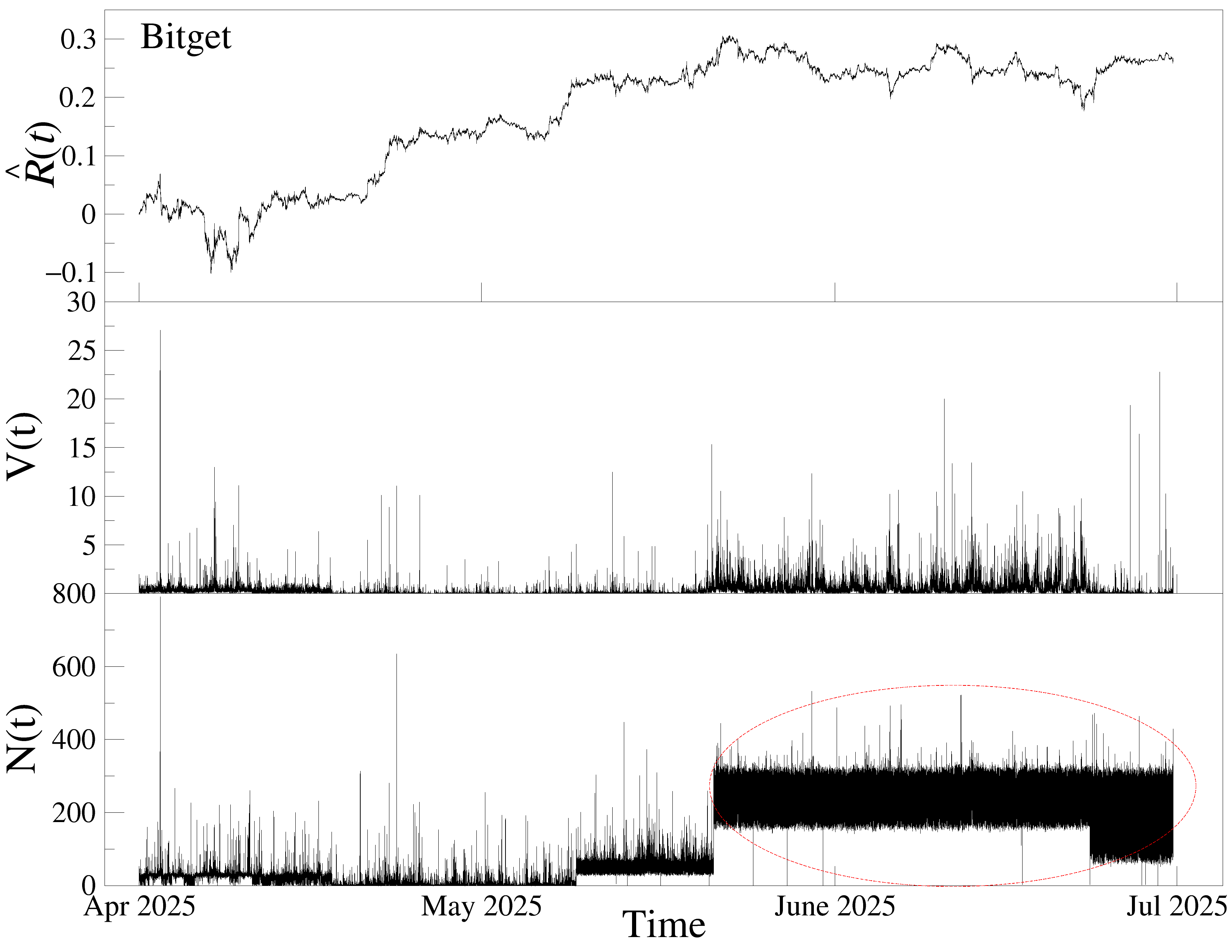}
\includegraphics[width=0.49\textwidth]{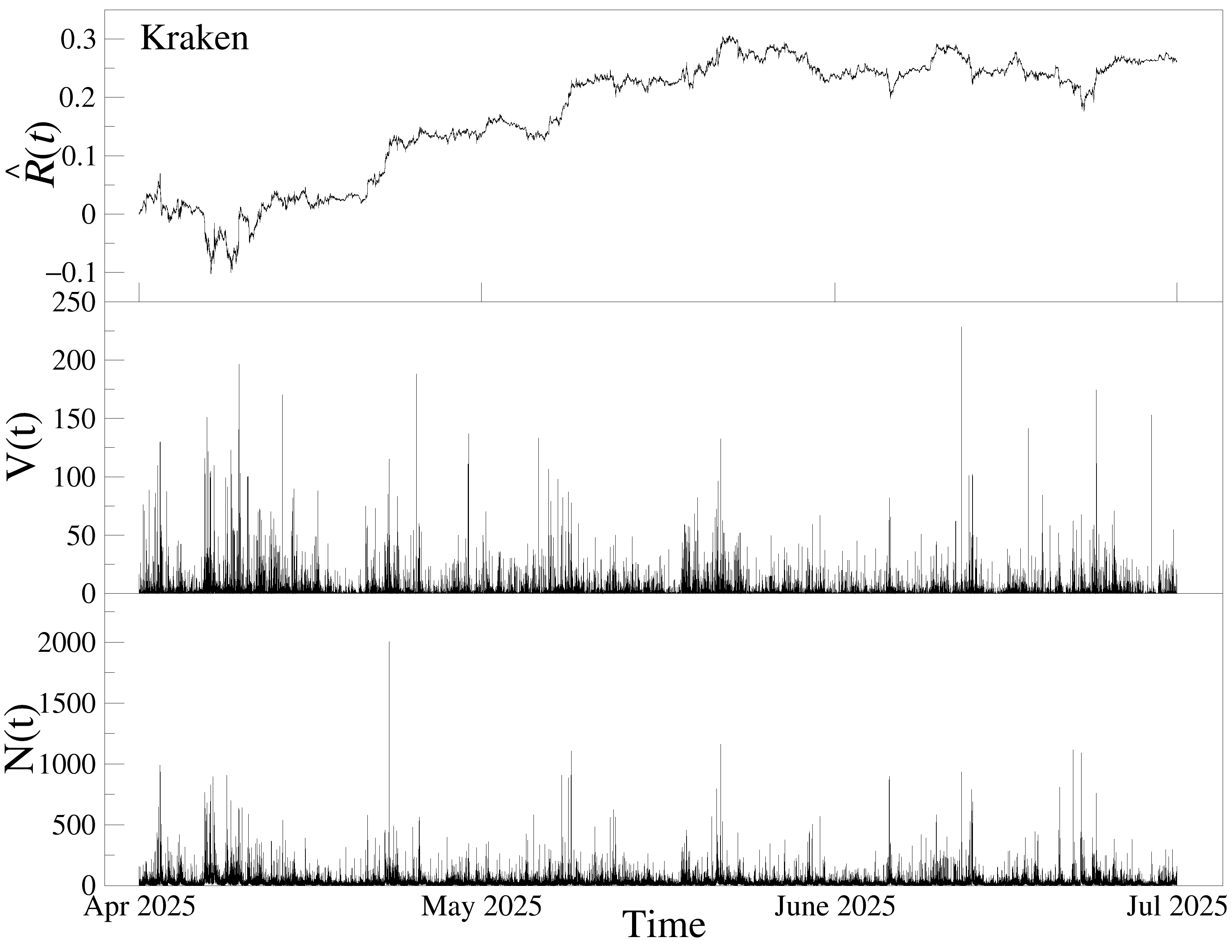}
\includegraphics[width=0.49\textwidth]{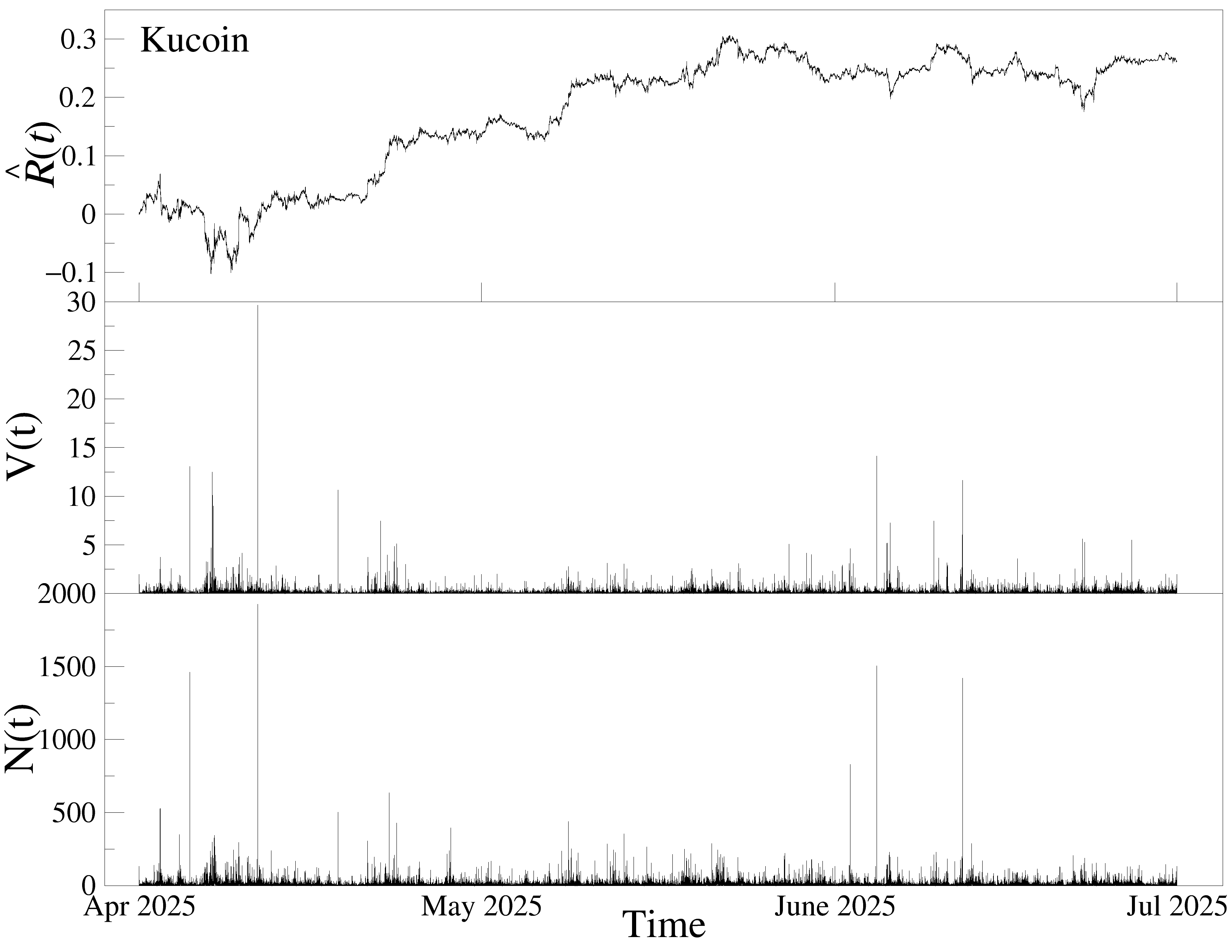}
\caption{Evolution of the cumulative log-returns $\hat{R}(t)$, trading volume $V_{\Delta t=1\mathrm{min}}(t)$, and the number of transactions $N_{\Delta t=1\mathrm{min}}(t)$ for BTC on Binance (top left), Bitget (top right), Kraken (bottom left), and KuCoin (bottom right). The period of increased transaction activity is marked by a red dashed ellipse.}
\label{fig::szeregiBTC}
\end{figure}

\begin{figure}[ht!]
\centering
\includegraphics[width=0.49\textwidth]{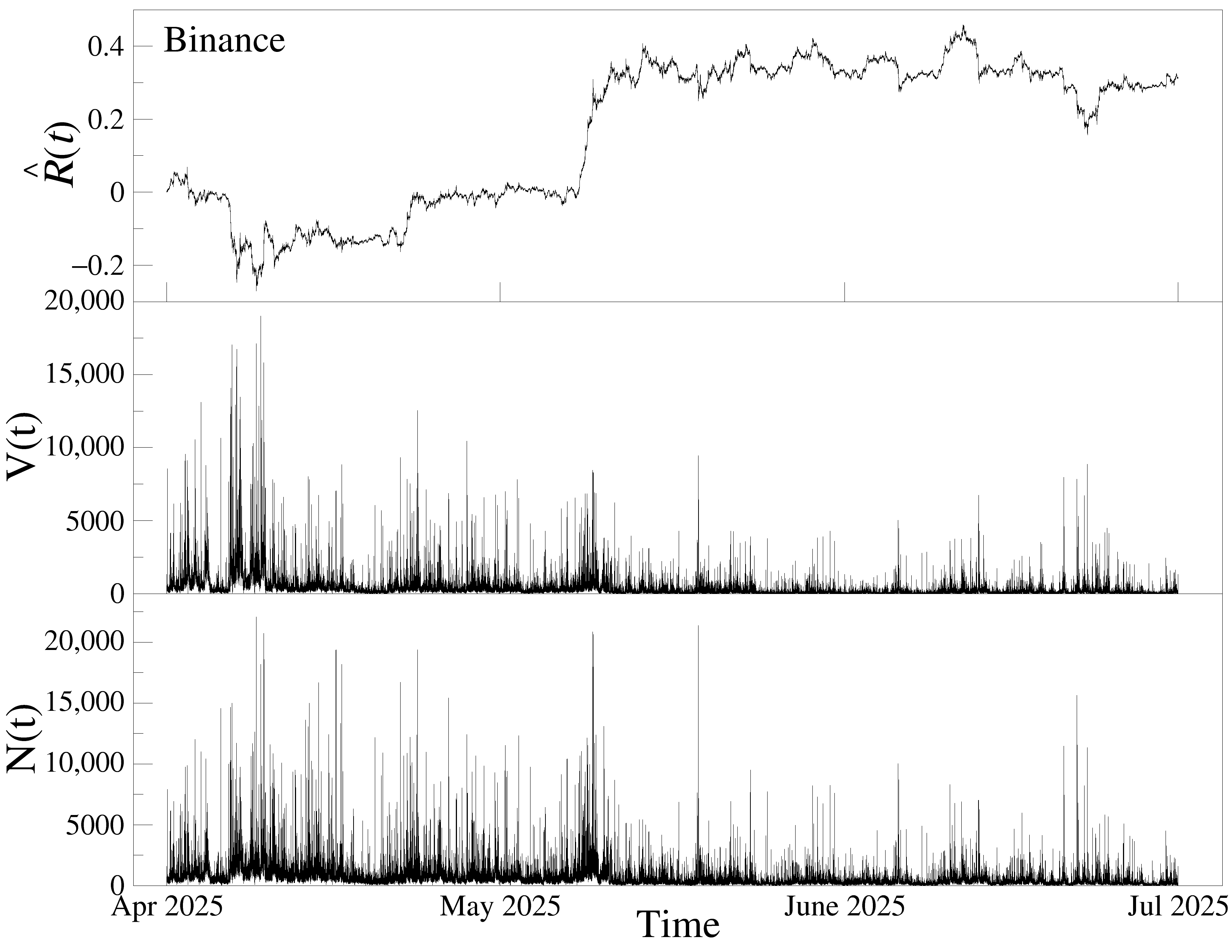}
\includegraphics[width=0.49\textwidth]{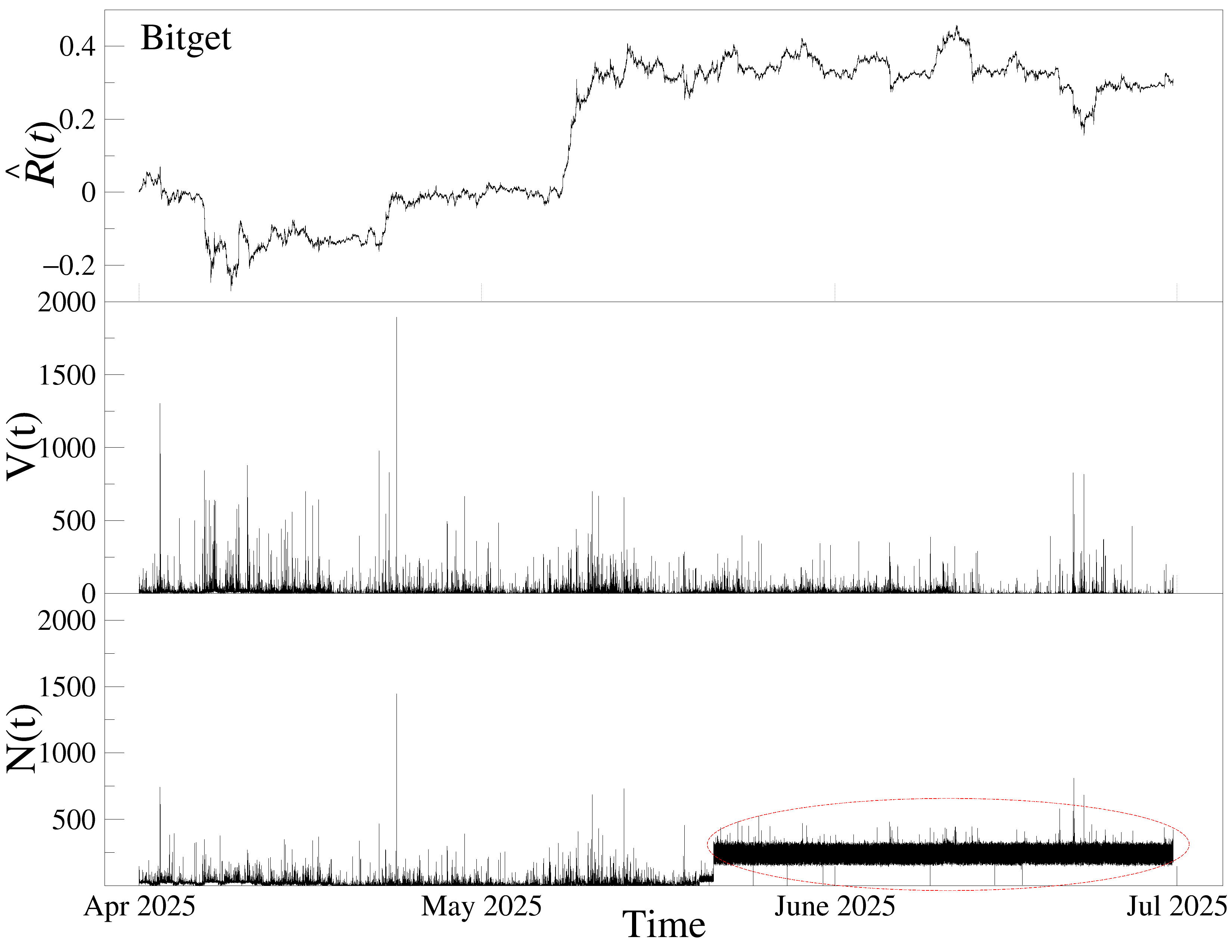}
\includegraphics[width=0.49\textwidth]{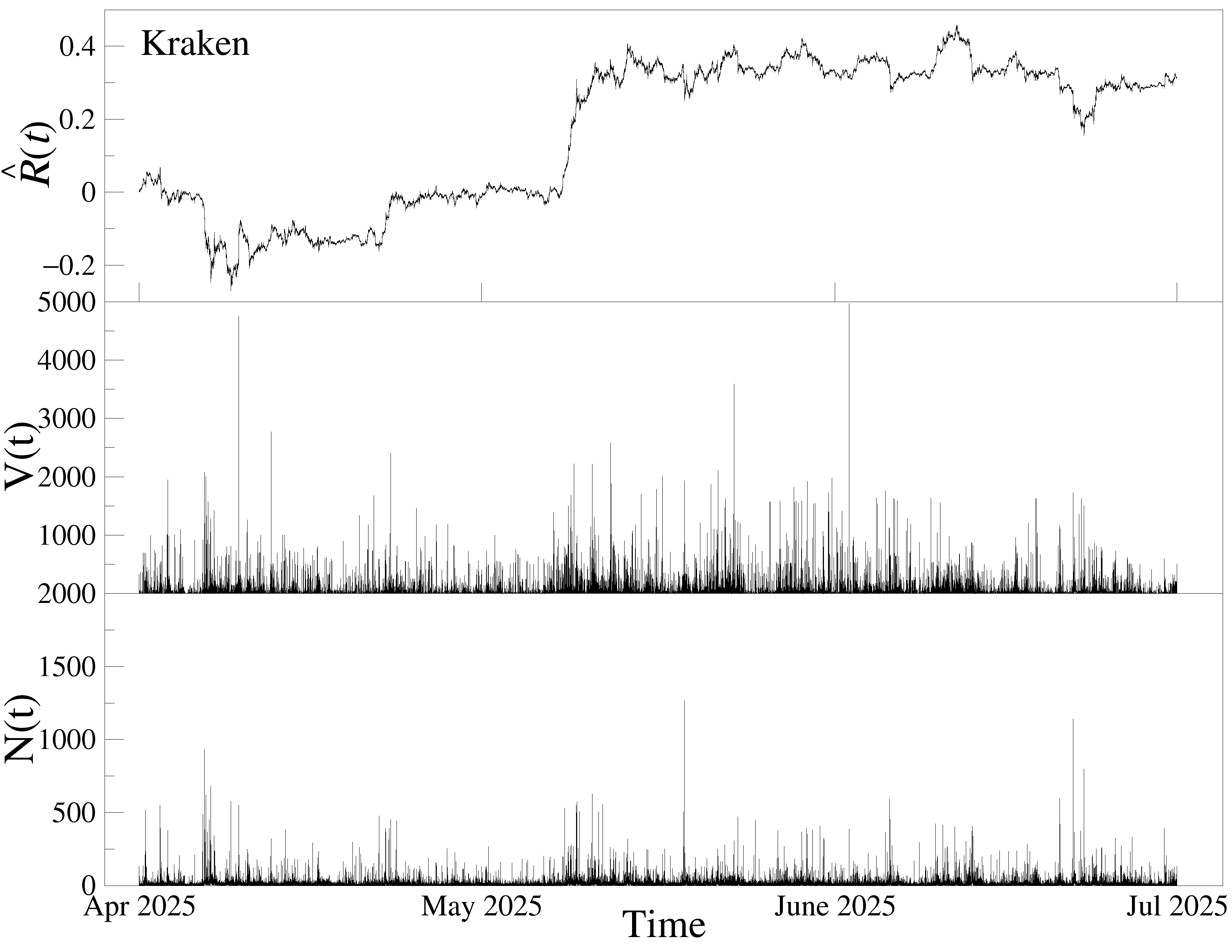}
\includegraphics[width=0.49\textwidth]{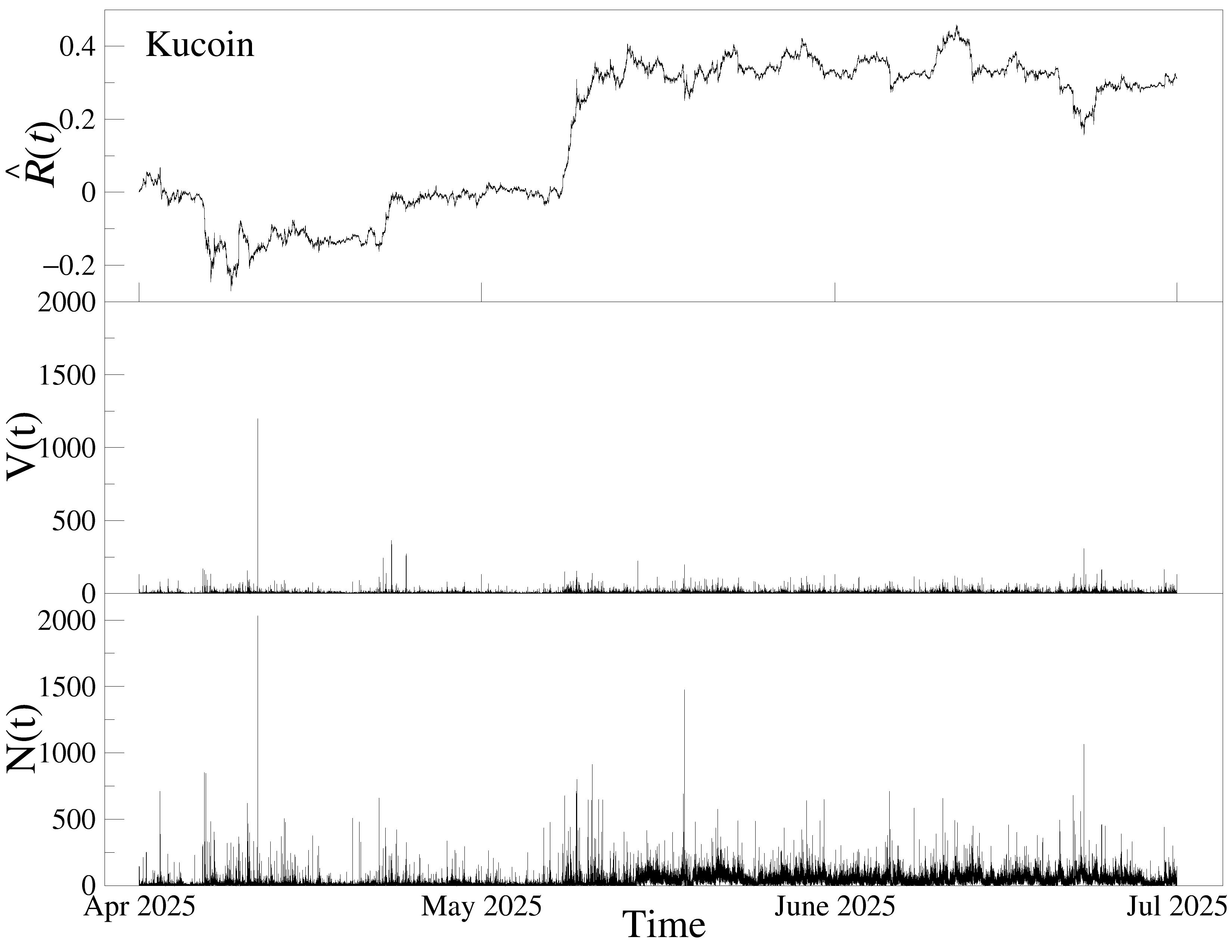}
\caption{The same as in Fig.~\ref{fig::szeregiBTC}, but for ETH.}
\label{fig::szeregiETH}
\end{figure}

\begin{figure}[ht!]
\centering
\includegraphics[width=0.49\textwidth]{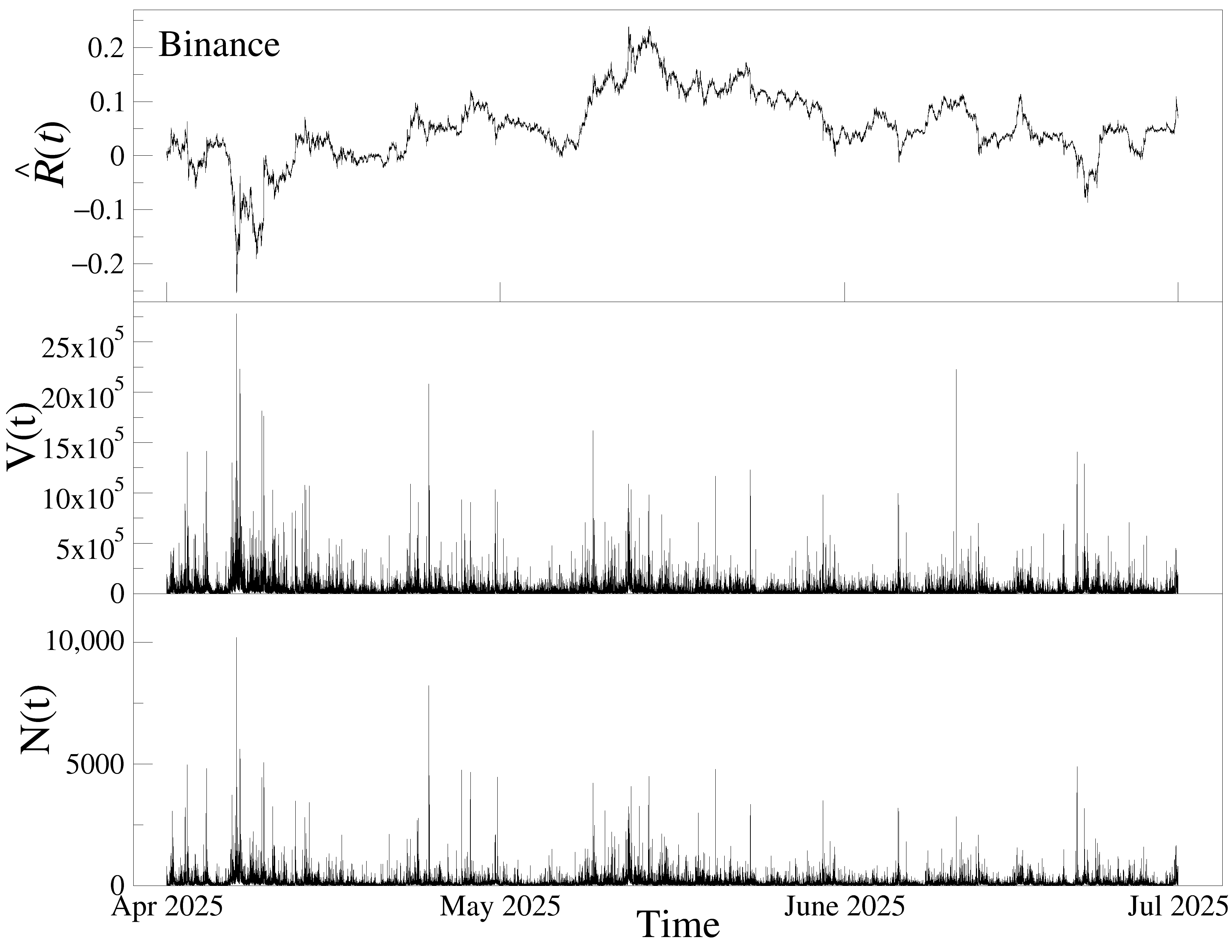}
\includegraphics[width=0.49\textwidth]{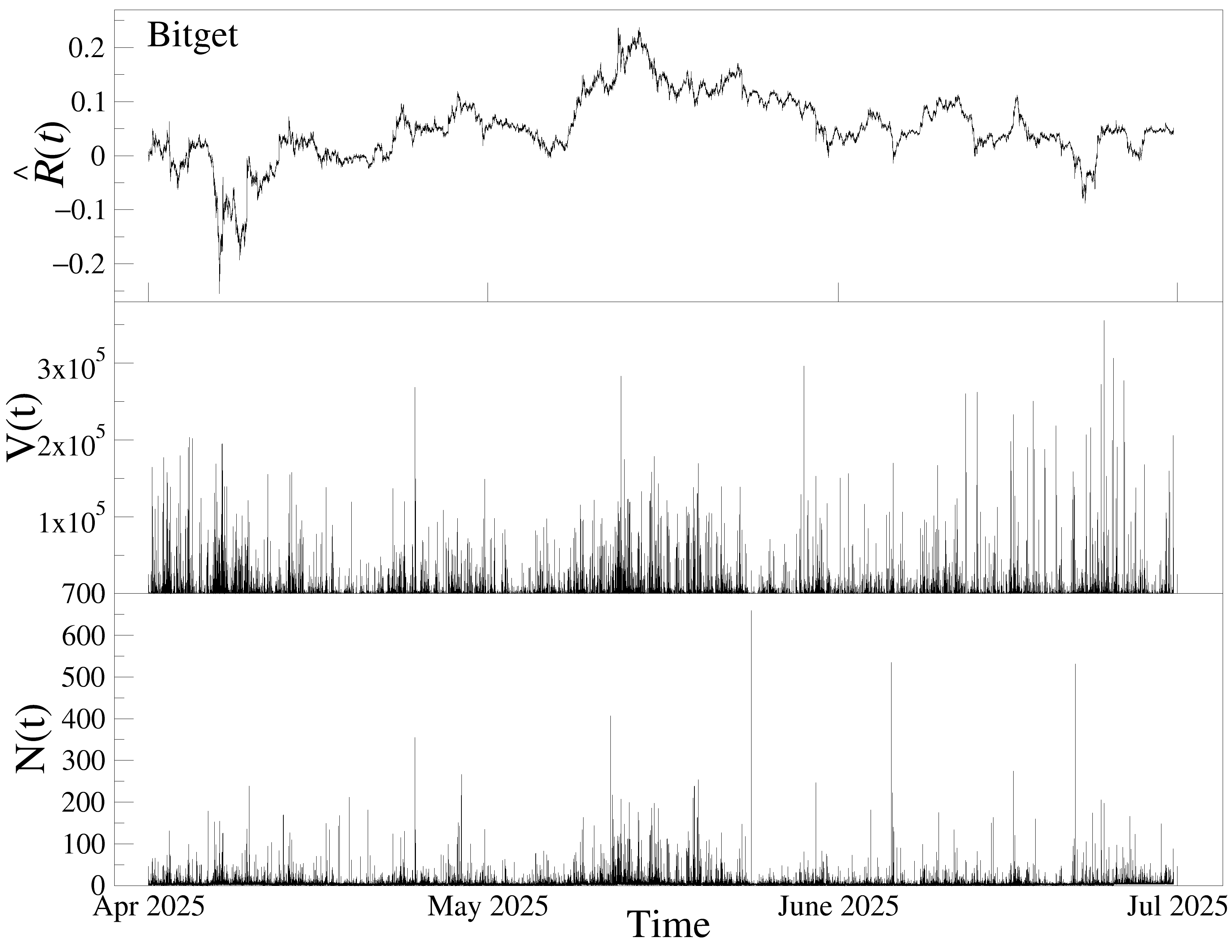}
\includegraphics[width=0.49\textwidth]{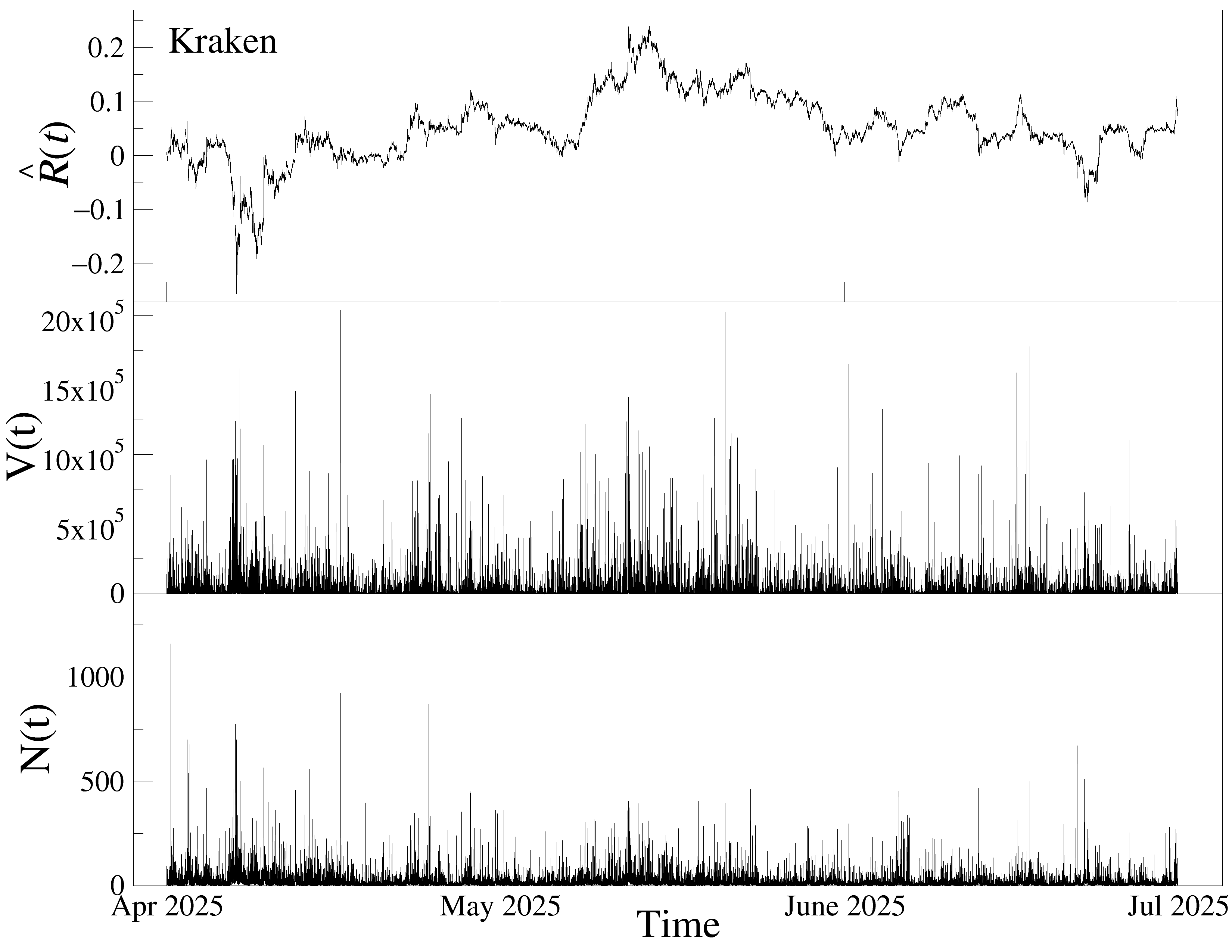}
\includegraphics[width=0.49\textwidth]{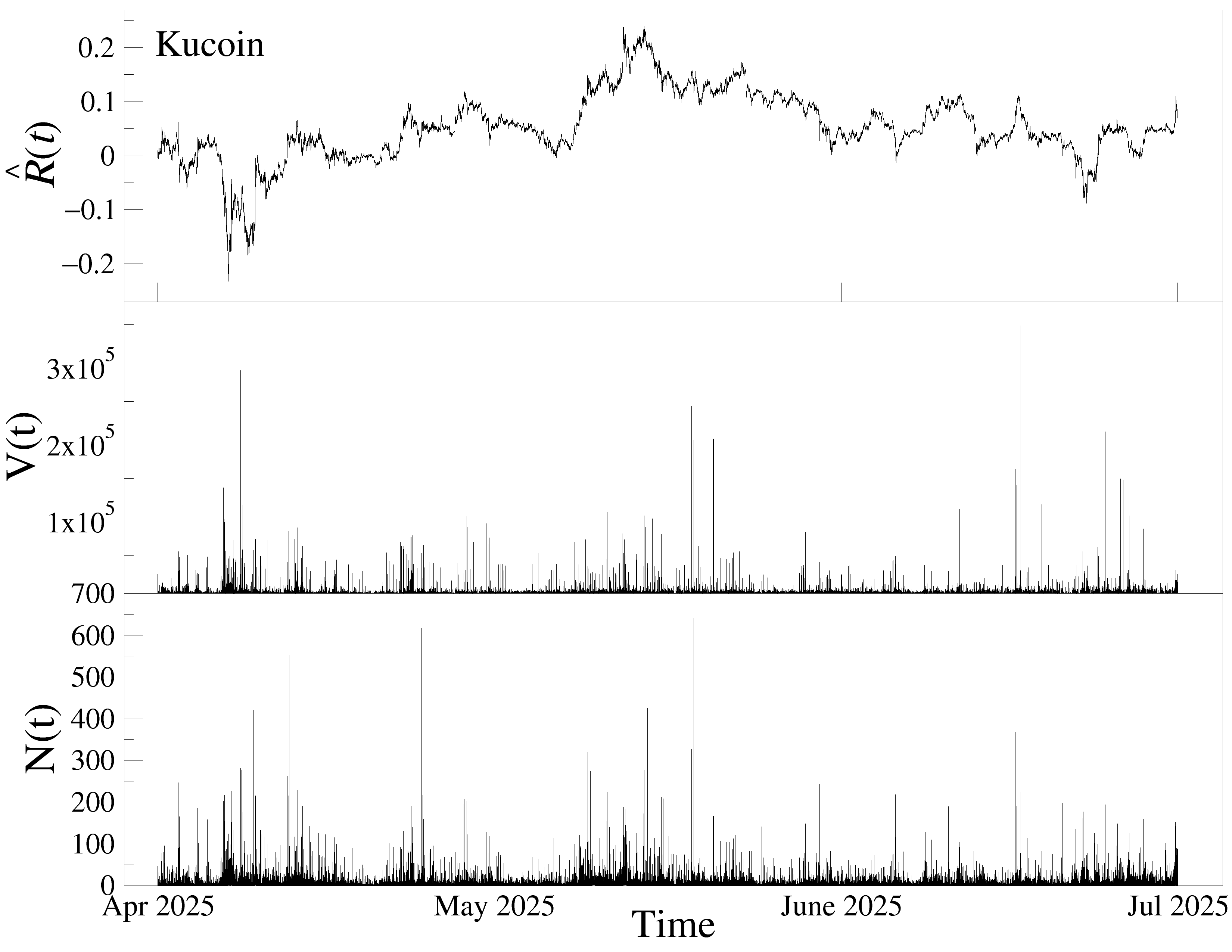}
\caption{The same as in Fig.~\ref{fig::szeregiBTC}, but for XRP.}
\label{fig::szeregiXRP}
\end{figure}

The temporal evolution of cumulative log-returns, ${\hat R}(t_i) = \sum_{k=1}^{i} R(t_k)$, $V_{\Delta t=1\mathrm{min}}(t)$, and $N_{\Delta t=1\mathrm{min}}(t)$ is presented for BTC, ETH, and XRP in Figs.~\ref{fig::szeregiBTC},~\ref{fig::szeregiETH}, and~\ref{fig::szeregiXRP}, respectively. The figures reveal that market activity is strongly intermittent. Large values of $V_{\Delta t=1\mathrm{min}}(t)$ and $N_{\Delta t=1\mathrm{min}}(t)$ occur in short bursts rather than being evenly distributed over time. These bursts often coincide with periods of more pronounced changes in cumulative returns, suggesting that stronger price movements are accompanied by increased trading activity. At the same time, this relationship is not homogeneous across exchanges or cryptocurrencies. For a given cryptocurrency, there are no substantial differences in the levels of cumulative returns between exchanges, whereas the corresponding volume and transaction-count series exhibit clear exchange-specific features. In particular, the temporal patterns of transaction counts and trading volume are similar across Binance and Kraken, although their magnitudes differ. By contrast, the corresponding patterns differ for KuCoin and, especially, for Bitget, where distinct regimes can be identified in the transaction-count series. This effect is particularly pronounced for BTC and ETH traded on Bitget. For these two cryptocurrencies, a significant increase in transactions per minute began on May 21, 2025. From that date until the end of the sample period, the number of transactions per minute does not fall below 150, except in a few cases. Interestingly, three of the moments with $N_{\Delta t=1\mathrm{min}}(t)<5$ occur simultaneously across both cryptocurrencies, suggesting a coordinated or systematic effect. This issue is examined in more detail in the following sections.

\section{Statistical characteristics}

\subsection{Return distributions}

The analysis begins with a comparison of the complementary cumulative distribution functions (CCDFs) of absolute log-returns, trading volume, and the number of transactions for BTC, ETH, and XRP on the four exchanges under consideration. For a random variable $X$, the CCDF is defined as $P(X>x)$. In the figures, empirical CCDFs are compared with reference heavy-tailed functions, including a power-law decay, $P(X>x)\sim x^{-\gamma}$, and a stretched exponential form, $P(X>x)\sim \exp(-x^\beta)$, where $\gamma$ is the tail exponent and $\beta$ is the stretching exponent~\cite{LaherrereJ-1998a}.

\begin{figure}[ht!]
\centering
\includegraphics[width=0.99\textwidth]{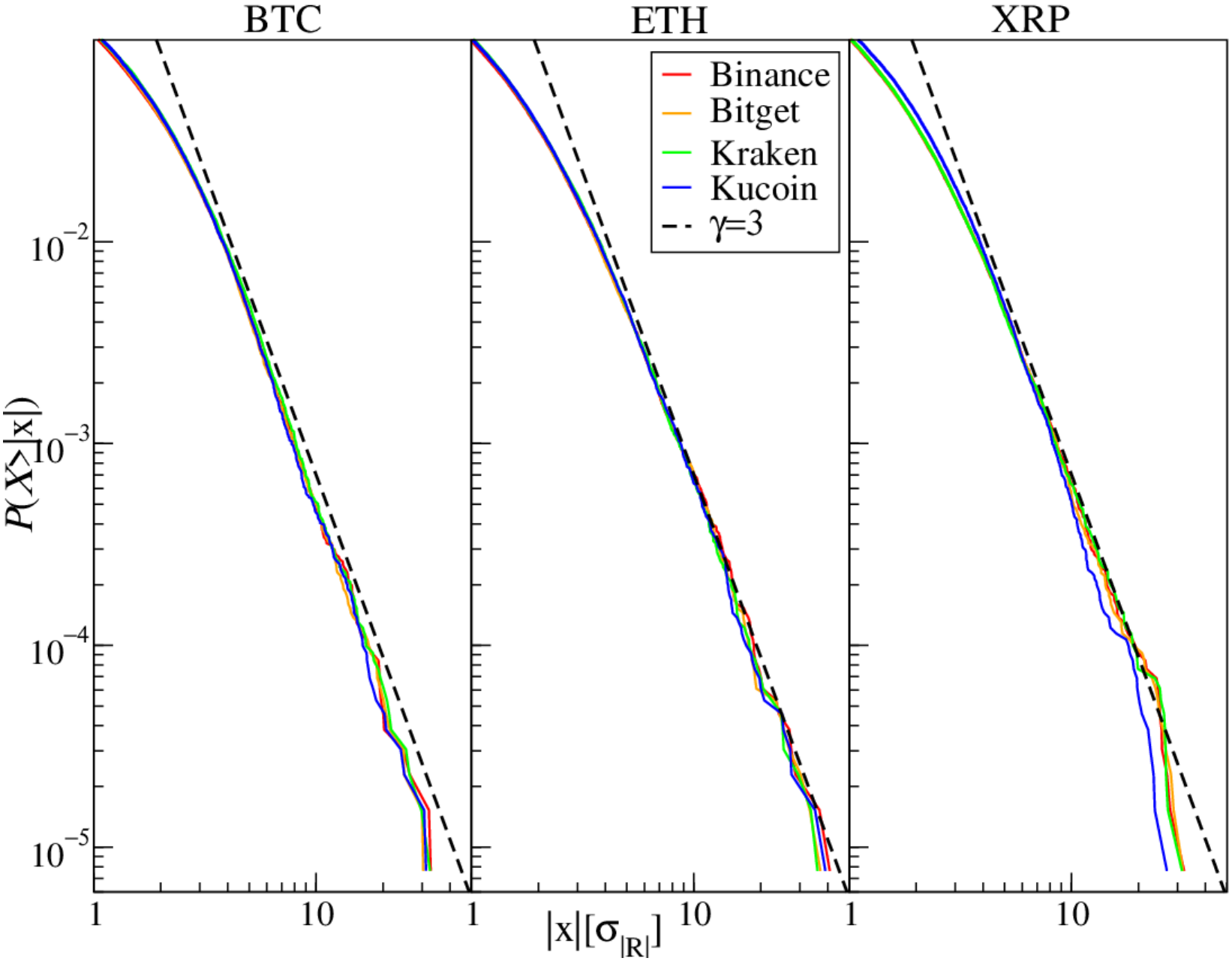}
\caption{Complementary cumulative distribution functions of standardised absolute log-returns $|R|$ for BTC, ETH, and XRP on Binance, Bitget, Kraken, and KuCoin, together with the Gaussian and power-law ($\gamma=3$) reference distributions.}
\label{fig::returnsR.pdf}
\end{figure}

\begin{figure}[ht!]
\centering
\includegraphics[width=0.99\textwidth]{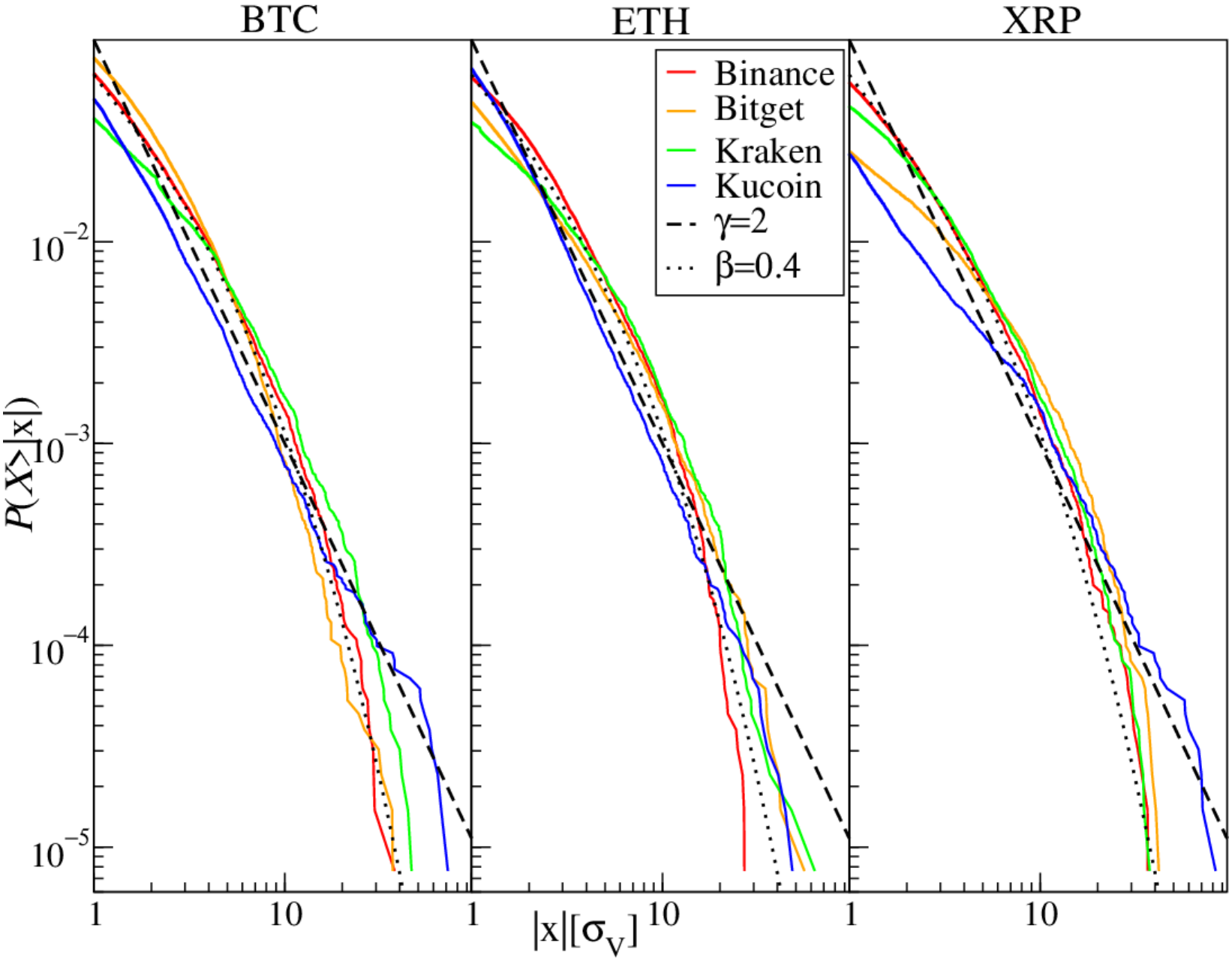}
\caption{Complementary cumulative distribution functions of standardised trading volume $V$ for BTC, ETH, and XRP on Binance, Bitget, Kraken, and KuCoin, together with reference distributions: the power-law distribution with $\gamma=2$ and the stretched exponential distribution with $\beta=0.4$.}
\label{fig::returnsV.pdf}
\end{figure}

\begin{figure}[ht!]
\centering
\includegraphics[width=0.99\textwidth]{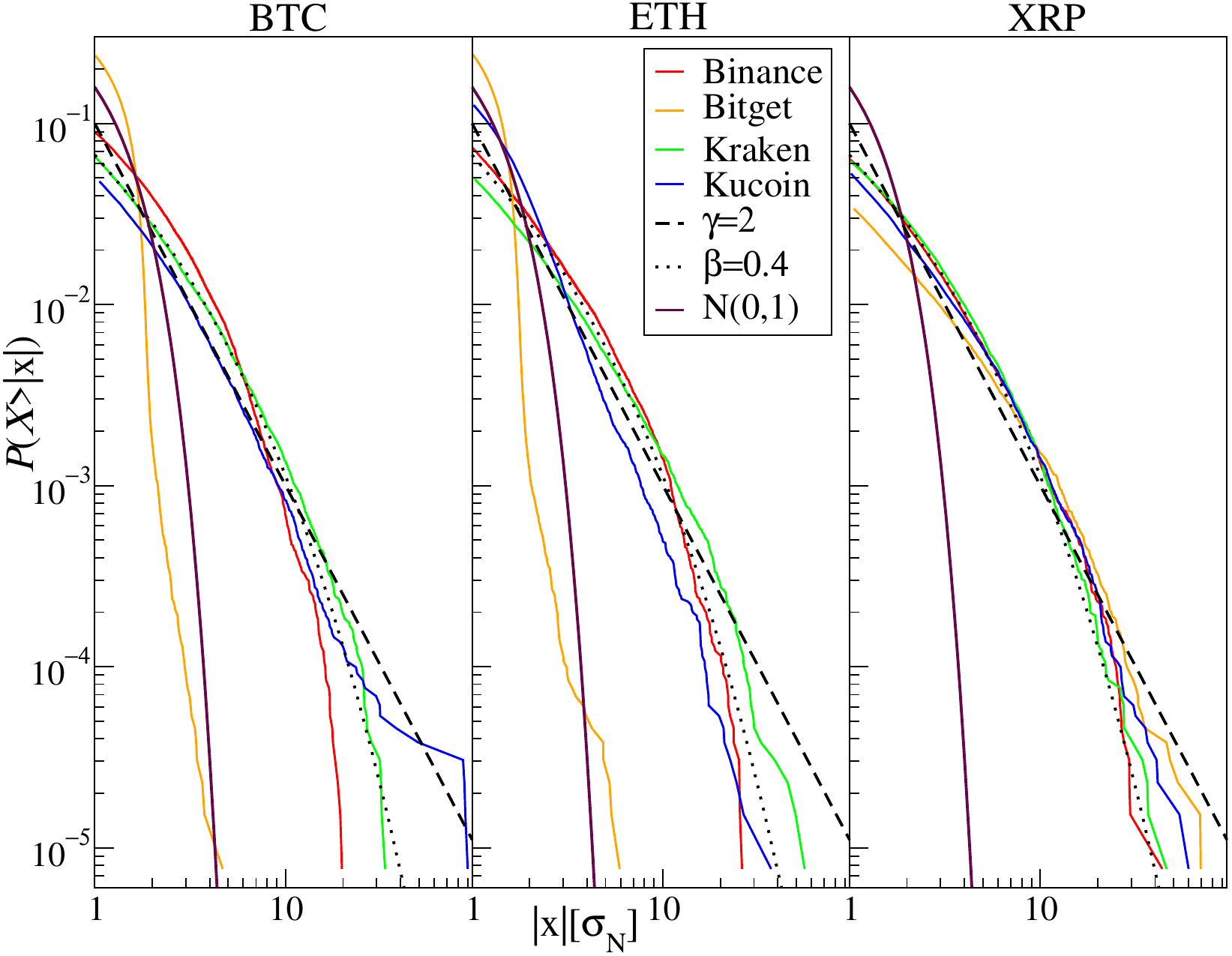}
\caption{Complementary cumulative distribution functions of the standardised number of transactions $N$ for BTC, ETH, and XRP on Binance, Bitget, Kraken, and KuCoin, together with the reference distributions: Gaussian, power law with $\gamma=2$, and stretched exponential with $\beta=0.4$.}
\label{fig::returnsN.pdf}
\end{figure}

Despite the differences in liquidity across exchanges reported in Tab.~\ref{tab::stats}, the 1-min distributions of standardised absolute log-returns are similar for each cryptocurrency on all exchanges, as shown in Fig.~\ref{fig::returnsR.pdf}. This indicates that, at the 1-min resolution, exchange-specific differences in price dynamics are largely suppressed. Such behaviour is consistent with the presence of arbitrage mechanisms that reduce discrepancies in exchange rates. Moreover, the tails of the return distributions are much heavier than those of the Gaussian distribution and are close to the power-law decay with exponent $\gamma=3$. This is consistent with the well-known heavy-tailed character of financial returns~\cite{MandelbrotBB-1963a,MantegnaStanley1995,Cont2001,WatorekM-2021a} and with the occurrence of the inverse cubic law at short time resolutions~\cite{GopikrishnanP-1998a,GopikrishnanP-1999a}.

The standardised volume distributions, presented in Fig.~\ref{fig::returnsV.pdf}, display more visible exchange-specific differences. Although their overall shapes are broadly similar, deviations are observed mainly in the tails. In particular, KuCoin exhibits a slightly thicker tail, which may be related to its significantly lower average volume $\langle V_{\Delta t} \rangle$ and to periods without trading, especially for BTC and XRP. In general, the volume distributions are clearly non-Gaussian and are better described by heavy-tailed reference functions, such as a power-law distribution with $\gamma=2$ or a stretched exponential distribution with $\beta=0.4$.

The largest differences between exchanges are observed for the distributions of the number of transactions, shown in Fig.~\ref{fig::returnsN.pdf}. In this case, Bitget stands out most clearly. For BTC and ETH, the distributions of $N_{\Delta t=1\mathrm{min}}(t)$ on Bitget are much narrower and closer to the Gaussian distribution than the corresponding distributions for the other exchanges. This reflects the specific temporal behaviour of $N_{\Delta t=1\mathrm{min}}(t)$ discussed in Sec.~\ref{Dataspec}, where a pronounced regime-like change in the number of transactions was observed. By contrast, the transaction-count distributions for Binance, Kraken, and KuCoin are heavy-tailed, indicating more intermittent trading activity with occasional bursts.

\subsection{Autocorrelation}

The specific behaviour of the transaction-count series on Bitget is further confirmed by the autocorrelation analysis. The autocorrelation function (ACF) is defined as
\begin{equation}
A(x,\tau) =\frac{\frac{1}{T-\tau}\sum_{i=1}^{T-\tau}\left[x(i)-\langle x\rangle\right]\left[x(i+\tau)-\langle x\rangle\right]}{\sigma_x^2},
\label{eq::acf}
\end{equation}
where $\langle x\rangle$ and $\sigma_x$ denote the estimated mean and standard deviation of the time series $x(i)$, respectively. The parameter $\tau$ is the lag expressed in data points, which corresponds to actual time $\tau\Delta t$.

\begin{figure}[ht!]
\centering
\includegraphics[width=0.99\textwidth]{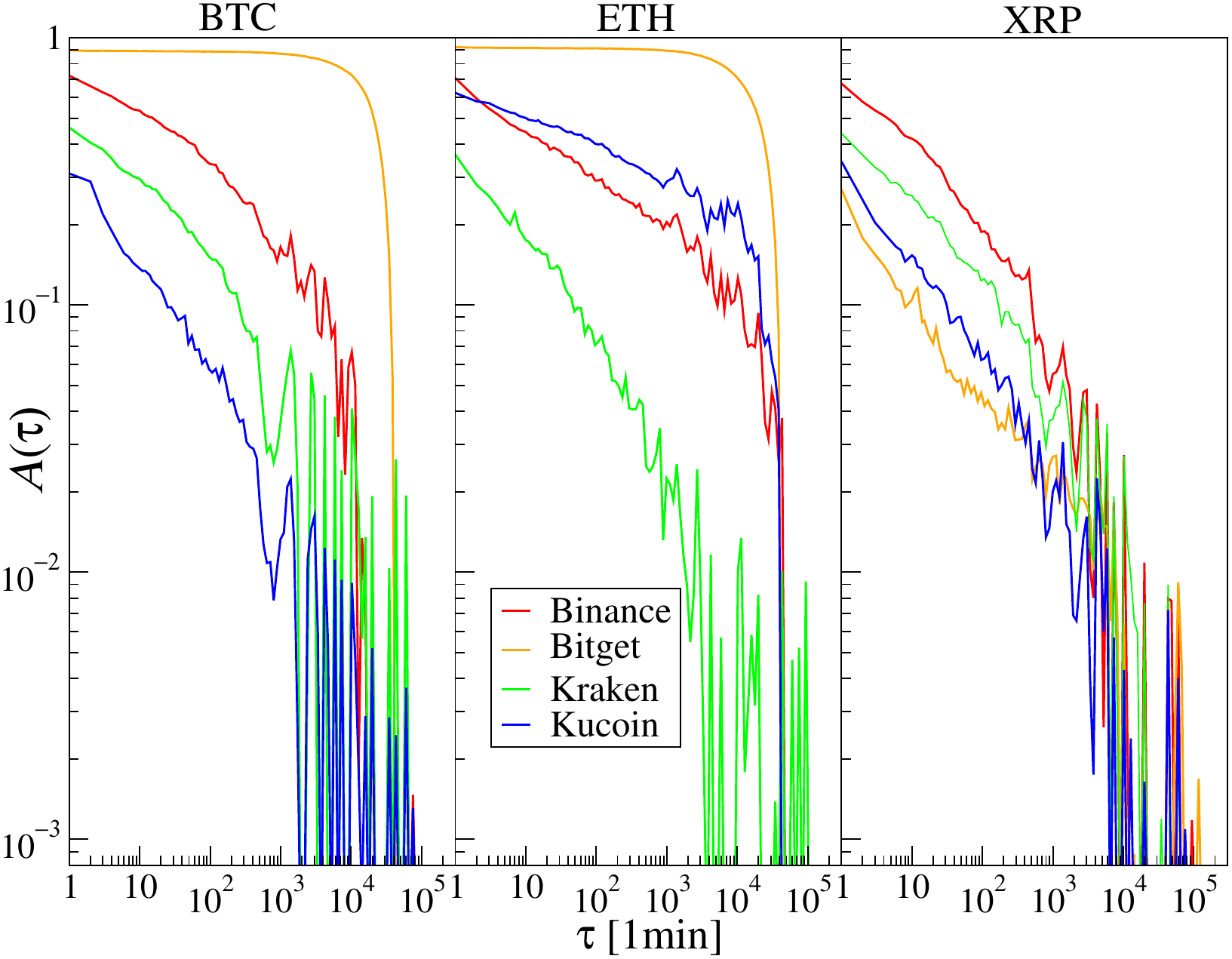}
\caption{Autocorrelation functions of the standardised number of transactions $N$ for BTC, ETH, and XRP on Binance, Bitget, Kraken, and KuCoin.}
\label{fig::acfN.pdf}
\end{figure}

As shown in Fig.~\ref{fig::acfN.pdf}, the ACF of $N_{\Delta t=1\mathrm{min}}(t)$ for BTC and ETH decays much more slowly on Bitget than on the other exchanges. In both cases, elevated correlations persist up to time lags of the order of $10^4$ min. This indicates the presence of a persistent temporal structure in the transaction process, which is consistent with the existence of the regimes observed in Figs.~\ref{fig::szeregiBTC},\ref{fig::szeregiETH}. This effect is not observed for XRP, for which the Bitget curve is close to those obtained for the remaining exchanges.

\begin{figure}[ht!]
\centering
\includegraphics[width=0.99\textwidth]{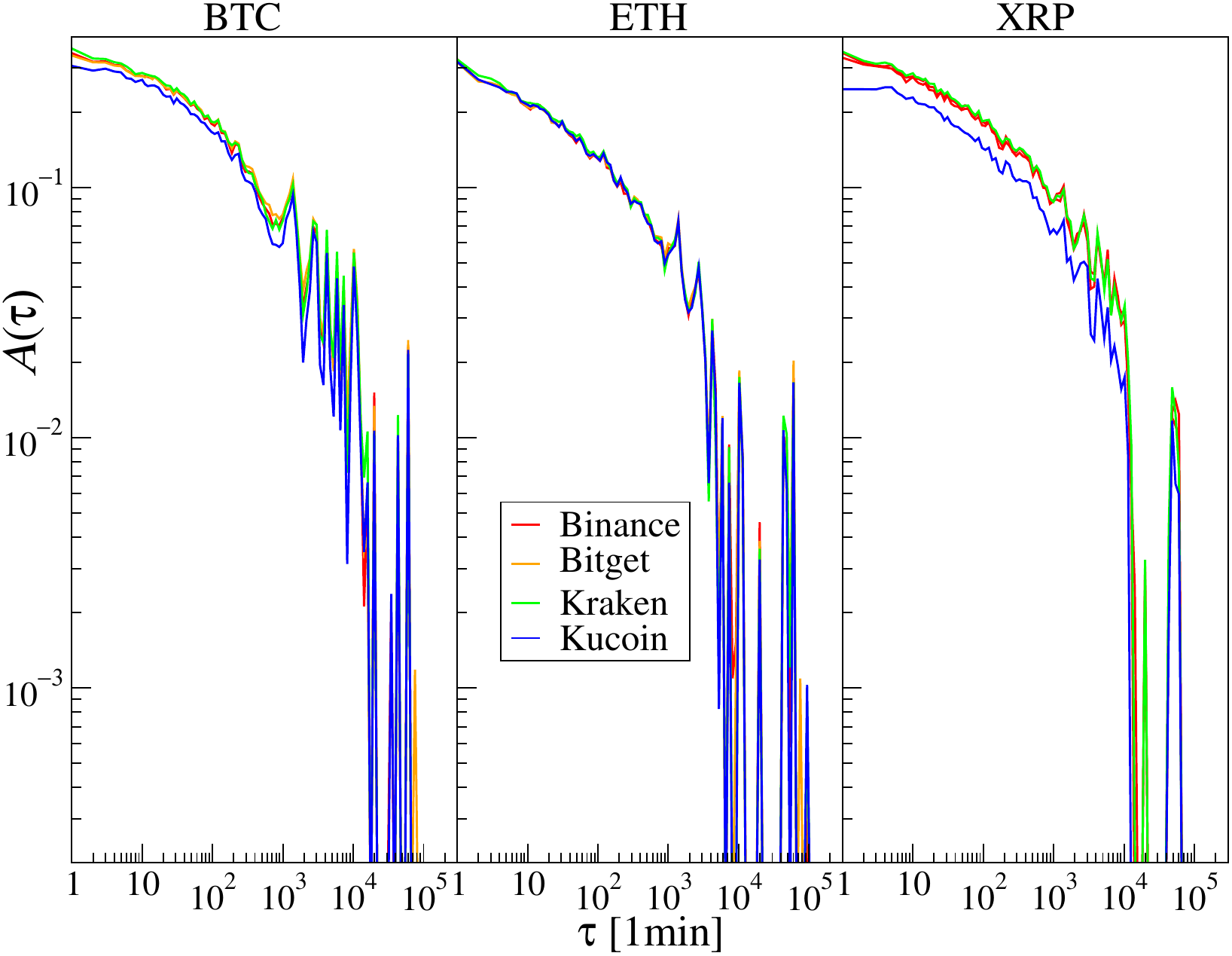}
\caption{Autocorrelation functions of standardised absolute log-returns $|R|$ for BTC, ETH, and XRP on Binance, Bitget, Kraken, and KuCoin.}
\label{fig::acfR.pdf}
\end{figure}

In contrast, the autocorrelation functions of absolute log-returns, shown in Fig.~\ref{fig::acfR.pdf}, are very similar across exchanges for each cryptocurrency. This suggests that, despite differences in liquidity and trading activity, the temporal structure of volatility is largely exchange-independent. The decay of ACF for $|R(t)|$ is relatively slow, which is consistent with the well-known volatility clustering effect~\cite{Ding1993,Cont2001}.

\begin{figure}[ht!]
\centering
\includegraphics[width=0.99\textwidth]{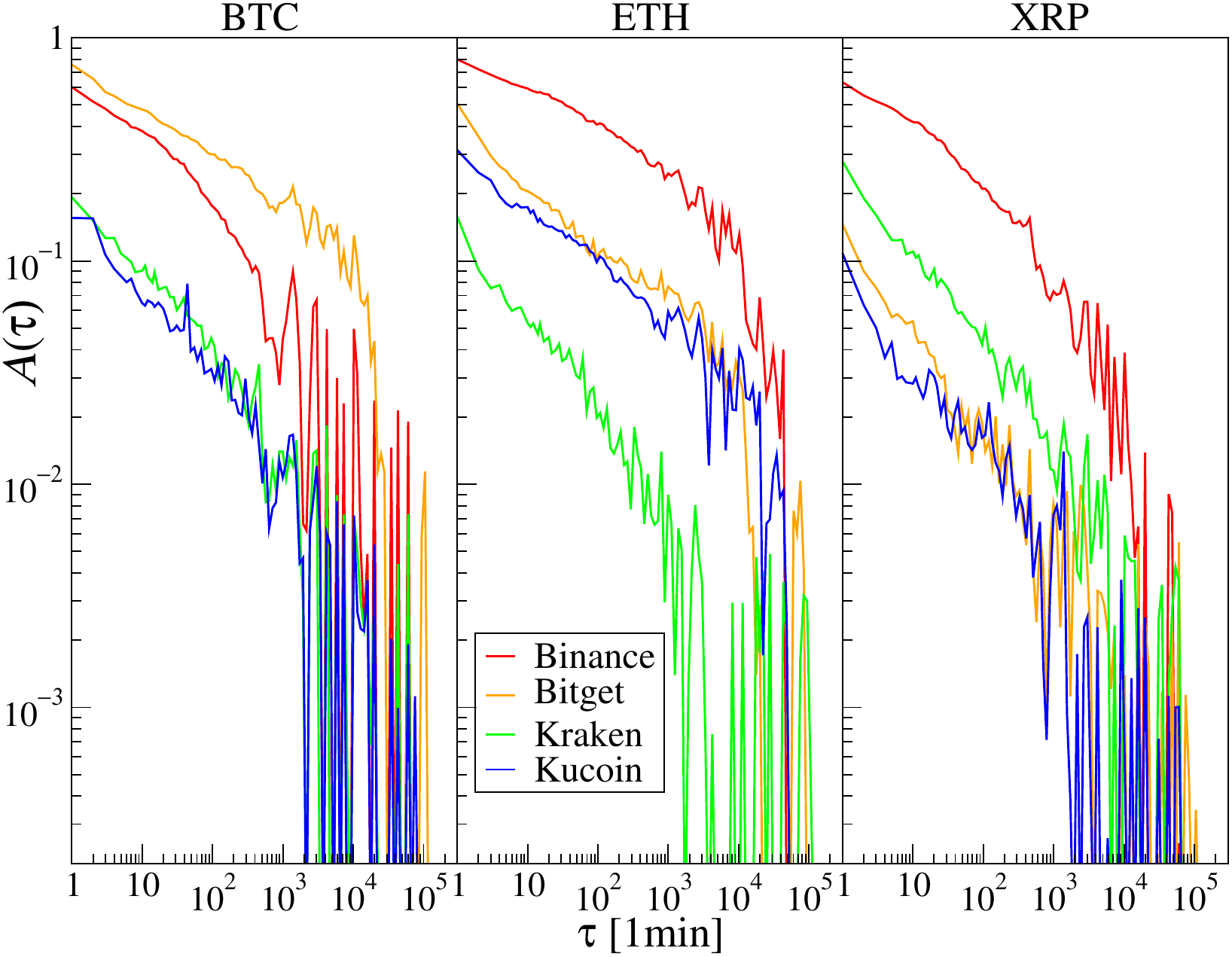}
\caption{Autocorrelation functions of standardised trading volume $V$ for BTC, ETH, and XRP on Binance, Bitget, Kraken, and KuCoin.}
\label{fig::acfV.pdf}
\end{figure}

The volume autocorrelation functions, presented in Fig.~\ref{fig::acfV.pdf}, show moderate differences between exchanges. Binance exhibits the strongest volume correlations for ETH and XRP, and the second strongest for BTC after Bitget. This indicates that trading volume is more sensitive to exchange-specific liquidity conditions than absolute log-returns. However, the exchange-specific differences in volume autocorrelations are less pronounced than those observed for the number of transactions.

\section{Multifractal characteristics}

The power-law decay of the autocorrelation functions discussed in the previous section suggests the presence of a non-trivial temporal organisation in the analysed time series. Such long-range dependence may indicate nonlinear correlations in the amplitudes of fluctuations, which are a necessary condition for multifractality in time series~\cite{KwapienJ-2023a}. Multifractality has been widely reported for financial time series~\cite{JiangZQ-2019a}, including cryptocurrencies~\cite{JamesN-2021a,JamesN-2022a,watorekfutnet2022,Brouty2024,BuiHQ-2025a,Drozdz2025futnet}, and is often interpreted as an indicator of complex market dynamics and market maturity~\cite{DrozdzS-2018a,DrozdzS-2023a}.

\begin{figure}[ht!]
\centering
\includegraphics[width=0.99\textwidth]{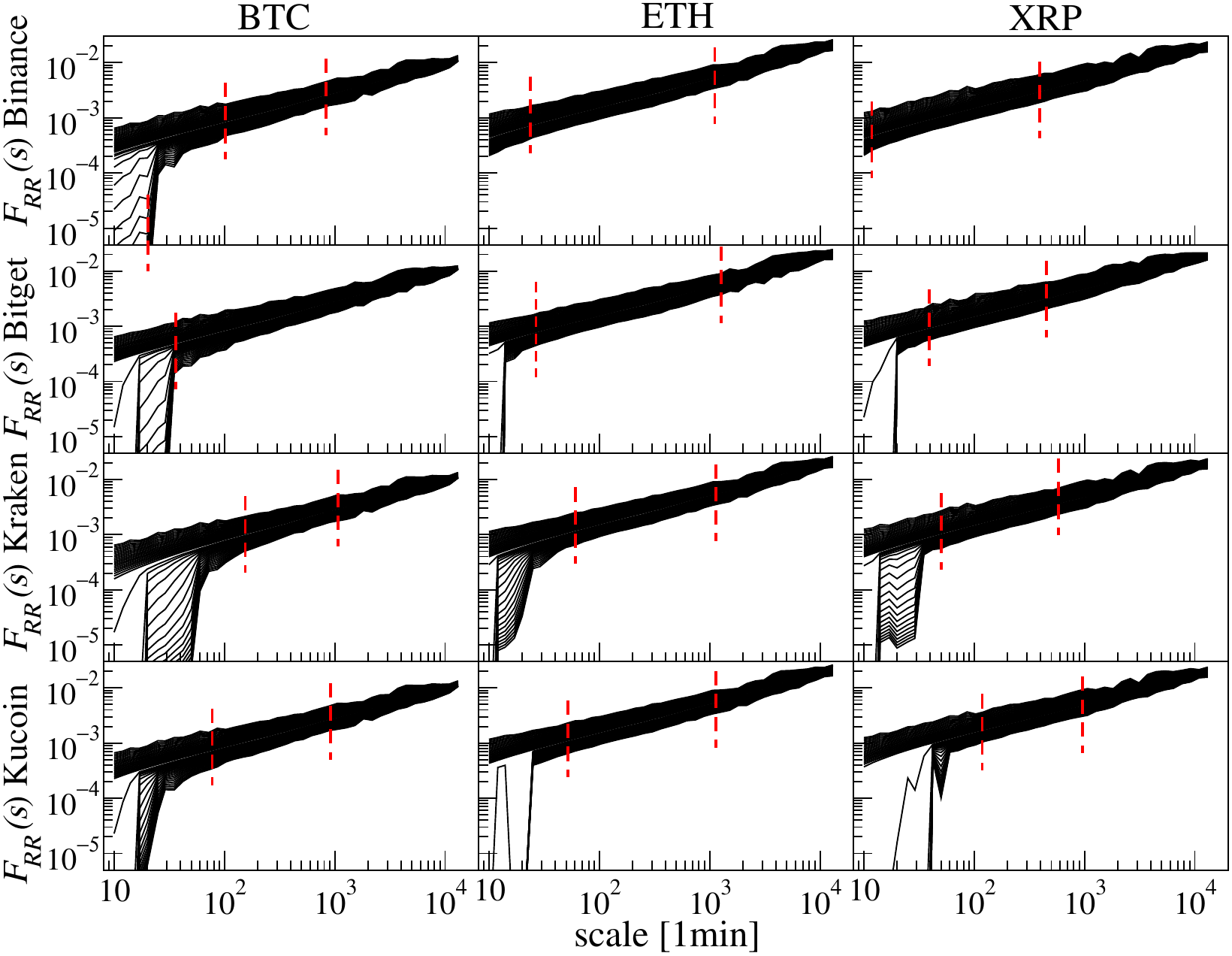}
\caption{Univariate fluctuation functions $F_{RR}(s)$ for BTC, ETH, and XRP log-returns $R$ on Binance, Bitget, Kraken, and KuCoin. Dashed red lines indicate the scale range selected for determining the multifractal spectra.}
\label{fig::FqR}
\end{figure}

\begin{figure}[ht!]
\centering
\includegraphics[width=0.99\textwidth]{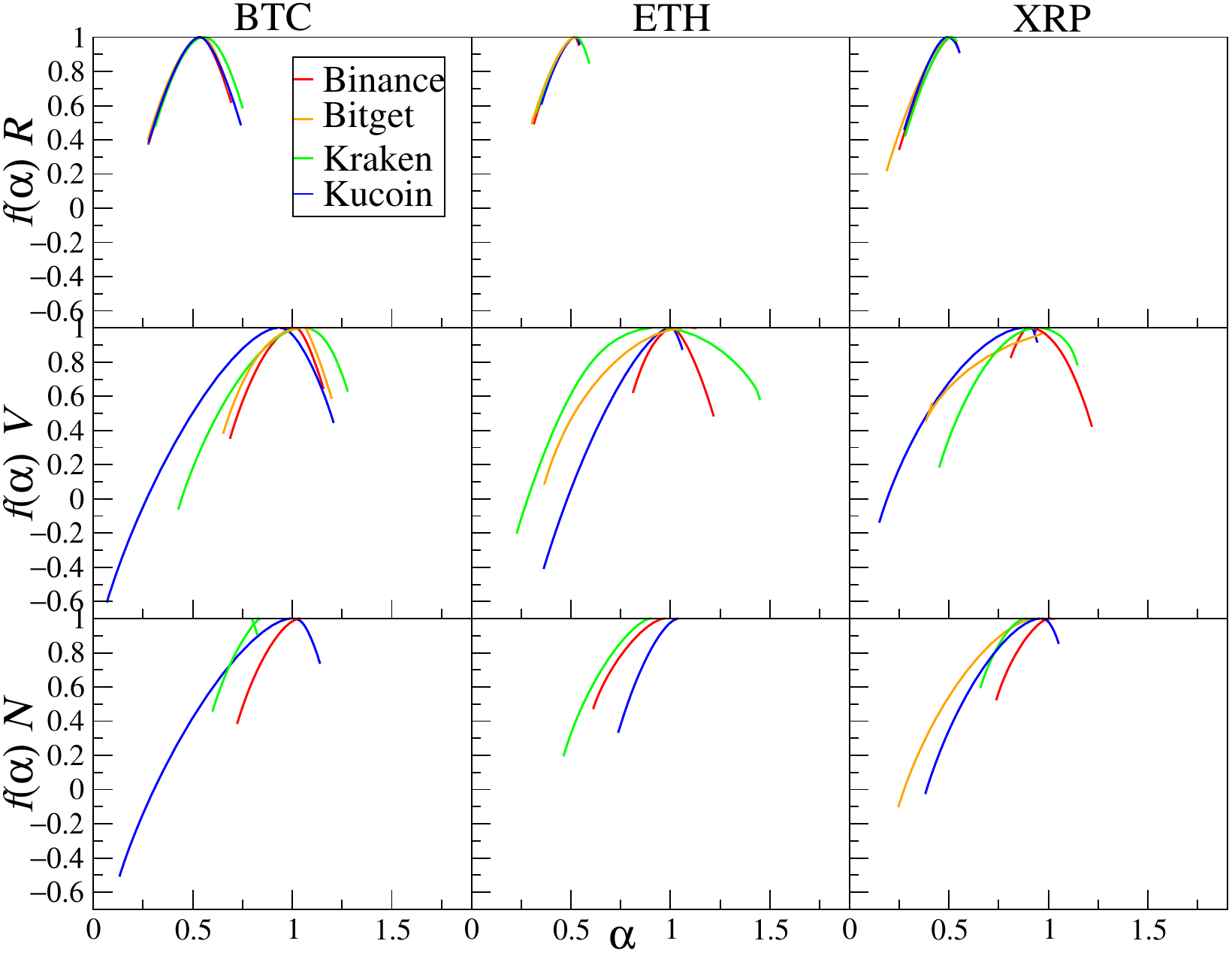}
\caption{Multifractal spectra for BTC, ETH, and XRP log-returns $R$ (top panels), trading volume $V$ (middle panels), and number of transactions $N$ (bottom panels) on Binance, Bitget, Kraken, and KuCoin.}
\label{fig::spectra}
\end{figure}

For the log-return series, the fluctuation functions $F_{RR}(s)$ shown in Fig.~\ref{fig::FqR} display clear scaling over the selected range of time scales. This allows for a reliable estimation of the multifractal spectra, presented in the top panels of Fig.~\ref{fig::spectra}. The spectra obtained for a given cryptocurrency are highly similar across exchanges, which indicates that the organisation of log-return fluctuations is largely exchange-independent. This result is consistent with the similarity of the return distributions and autocorrelation functions discussed in the previous sections. The spectra are left-sided asymmetric, as is typically observed for financial time series~\cite{JiangZQ-2019a,DrozdzS-2023a}. Such asymmetry is mainly associated with the heavy tails of the return distributions, which enhance the multifractality of large fluctuations represented by the left arm of the spectrum~\cite{Drozdz2025futnet,kluszczynski2025}. Among the analysed cryptocurrencies, BTC exhibits the widest right arm of the spectrum on all exchanges, suggesting the strongest organisation of small return fluctuations.

\begin{figure}[ht!]
\centering
\includegraphics[width=0.99\textwidth]{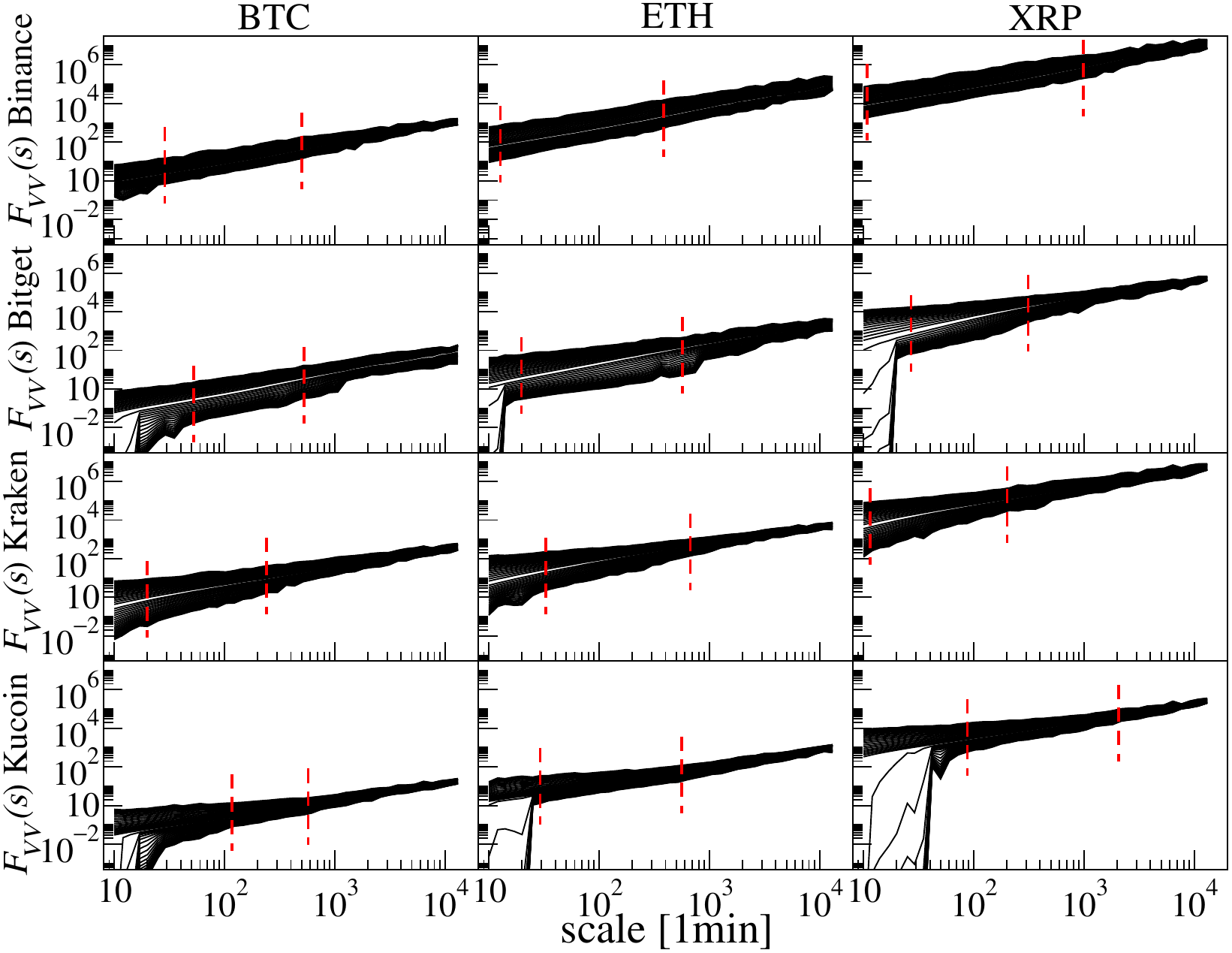}
\caption{Univariate fluctuation functions $F_{VV}(s)$ for BTC, ETH, and XRP trading volume $V$ on Binance, Bitget, Kraken, and KuCoin. Dashed red lines indicate the scale range selected for determining the multifractal spectra.}
\label{fig::FqV}
\end{figure}

More pronounced exchange-specific differences are observed for trading volume. The corresponding fluctuation functions $F_{VV}(s)$ are shown in Fig.~\ref{fig::FqV}, while the multifractal spectra are presented in the middle panels of Fig.~\ref{fig::spectra}. For $q<0$, the scaling quality is weaker for Bitget and KuCoin, especially for ETH and XRP. As a consequence, the right arm of the spectrum is poorly developed or absent in these cases, which indicates that small volume fluctuations are less regularly organised. In contrast, the left arm of the volume spectra is much longer than in the case of log-returns. This reflects the stronger contribution of large fluctuations and is consistent with the thicker tails of the volume distributions. This effect is particularly visible for KuCoin, whose volume distributions are close to a power-law form with exponent $\gamma \approx 2$ (Fig.~\ref{fig::returnsV.pdf}), whereas the return distributions are closer to a power law with exponent near $\gamma=3$ (Fig.~\ref{fig::returnsR.pdf}). In general, a thinner distribution tail corresponds to a shorter left arm of the multifractal spectrum, which is consistent with the role of broad distributions in strengthening multifractality~\cite{Drozdz2025futnet,kluszczynski2025}. The shortest left arm is observed for Binance, where the volume distributions are the thinnest and are close to a stretched exponential form with $\beta \approx 0.4$. At the same time, Binance exhibits the longest right arm of the volume spectrum, suggesting a more pronounced organisation of small volume fluctuations. This may be related to the substantially higher liquidity of this exchange as reported in Tab.~\ref{tab::stats}.

\begin{figure}[ht!]
\centering
\includegraphics[width=0.99\textwidth]{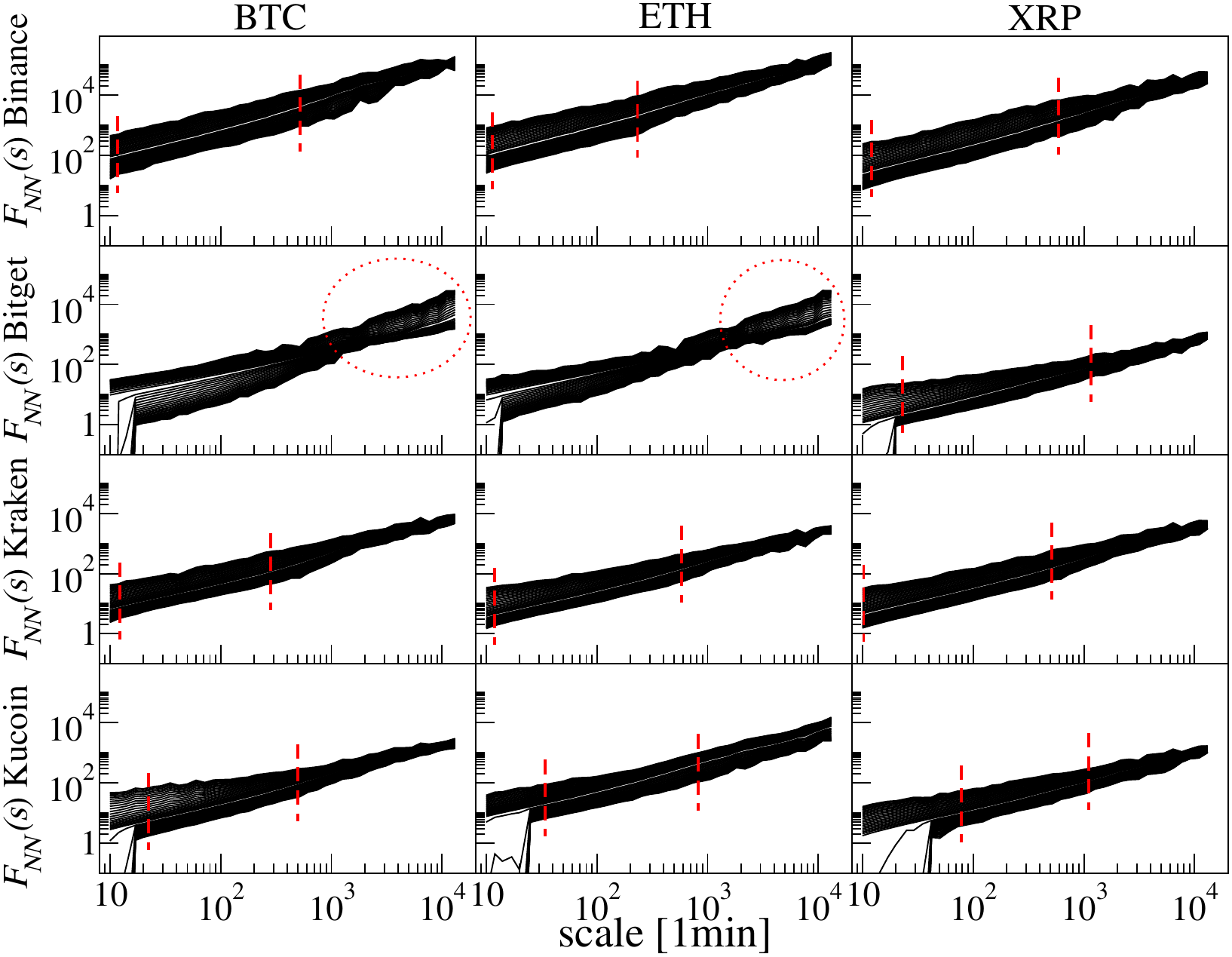}
\caption{Univariate fluctuation functions $F_{NN}(s)$ for BTC, ETH, and XRP number of transactions $N$ on Binance, Bitget, Kraken, and KuCoin. Dashed red lines indicate the scale range selected for determining the multifractal spectra. Red dashed ellipses mark the distorted scaling region observed for BTC and ETH on Bitget.}
\label{fig::FqN}
\end{figure}

The multifractal properties of the number of transactions are particularly informative. As shown in the bottom panels of Fig.~\ref{fig::spectra}, the spectra differ strongly between the exchanges. Despite the exceptionally slow decay of the autocorrelation function of $N_{\Delta t=1\mathrm{min}}(t)$ for BTC and ETH on Bitget (Fig.~\ref{fig::acfN.pdf}), the scaling quality of the corresponding fluctuation functions $F_{NN}(s)$ is the weakest (Fig.~\ref{fig::FqN}). In these two cases, the scaling is distorted at larger time scales, preventing a reliable estimate of the multifractal spectrum. The deformation of $F_{NN}(s)$, marked by the red dashed ellipses in Fig.~\ref{fig::FqN}, suggests a systematic disturbance in the temporal organisation of the transaction-count process. This is consistent with the regime-change behaviour of $N_{\Delta t=1\mathrm{min}}(t)$ observed earlier for Bitget. Interestingly, the same effect is not observed for XRP on Bitget. In this case, the scaling of $F_{NN}(s)$ and the resulting multifractal spectrum are similar to those obtained for the other exchanges. This suggests that the disturbance is not a general property of Bitget as a whole, but rather a token-specific effect affecting BTC and ETH. For all other transaction-count spectra, the length of the left arm of the spectrum is again related to the thickness of the distribution tails. The longest left arms are observed for the tokens and exchanges with the broadest transaction-count distributions: BTC on KuCoin, ETH on Kraken, and XRP on Bitget. In contrast, the right-hand side of the spectrum is very short in almost all cases, indicating that the organisation of small fluctuations in the number of transactions is close to monofractal.

\section{Cross-correlations}

\subsection{Scatter plots}

In the previous sections, the statistical properties of the individual time series $R_{\Delta t=1\mathrm{min}}(t)$, $V_{\Delta t=1\mathrm{min}}(t)$, and $N_{\Delta t=1\mathrm{min}}(t)$ were analysed separately. In this section, the relationships between these variables are examined. It is well known that, in financial markets, returns, trading volume, and the number of transactions are interrelated through trading activity, liquidity conditions, and the price formation process~\cite{Farmer2004,RakR-2015a}. In particular, periods of high trading intensity are typically associated with larger trading volume and with an increased probability of larger absolute price changes~\cite{gillemot2006there,podobnik2009cross}. One of the mechanisms behind this dependence is price impact, i.e., the effect whereby incoming order flow affects transaction prices~\cite{BouchaudJP-2010a}. Large or persistent buy and sell pressure may consume liquidity available at the best quotes and move prices, implying that intense trading activity can be linked to both larger turnover and stronger return fluctuations~\cite{Farmer2004}. A positive relationship between trading volume and the number of transactions is therefore also expected.

\begin{figure}[ht!]
    \centering
    \includegraphics[width=0.49\textwidth]{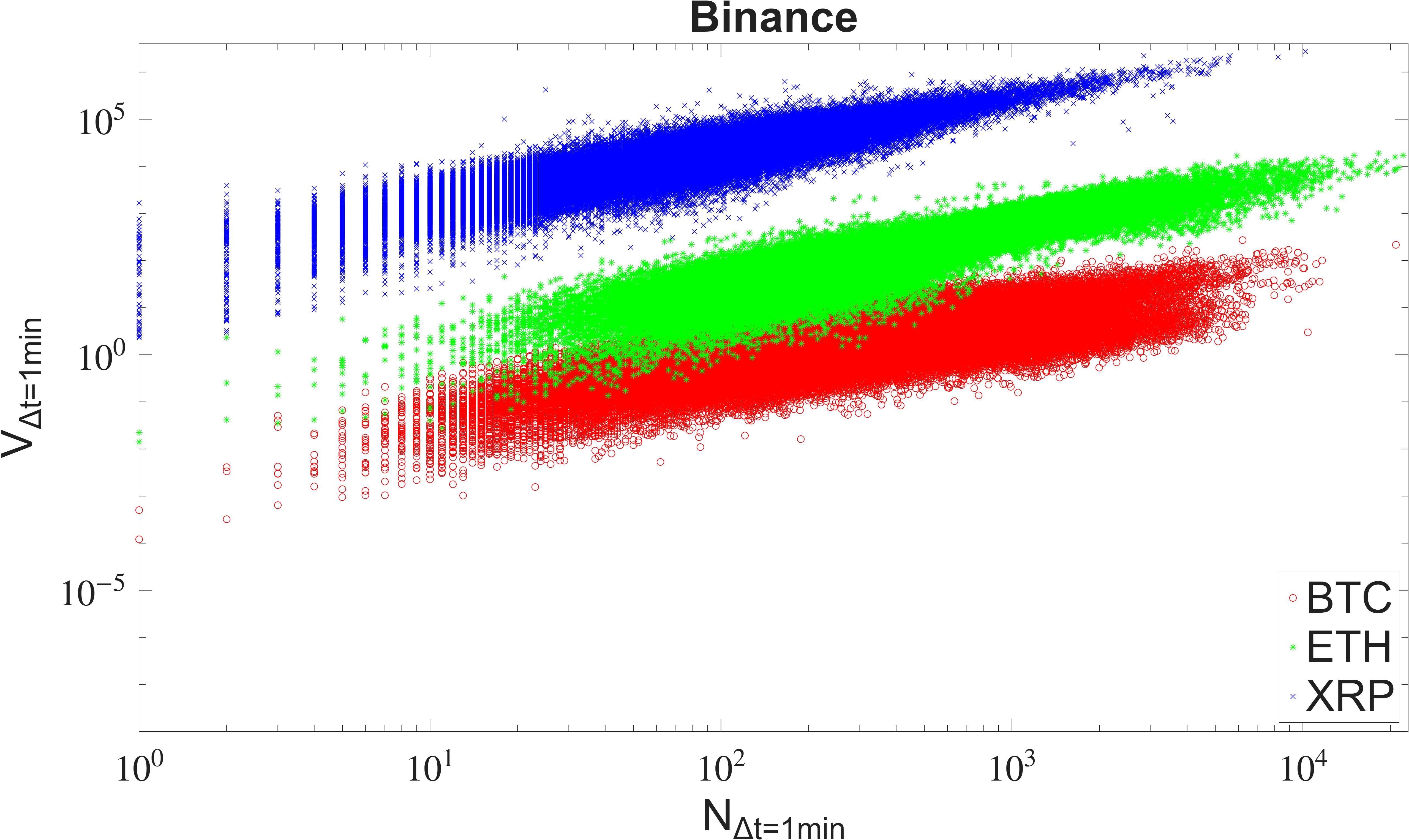}
     \includegraphics[width=0.49\textwidth]{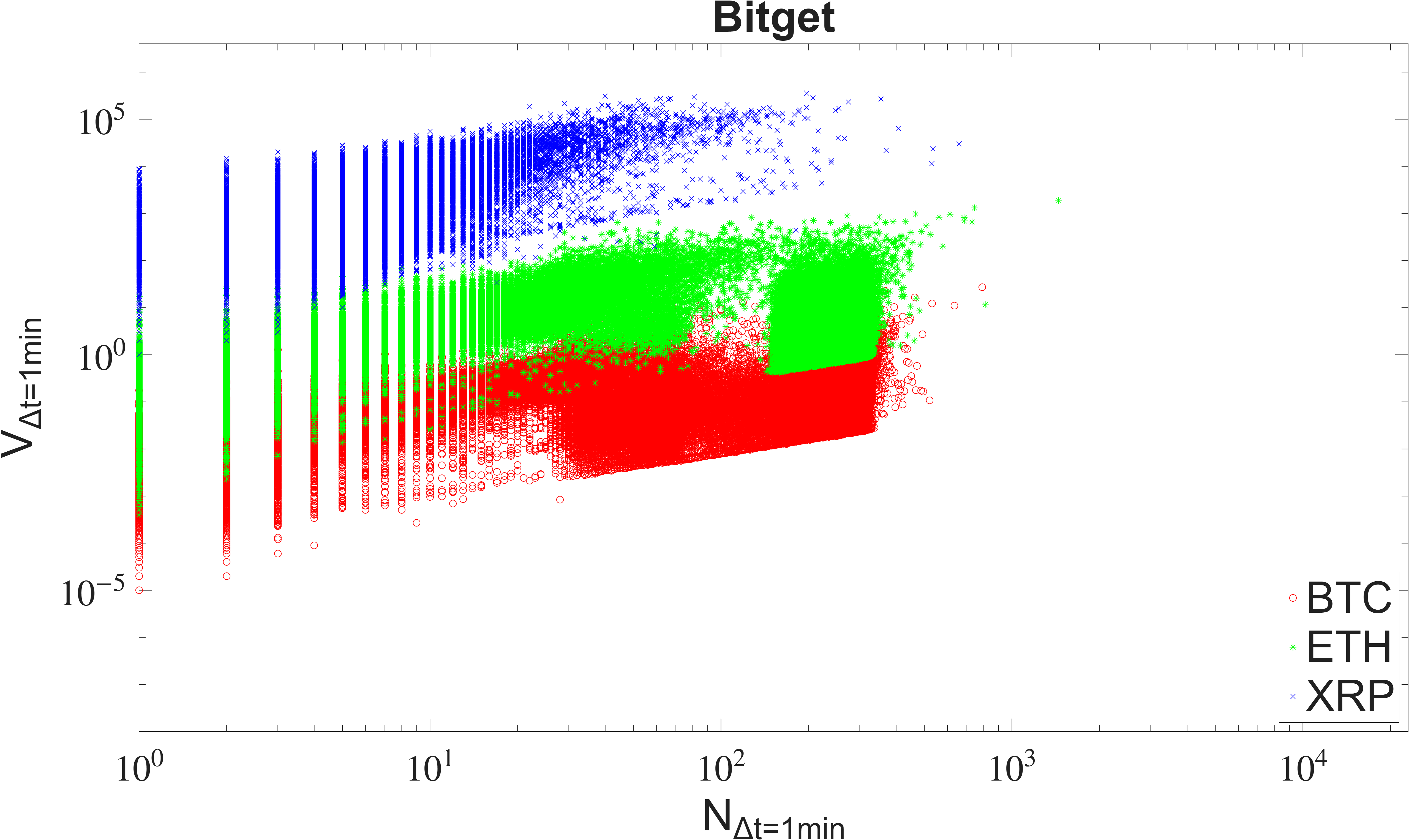}
      \includegraphics[width=0.49\textwidth]{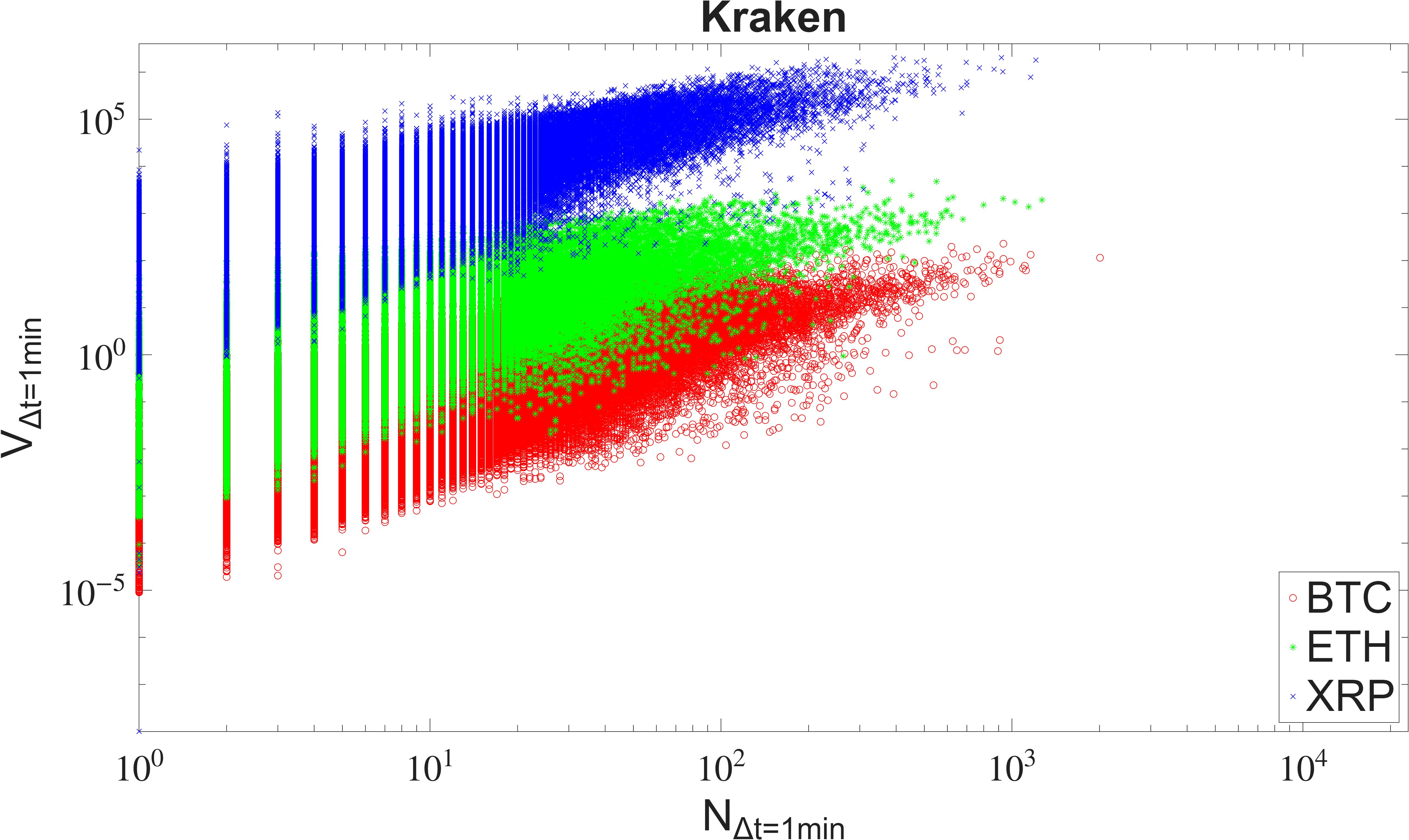}
       \includegraphics[width=0.49\textwidth]{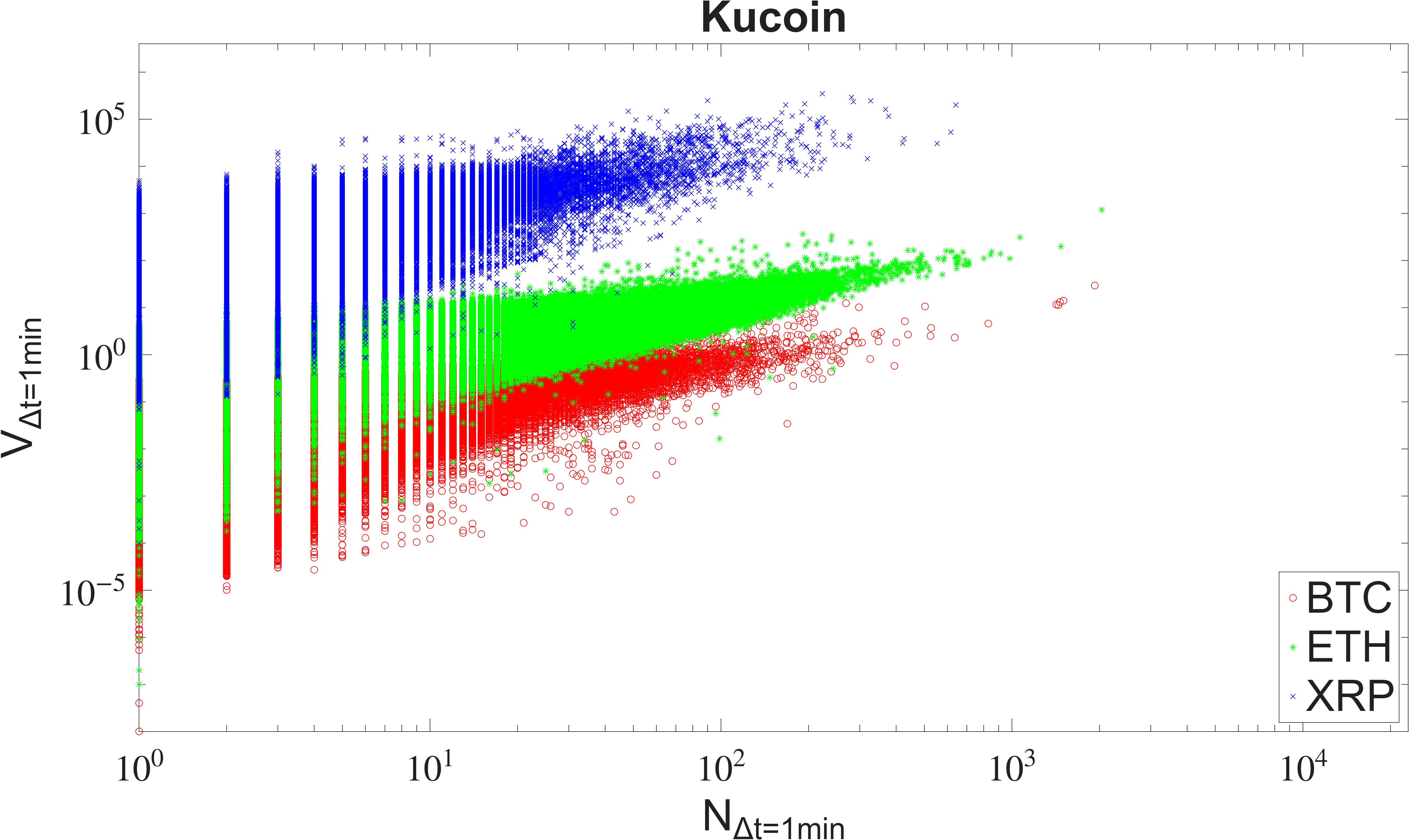}
       \caption{Relationship between trading volume $V_{\Delta t=1\mathrm{min}}(t)$ and the number of transactions $N_{\Delta t=1\mathrm{min}}(t)$ on Binance, Bitget, Kraken, and KuCoin for BTC, ETH, and XRP. Each point corresponds to a 1-min interval.}
    \label{fig:N_V}
\end{figure}

\begin{figure}[ht!]
    \centering
    \includegraphics[width=0.49\textwidth]{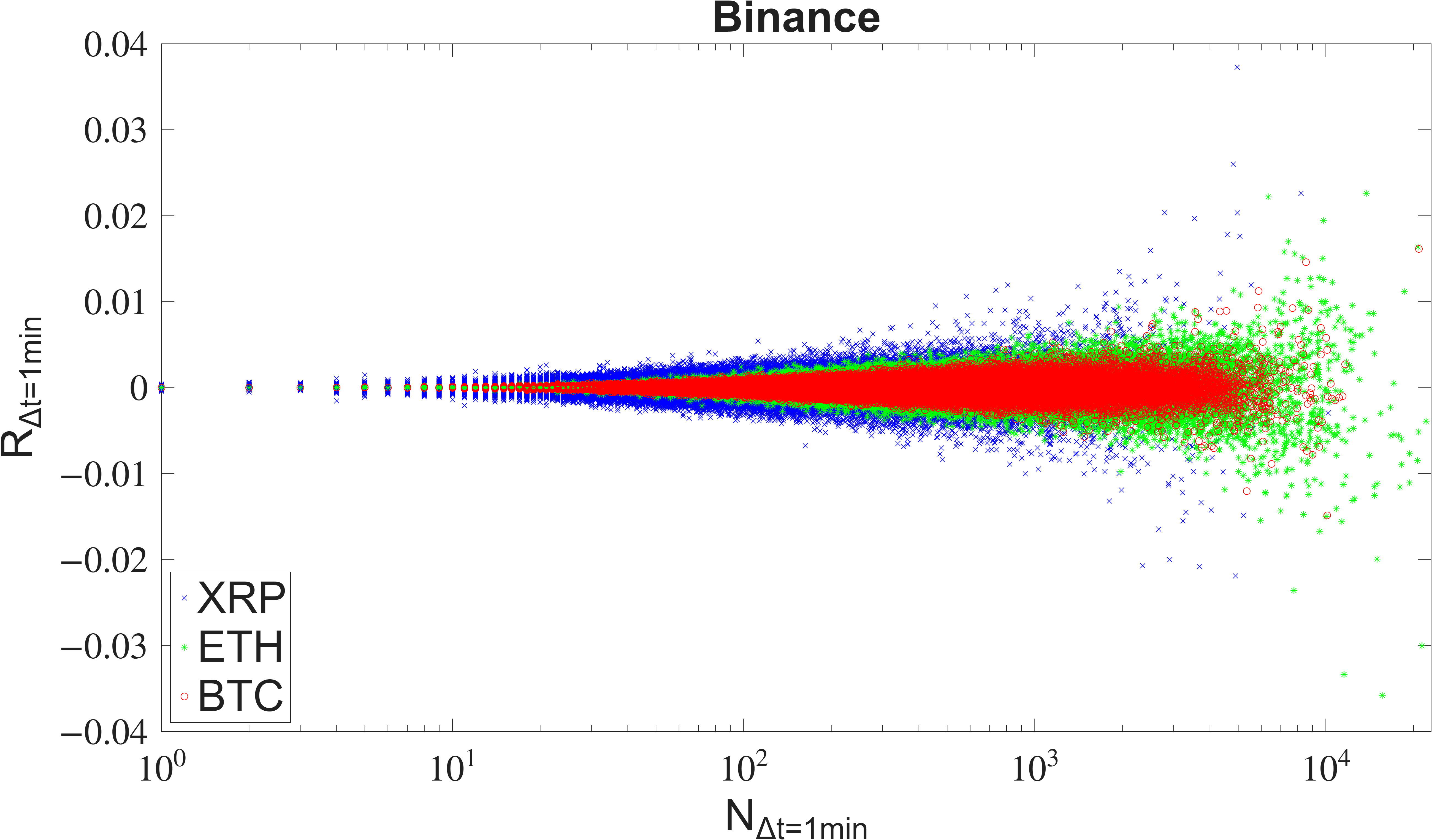}
     \includegraphics[width=0.49\textwidth]{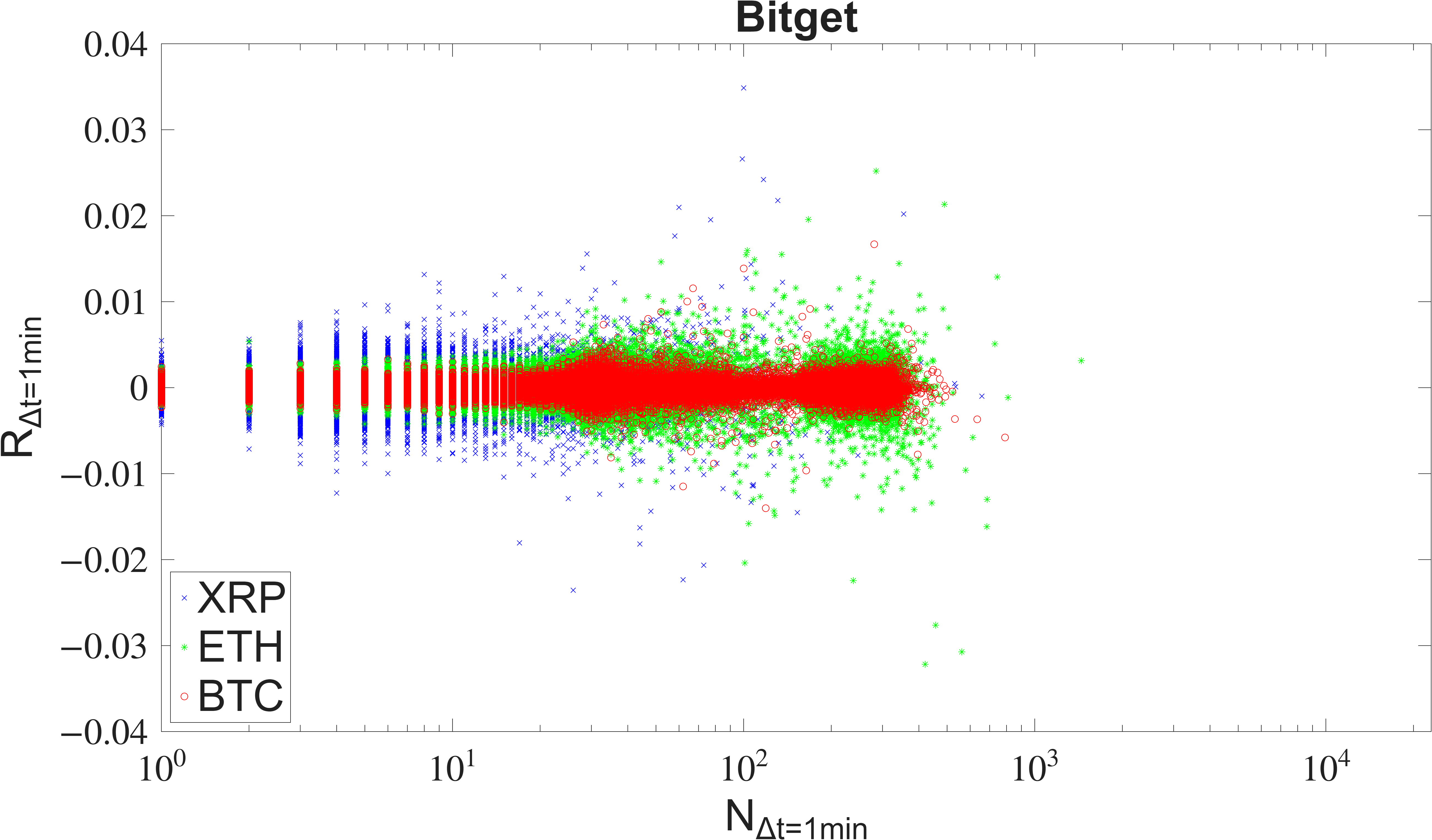}
      \includegraphics[width=0.49\textwidth]{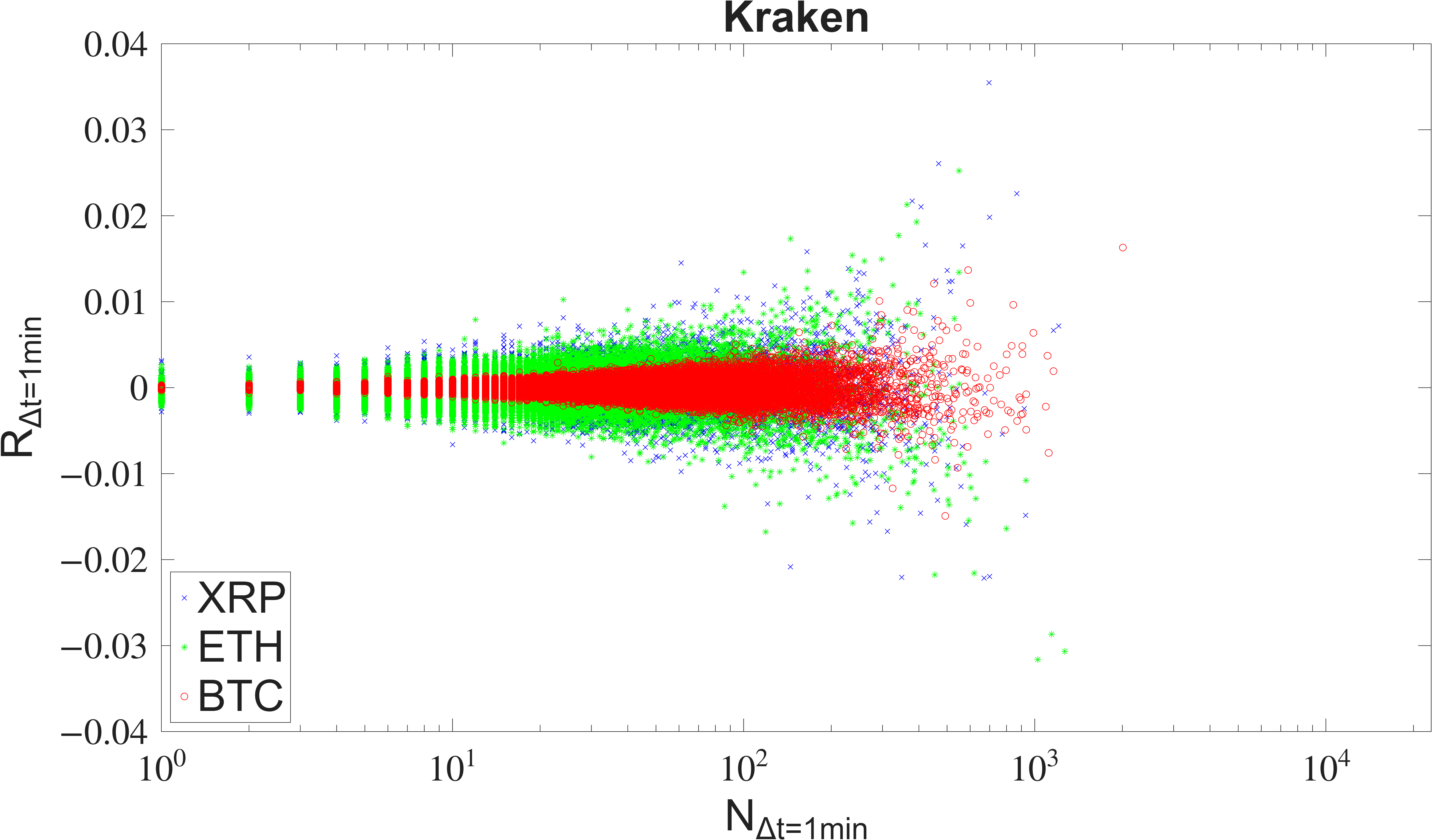}
       \includegraphics[width=0.49\textwidth]{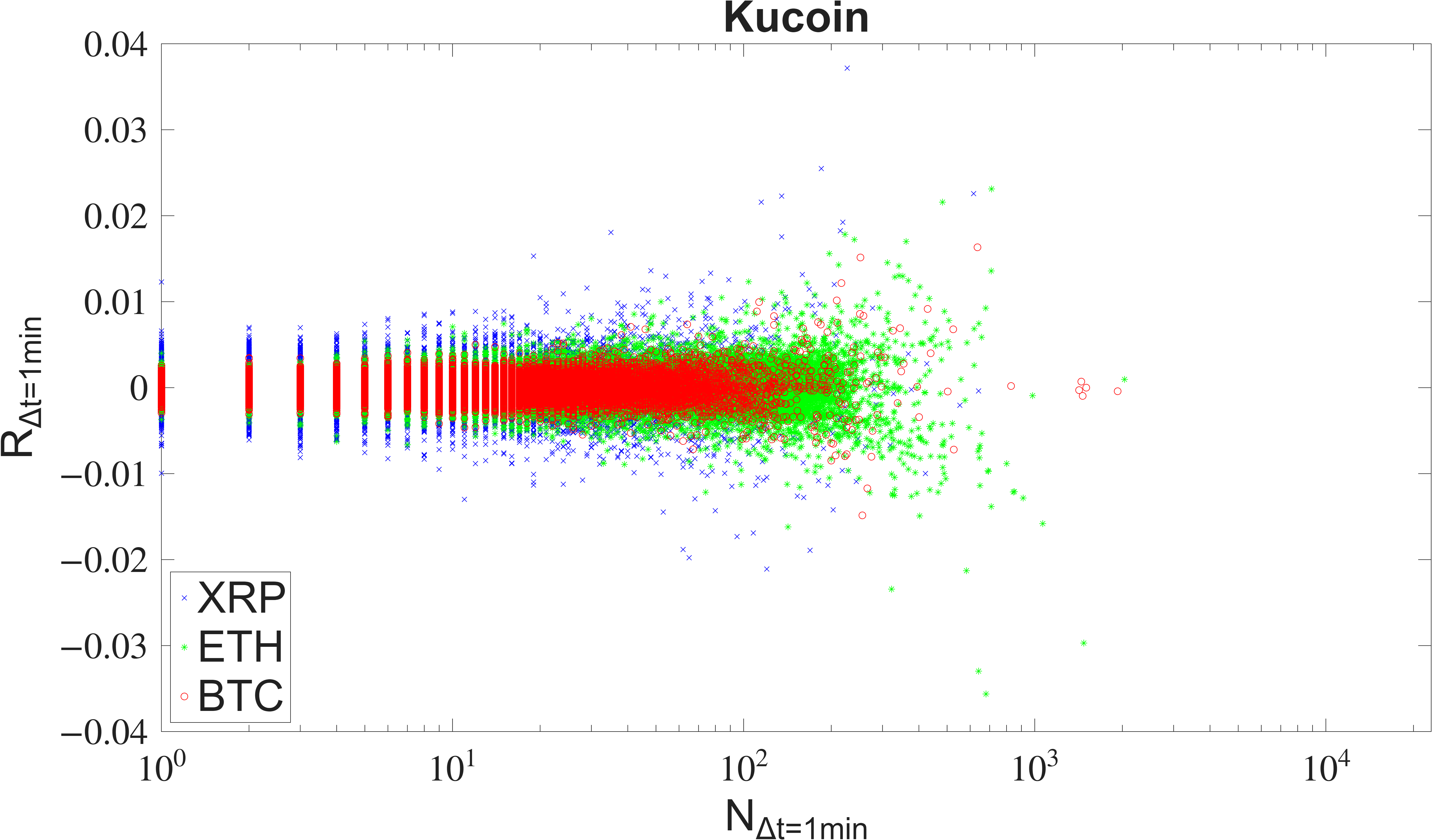}
       \caption{Relationship between log-returns $R_{\Delta t=1\mathrm{min}}(t)$ and the number of transactions $N_{\Delta t=1\mathrm{min}}(t)$ on Binance, Bitget, Kraken, and KuCoin for BTC, ETH, and XRP. Each point corresponds to a 1-min interval.}
    \label{fig:R_N}
\end{figure}

\begin{figure}[ht!]
    \centering
    \includegraphics[width=0.49\textwidth]{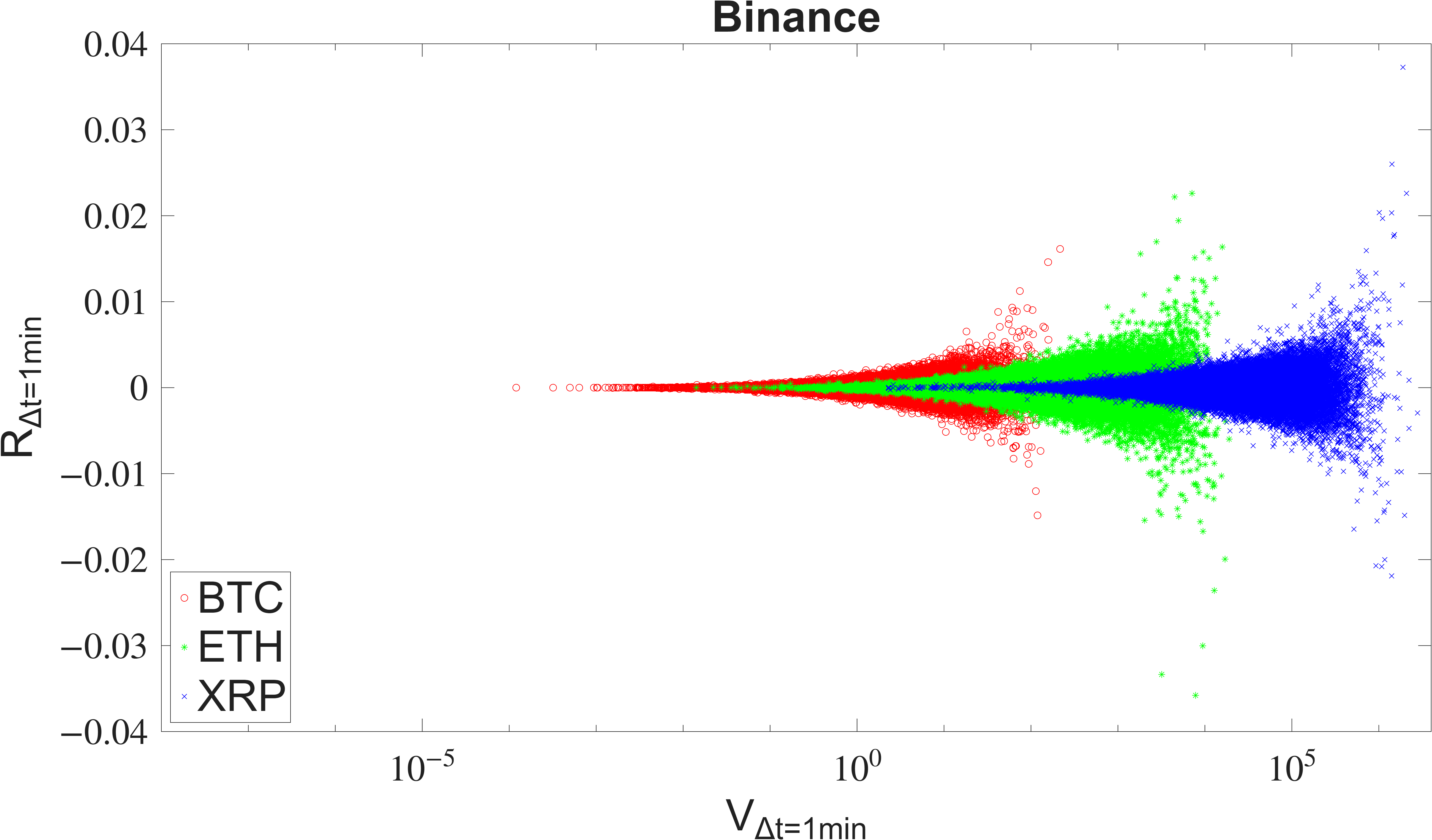}
     \includegraphics[width=0.49\textwidth]{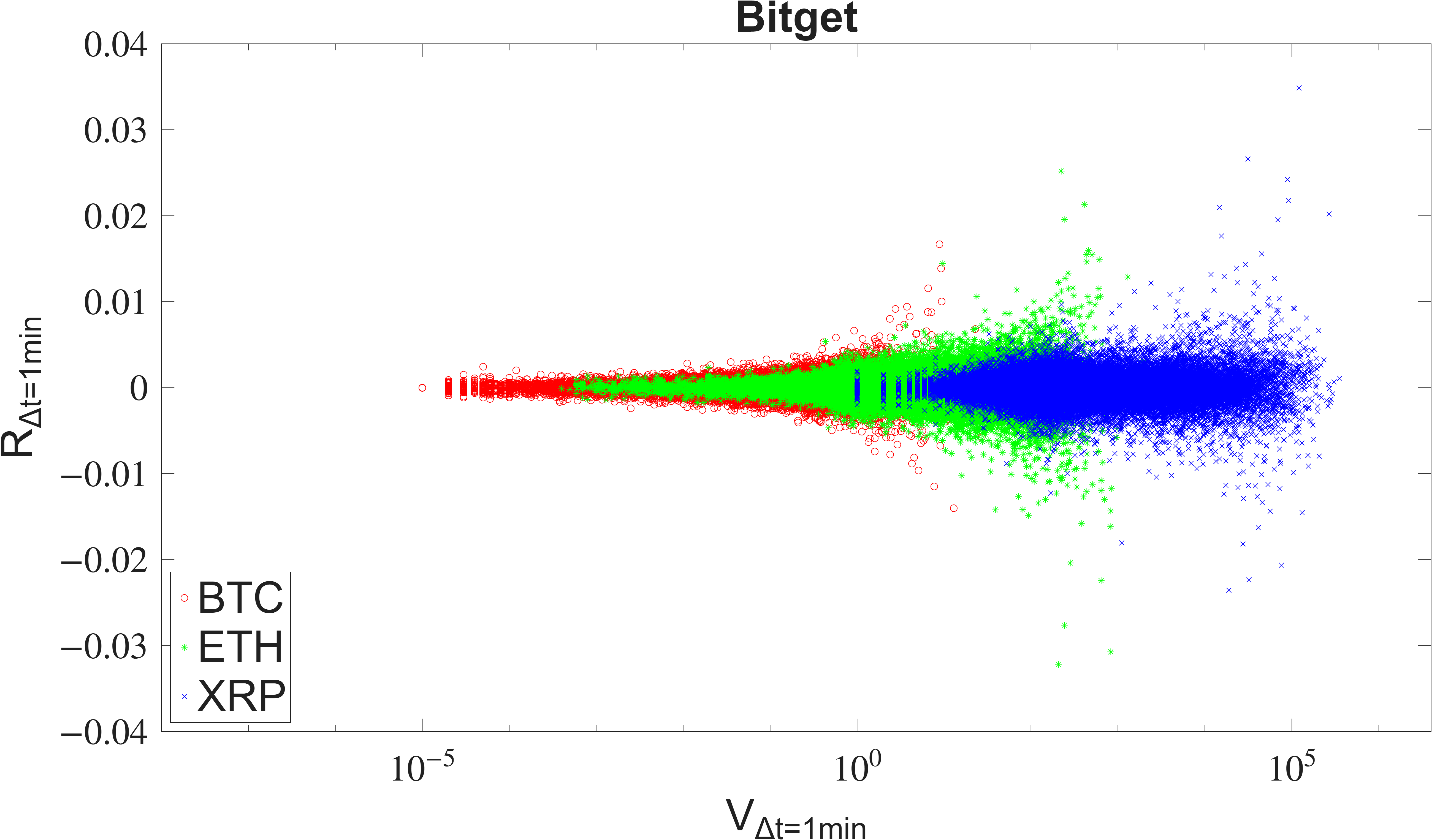}
      \includegraphics[width=0.49\textwidth]{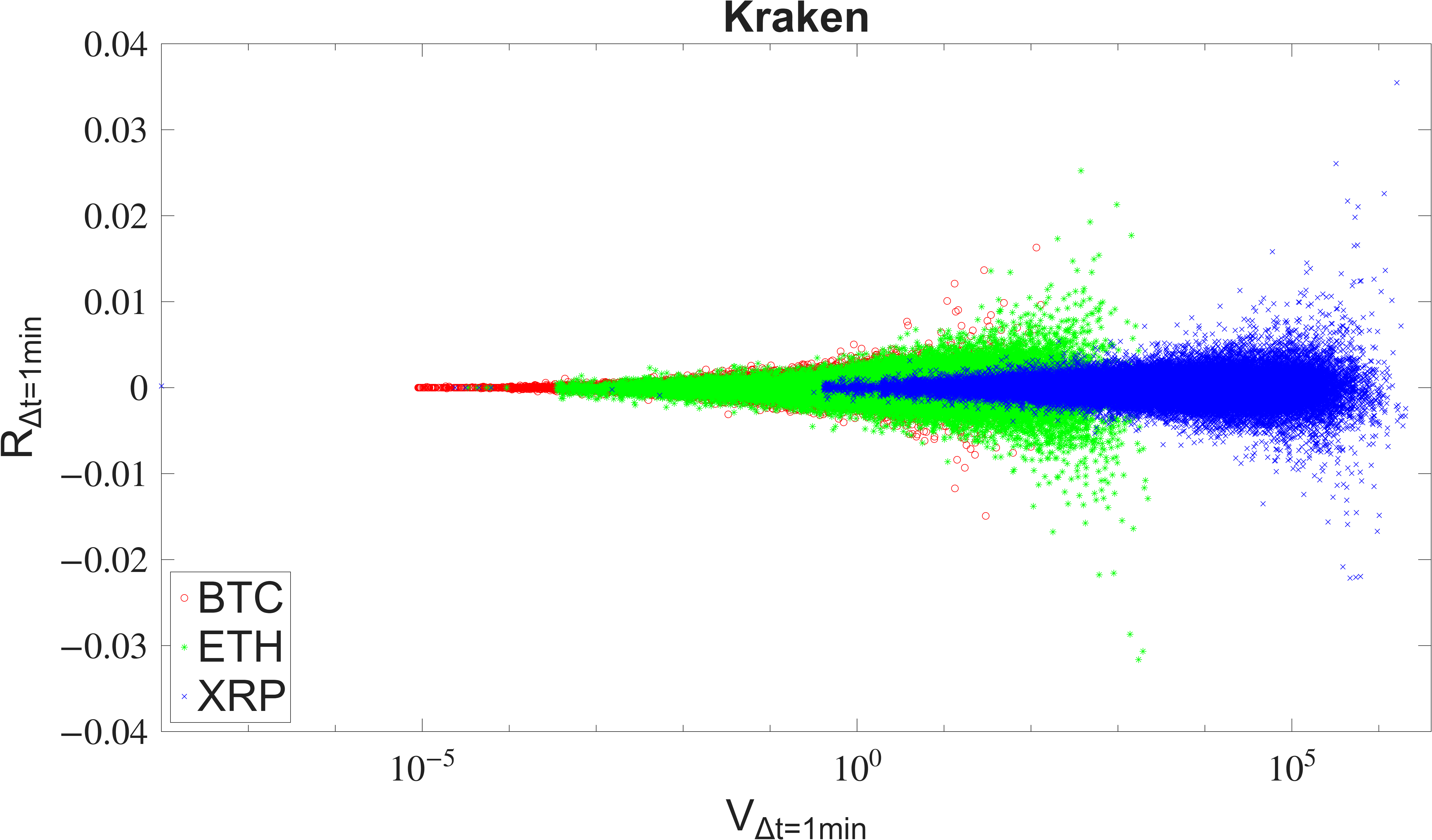}
       \includegraphics[width=0.49\textwidth]{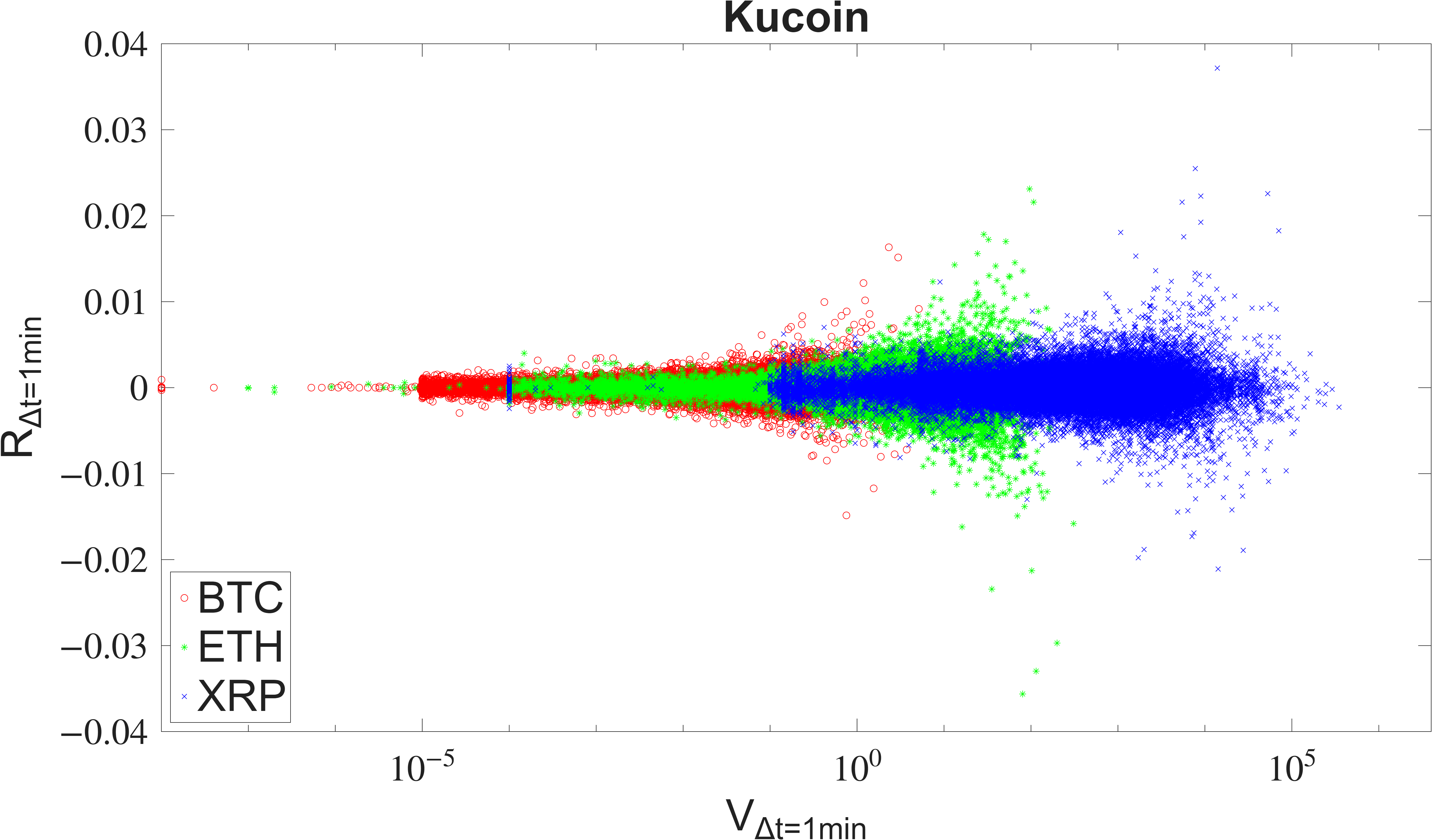}
  \caption{Relationship between log-returns $R_{\Delta t=1\mathrm{min}}(t)$ and trading volume $V_{\Delta t=1\mathrm{min}}(t)$ on Binance, Bitget, Kraken, and KuCoin for BTC, ETH, and XRP. Each point corresponds to a 1-min interval.}
    \label{fig:R_V}
\end{figure}

Figs.~\ref{fig:N_V},~\ref{fig:R_N}, and~\ref{fig:R_V} show the relationships between $N_{\Delta t=1\mathrm{min}}(t)$, $V_{\Delta t=1\mathrm{min}}(t)$, and $R_{\Delta t=1\mathrm{min}}(t)$ for BTC, ETH, and XRP on Binance, Bitget, Kraken, and KuCoin. The scatter plots reveal several robust regularities that are common across exchanges, but they also expose meaningful exchange-specific differences and anomalies, particularly on Bitget.

The clearest and most stable pattern is observed for the relationship between trading volume and the number of transactions, shown in Fig.~\ref{fig:N_V}. Across all exchanges and all three cryptocurrencies, $V_{\Delta t=1\mathrm{min}}(t)$ increases with $N_{\Delta t=1\mathrm{min}}(t)$. On the log-log scale, the points form upward-sloping bands, indicating a strong positive association between trading volume and the number of transactions. This confirms that periods with more intensive trading activity are naturally associated with larger aggregate turnover. At the same time, the position and shape of the point clouds differ substantially across exchanges and assets. Binance displays the smoothest and most regular relationship, with clearly separated clouds for BTC, ETH, and XRP. Kraken shows a similar pattern, although the range of observations is somewhat narrower. KuCoin also preserves the positive dependence, but the dispersion is visibly larger. Bitget stands out as the most irregular case. Instead of a single smooth scaling relation, the scatter plots reveal segmented clusters and abrupt changes in the shape of the cloud for BTC and ETH. This fragmentation is consistent with the regime-like behaviour of $N_{\Delta t=1\mathrm{min}}(t)$ discussed in the previous sections. In contrast, the corresponding relationship for XRP on Bitget is more similar to those observed on the other exchanges.

The relationship between returns and the number of transactions presented in Fig.~\ref{fig:R_N} is weak in terms of direction but informative in terms of volatility. The point clouds are centred around zero, which indicates that a larger number of transactions is not associated with a systematic tendency towards either positive or negative returns. However, the dispersion of returns clearly increases with $N_{\Delta t=1\mathrm{min}}(t)$. For small values of $N_{\Delta t=1\mathrm{min}}(t)$, returns are tightly concentrated around zero, whereas for larger values of $N_{\Delta t=1\mathrm{min}}(t)$, the range of observed returns expands. Thus, what the figures indicate is not a strong correlation between $R_{\Delta t=1\mathrm{min}}(t)$ and $N_{\Delta t=1\mathrm{min}}(t)$ themselves, but rather a positive correlation between transaction intensity and return volatility, i.e., between $N_{\Delta t=1\mathrm{min}}(t)$ and $|R_{\Delta t=1\mathrm{min}}(t)|$. This funnel-shaped structure is particularly regular for Binance and Kraken. KuCoin shows the same qualitative pattern, although with more outliers and stronger dispersion. Bitget again departs from the general picture for BTC and ETH: the return--transaction count relation is less smooth and appears segmented rather than continuously expanding. This suggests that the anomaly observed on Bitget is linked primarily to the transaction-count process.

A similar conclusion emerges from the relationship between returns and trading volume shown in Fig.~\ref{fig:R_V}. The spread of returns increases markedly with trading volume. Large-volume intervals are therefore associated with greater short-term price variability, but not with a systematic direction of price changes. This is consistent with the price-impact mechanism: larger traded volume increases the likelihood that available liquidity is consumed and that prices adjust more strongly. Binance again provides the cleanest example of this effect. The return cloud widens gradually as volume increases, and the three assets are well separated along the volume axis. Kraken exhibits a comparable but somewhat more compressed pattern. KuCoin and Bitget show the same qualitative dependence, although with stronger noise and more dispersed observations.

Taken together, the scatter plots indicate that Binance and Kraken exhibit the most regular market structure. On both exchanges, the relationships between trading volume, transaction intensity, and return dispersion are smooth and conform to the standard stylised facts of financial market microstructure: (i) trading volume increases with the number of transactions, and (ii) return volatility increases with both trading volume and transaction intensity. KuCoin follows the same general structure, but the relationships are visibly noisier and more dispersed, which may reflect lower liquidity or a less stable trading environment.

Bitget differs from the other exchanges mainly in the relationships involving the number of transactions for BTC and ETH. The return--volume relation is broadly comparable to that observed on the other platforms, suggesting that the price response to traded volume is not the primary source of the anomaly. The main deviations occur in the volume--transaction count and return--transaction count relations, where the observations form separate clusters rather than a continuous cloud. This suggests that the anomaly is related to how trading volume is distributed across individual transactions. A plausible interpretation is that, on Bitget for BTC and ETH, similar levels of traded volume may be executed through structurally different numbers of trades. This may result from time-varying average trade size, changes in order fragmentation or shifts between liquidity regimes, or potentially from artificial inflation of transaction counts, including wash-trading-like activity, which could distort the relationship between traded volume and the number of recorded transactions. 

The scatter-plot analysis is qualitative, but it provides a useful diagnostic tool for identifying nonlinear dependencies, heterogeneity between assets, and exchange-specific anomalies. These observations motivate a more quantitative analysis based on detrended cross-correlation coefficients, which allows us to examine how the strength of the dependence between variables changes with the time scale.

\subsection{Detrended cross-correlations}

\begin{figure}[ht!]
\centering
\includegraphics[width=0.99\textwidth]{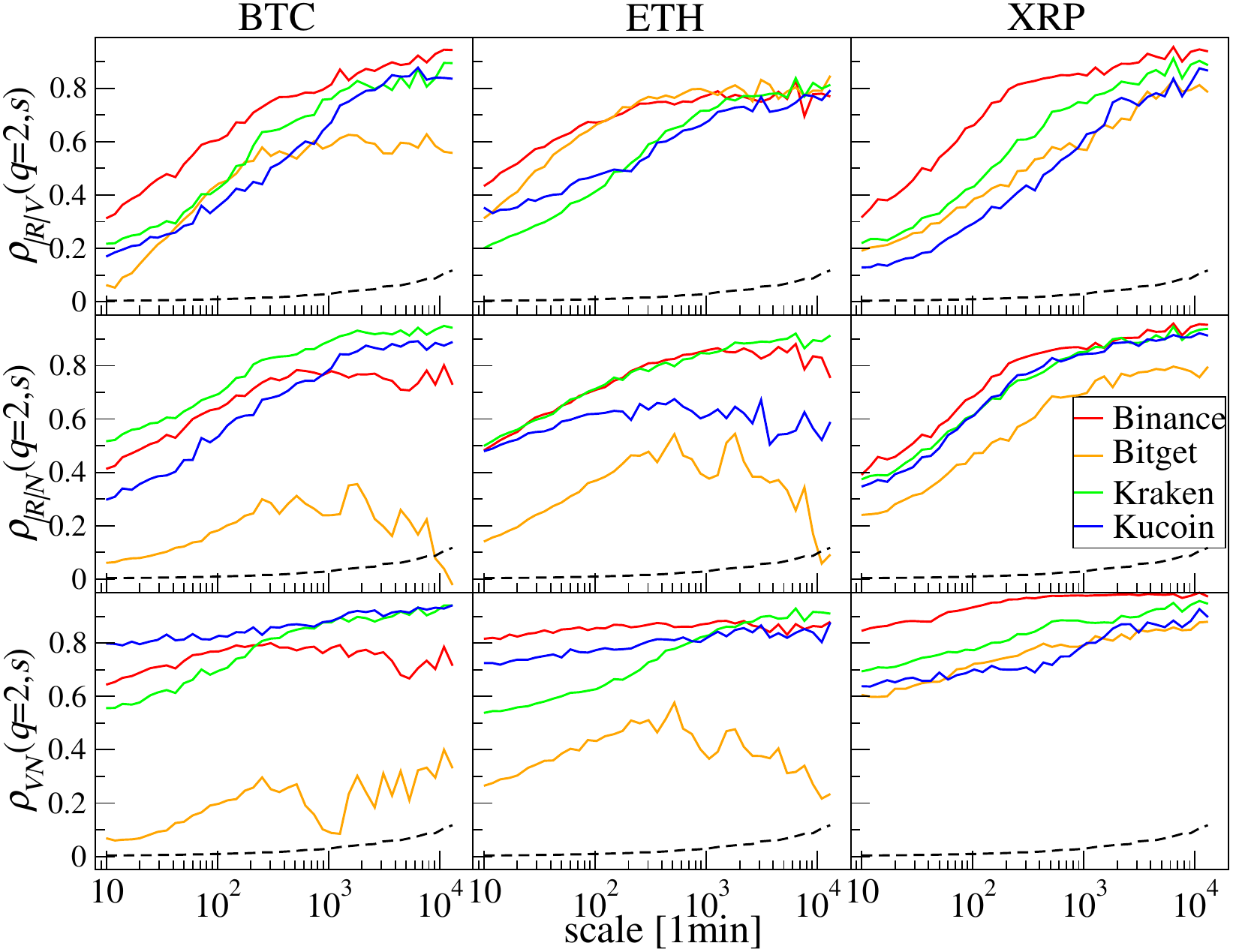}
\caption{Detrended cross-correlation coefficient $\rho(q=2,s)$ calculated between absolute log-returns $|R|$ and trading volume $V$ (top panels), absolute log-returns $|R|$ and the number of transactions $N$ (middle panels), and trading volume $V$ and the number of transactions $N$ (bottom panels) for BTC, ETH, and XRP traded on each considered exchange. The dashed curve denotes the scale-dependent standard deviation of $\rho(q=2,s)$ obtained from 100 independently shuffled surrogate realisations preserving the distributions of the original time series.}
\label{fig::rhor.XY}
\end{figure}

To quantify the dependencies observed in the previous section, the detrended cross-correlation coefficient $\rho(q,s)$ was calculated for the pairs formed by absolute log-returns $|R|$, trading volume $V$, and the number of transactions $N$. The results presented in Fig.~\ref{fig::rhor.XY} indicate that, overall, the cross-correlations increase systematically with the time scale $s$. This scale-dependent strengthening of correlations is consistent with effects commonly observed in financial markets, where correlations become more pronounced after temporal aggregation ~\cite{Watorek2019,Drozdz2025entr}. At the same time, the magnitude and stability of this increase depend on both the exchange and the pair of variables considered. To assess whether the observed values of $\rho(q=2,s)$ exceed the level expected from finite-sample fluctuations, a shuffled-surrogate reference was constructed. For each pair of series and each scale $s$, 100 independent surrogate realisations were generated by randomly shuffling the original time series while preserving its distribution. The resulting scale-dependent standard deviation is shown by dashed curves in Fig.~\ref{fig::rhor.XY}.

For the relation between absolute log-returns $|R|$ and trading volume $V$, shown in the top row of Fig.~\ref{fig::rhor.XY}, the cross-correlation coefficient generally increases from weak or moderate values on short scales to high values on longer scales. This pattern is visible for all three cryptocurrencies, confirming that large price fluctuations become increasingly coupled with traded volume when the data are aggregated over longer horizons. Among the exchanges, Binance typically exhibits the strongest correlations, reaching values of $\rho(q=2,s) \approx 0.9$ or higher for BTC and XRP and remaining consistently high for ETH. Kraken and KuCoin also show a clear monotonic increase, although with slightly lower levels at short and intermediate scales. Bitget differs somewhat from the other markets: while the correlations initially increase, they tend to saturate earlier and remain visibly weaker at the largest scales for BTC.

The correlations between absolute log-returns $|R|$ and the number of transactions $N$, presented in the middle row of Fig.~\ref{fig::rhor.XY}, reveal a similarly strong scale-dependent increase. In several cases, the coupling between $|R|$ and $N$ is even more pronounced than that observed between $|R|$ and $V$. For BTC and XRP, Binance, Kraken, and KuCoin all display a marked transition from moderate short-scale dependence to very strong long-scale correlations, frequently approaching $\rho(q=2,s) \approx 0.9$. ETH shows the same qualitative tendency, although with greater heterogeneity across exchanges. A notable exception is again Bitget, where the $|R|-N$ correlations remain substantially lower and become less stable at large scales for BTC and ETH. This suggests that, on Bitget, the number of transactions is less tightly synchronised with volatility than on the other exchanges, which supports the visual observations from the scatter plots.

The strongest and most persistent cross-correlations are observed between the trading volume $V$ and the number of transactions $N$, as shown in the bottom row of Fig.~\ref{fig::rhor.XY}. This is expected since traded volume and the number of transactions are directly linked through market activity. For nearly all assets and exchanges, $\rho(q=2,s)$ is already relatively high on short scales and increases further with scale, often reaching values above 0.8 and, in some cases, approaching unity. Binance shows the strongest correlations between $V$ and $N$, particularly for XRP, where the coefficient remains close to one on a wide range of scales. Kraken and KuCoin also exhibit robust long-scale synchronisation for all three cryptocurrencies. In contrast, Bitget again stands out with clearly weaker correlations, especially for BTC and ETH. In these two cases, the increase is slower, and the values of $\rho(q=2,s)$ remain close to 0.2 over a wide range of scales, well below those observed on the other exchanges.

These results quantitatively confirm the conclusions drawn from the scatter-plot analysis. The relation between trading volume and the number of transactions is the strongest overall, while the links involving absolute returns are slightly weaker but still substantial, especially on Binance, Kraken, and KuCoin. The comparatively weaker and less stable correlations observed for Bitget, particularly in the pairs involving $N$, indicate that the anomalous behaviour identified earlier is primarily associated with the transaction-count process rather than with the return--volume relation. This further supports the interpretation that, for BTC and ETH on Bitget, the transaction dynamics differ from those observed on the remaining exchanges.

\section{Rolling window analysis}

In the previous section, the differences in the cross-correlations between $|R|$, $V$, and $N$ were observed. For BTC and ETH, the correlations involving the number of transactions were clearly weaker on Bitget than on the other exchanges. Therefore, the temporal evolution of these correlations is examined here, with particular attention paid to the change in the behaviour of the transaction-count series on Bitget after mid-May, as shown in Figs.~\ref{fig::szeregiBTC} and~\ref{fig::szeregiETH}. For $s=10$, which is used in the rolling-window analysis below, the surrogate-based reference level estimated in the previous section is close to zero.

\subsection{$\rho(q,s)$ coefficient}

\begin{figure}[ht!]
\centering
\includegraphics[width=0.99\textwidth]{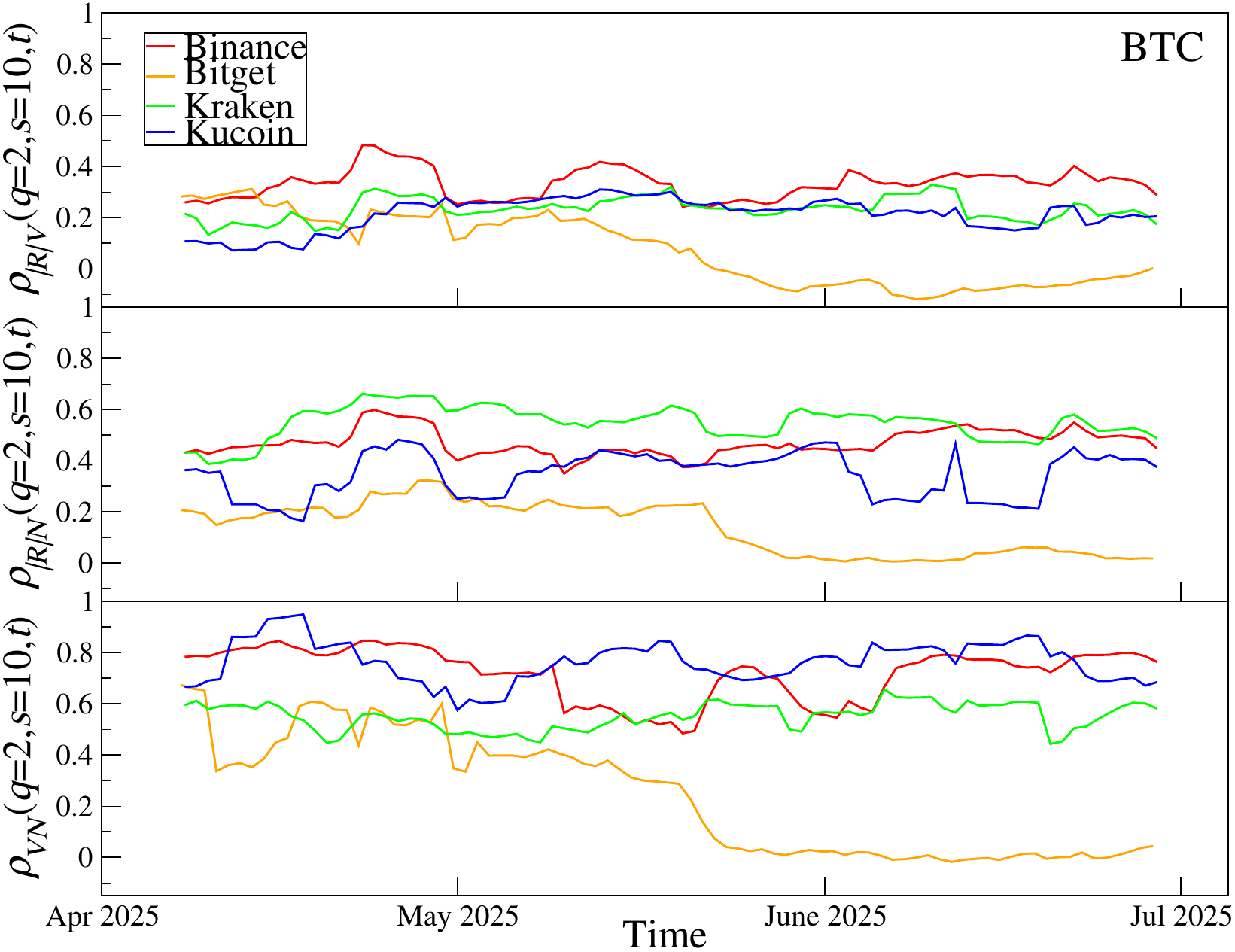}
\caption{Detrended cross-correlation coefficient $\rho(q=2,s=10)$ calculated in a rolling window of length 7 days, with a step of 24 hours, for BTC on Binance, Bitget, Kraken, and KuCoin. The top, middle, and bottom panels show the pairs $|R|-V$, $|R|-N$, and $V-N$, respectively. The time coordinate of each point denotes the end of the corresponding rolling window.}
\label{fig::Pq2s10oknoBTC}
\end{figure}

\begin{figure}[ht!]
\centering
\includegraphics[width=0.99\textwidth]{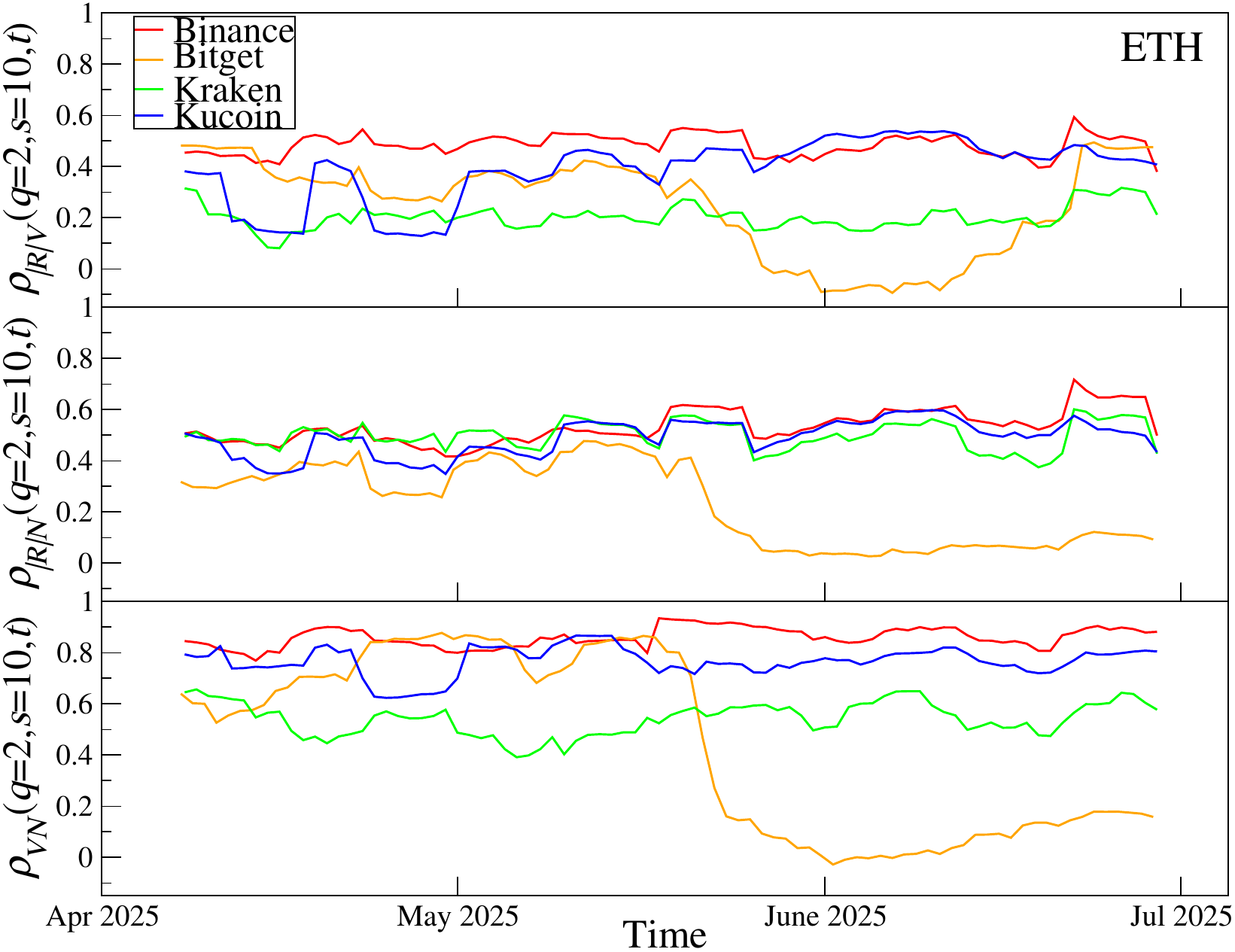}
\caption{The same as in Fig.~\ref{fig::Pq2s10oknoBTC}, but for ETH.}
\label{fig::Pq2s10oknoETH}
\end{figure}

\begin{figure}[ht!]
\centering
\includegraphics[width=0.99\textwidth]{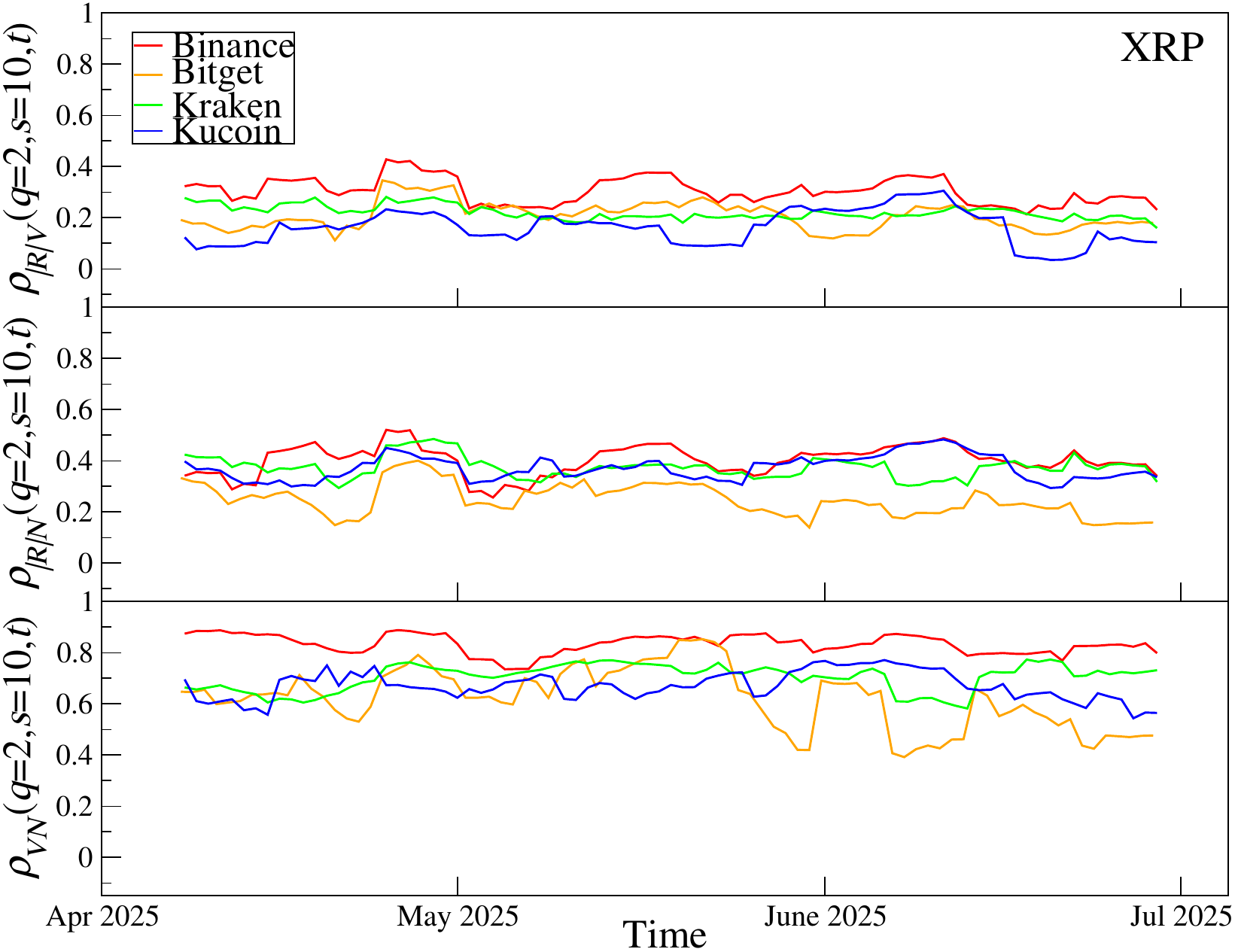}
\caption{The same as in Fig.~\ref{fig::Pq2s10oknoBTC}, but for XRP.}
\label{fig::Pq2s10oknoXRP}
\end{figure}

The analysed period was examined using a rolling window of 10,080 min (7 days) with a step of 1,440 min (1 day). The window length was chosen as a compromise between temporal resolution and statistical stability: it is long enough to estimate the coefficient reliably, while still allowing changes in the dependence structure to be localised in time. Figs.~\ref{fig::Pq2s10oknoBTC},~\ref{fig::Pq2s10oknoETH}, and~\ref{fig::Pq2s10oknoXRP} present the time evolution of $\rho(q=2,s=10)$ estimated for the pairs $|R|-V$, $|R|-N$, and $V-N$ for BTC, ETH, and XRP across Binance, Bitget, Kraken, and KuCoin. In general, the highest cross-correlation levels are most often observed on Binance, which therefore provides a useful benchmark against which the behaviour of the other exchanges can be compared. The most important result is the strong deviation of Bitget correlations after mid-May. This effect is clearly visible for BTC and ETH, but not for XRP.

For BTC shown in Fig.~\ref{fig::Pq2s10oknoBTC}, Binance usually exhibits the strongest correlations or remains among the exchanges with the strongest correlations, especially for the $|R|-V$ relation. Kraken and KuCoin usually remain at somewhat lower, but still clearly positive, levels. The main exception concerns the $V-N$ relation, where KuCoin is often comparable to, or even above, Binance during parts of the sample. Bitget, by contrast, starts to separate from the other exchanges around mid-May. This breakdown is most clearly visible in the $V-N$ relation, where the Bitget values drop sharply and remain close to zero for a significant portion of the remaining period. Declines in the $|R|-V$ and $|R|-N$ relations are also evident and are broadly similar in magnitude, although both are less extreme than for $V-N$.

A closely related pattern is observed for ETH in Fig.~\ref{fig::Pq2s10oknoETH}. The weakening of correlations on Bitget begins essentially at the same time as on BTC, but is even more pronounced. Here again, Binance usually records the strongest correlations, though KuCoin is at times very close. The most pronounced deterioration on Bitget is observed for the $|R|-N$ and $V-N$ relations, whose values fall sharply and remain close to zero over a substantial part of late May and June. A decrease is also observed in the $|R|-V$ relation; however, this correlation partially recovers toward the end of June and returns to a level comparable to that observed on the other exchanges.

By contrast, XRP does not display an analogous regime shift on Bitget, as shown in  Fig.~\ref{fig::Pq2s10oknoXRP}. For this asset, Binance again tends to remain at the upper end of the range, especially for the $|R|-V$ and $V-N$ relations, while Kraken and KuCoin usually remain at somewhat lower but stable levels. Bitget is generally weaker than Binance, but it does not experience the same persistent collapse toward zero as observed for BTC and ETH.

\begin{figure}[ht!]
\centering
\includegraphics[width=0.99\textwidth]{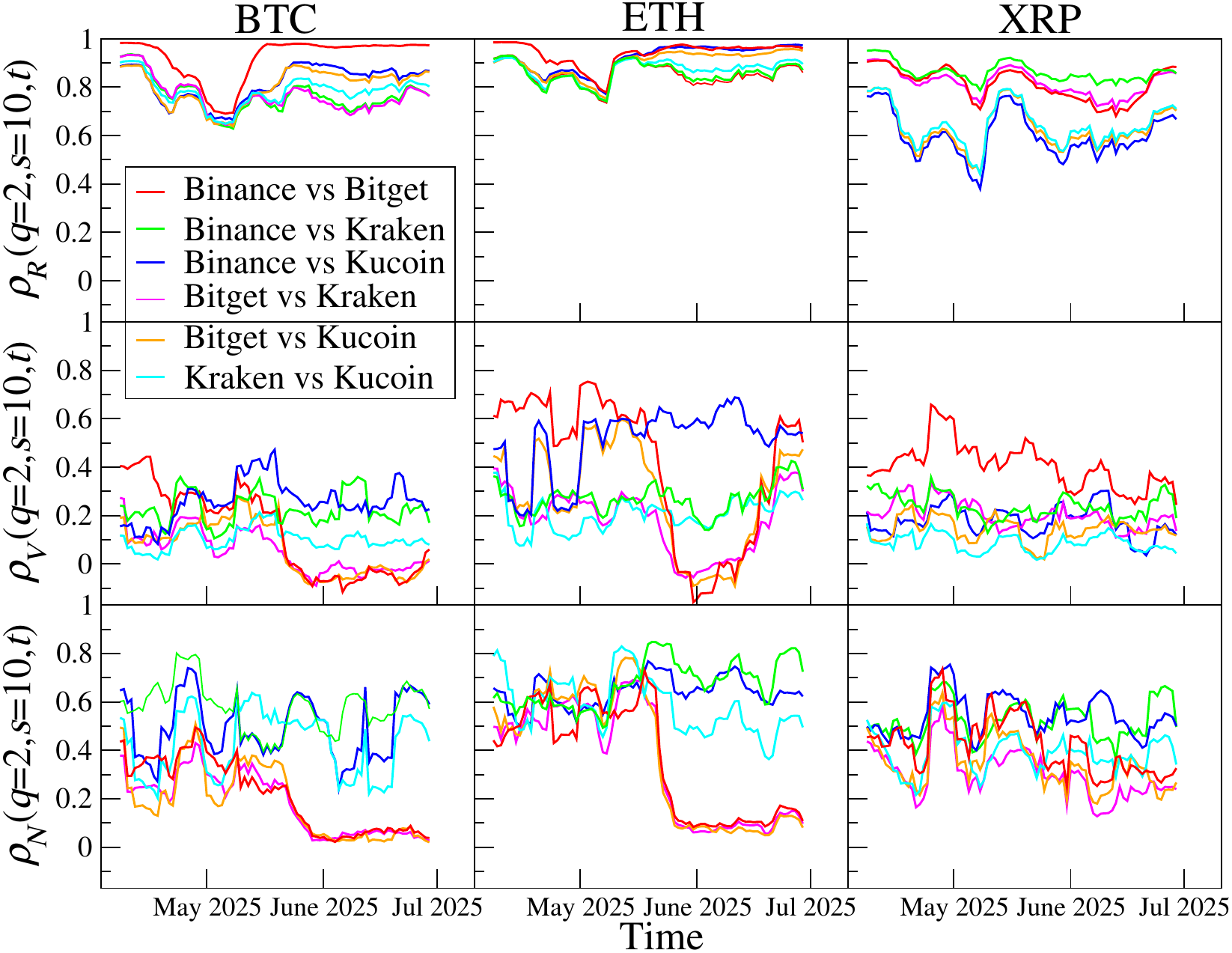}
\caption{Detrended cross-correlation coefficient $\rho(q=2,s=10)$ calculated in a rolling window of length 7 days, with a step of 24 hours, for BTC (left), ETH (middle) and XRP (right) $R$ (top), $V$ (middle), $N$ (bottom) between Binance, Bitget, Kraken, and KuCoin exchanges. The time coordinate of each point denotes the end of the corresponding rolling window.}
\label{fig::Pq2s10oknoex}
\end{figure}

The different behaviour of the transaction-count series for BTC and ETH on Bitget is also visible in cross-exchange correlations calculated for the same asset and the same variable. Fig.~\ref{fig::Pq2s10oknoex} shows the temporal evolution of $\rho(q=2,s=10)$ for $R$, $V$, and $N$ of BTC, ETH, and XRP between all pairs of Binance, Bitget, Kraken, and KuCoin.

For log-returns, shown in the top panels of Fig.~\ref{fig::Pq2s10oknoex}, cross-exchange correlations remain high throughout most of the analysed period. For BTC and ETH, they are generally close to each other and typically remain around 0.8 or above, with only temporary decreases in May 2025. This indicates that price dynamics remain strongly synchronised across exchanges. In particular, the Binance--Bitget pair exhibits the strongest cross-exchange return correlations often reaching values close to unity. For XRP, return correlations are somewhat lower and more heterogeneous across exchange pairs, reaching values close to 0.9 at the beginning of the sample and decreasing to around 0.4--0.6 during May. The strongest XRP return correlations are observed mainly for pairs involving Binance, Bitget, and Kraken, whereas correlations involving KuCoin are generally weaker.

Larger temporal variation is observed for traded volume, shown in the middle panels of Fig.~\ref{fig::Pq2s10oknoex}. For BTC and ETH, the Binance--Bitget volume correlation is among the highest at the beginning of the analysed period, but it decreases sharply after mid-May. Correlations between Bitget and the other exchanges, namely Kraken and KuCoin, also decline at the same time. For BTC, these correlations remain close to zero or slightly negative for a substantial part of the remaining period. For ETH, the decline is also clearly visible, although a rebound toward the pre-decline level is observed near the end of June. This recovery is consistent with the increase in the $|R|-V$ correlation on Bitget shown in Fig.~\ref{fig::Pq2s10oknoETH}. By contrast, volume correlations between exchange pairs not involving Bitget evolve more smoothly. For XRP, the Binance--Bitget volume correlation remains the strongest over the whole period, while the other exchange pairs fluctuate at lower but relatively stable levels.

The strongest effect is observed for the number of transactions, shown in the bottom panels of Fig.~\ref{fig::Pq2s10oknoex}. For BTC and ETH, the correlations between Bitget and other exchanges: Binance, Kraken, and KuCoin decrease after mid-May from moderate positive levels, approximately 0.4--0.6, to values close to zero, and they remain at this level until the end of the sample. This pattern is not observed for the other exchange pairs and is not present for XRP. Hence, the transaction-count series on Bitget becomes almost completely decorrelated from the corresponding series on the other exchanges for BTC and ETH. This confirms and extends the earlier observation that the anomalous behaviour of Bitget is mainly associated with the number of transactions and is specific to BTC and ETH.

 Overall, the rolling-window results show that the post-mid-May weakening of the dependence structure is both exchange-specific and asset-dependent. The effect primarily involves the number of transactions on Bitget: it is visible both in its correlations with $|R|$ and $V$ within the same exchange and in its cross-exchange correlations with the corresponding transaction-count series on the other exchanges. A similar, although slightly weaker and less persistent, effect is also observed for traded volume. The anomalous behaviour identified in the previous sections is therefore not only visible in static measures and scatter plots, but also appears as a time-localised structural change in the correlation dynamics. Binance, in contrast, is generally the exchange with the highest and most stable cross-correlation levels across the sample.

\subsection{Approximate and sample entropy}

To further investigate the weakening of cross-correlations observed on Bitget after mid-May 2025, approximate entropy (ApEn), defined in Section~\ref{Entr} by Eq.~\eqref{eq::apen.definition}, was calculated using the same rolling-window procedure as in the previous section. The aim of this analysis was to determine whether the decrease in dependence between $|R|$, $V$, and $N$ was accompanied by a change in the regularity of the underlying time series.

\begin{figure}[ht!]
\centering
\includegraphics[width=0.99\textwidth]{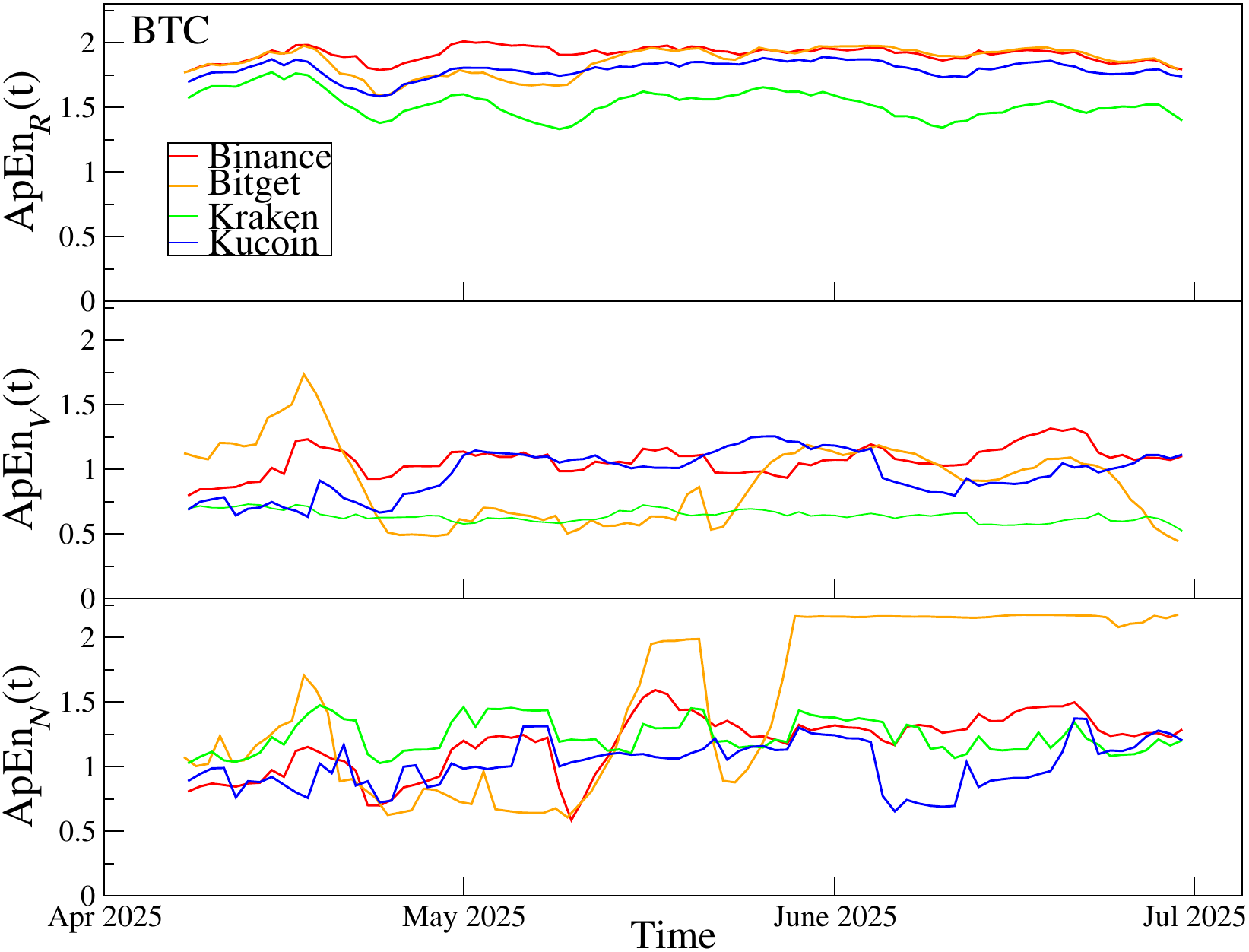}
\caption{Approximate entropy with $m=2$ and $\tau=1$, calculated in a rolling window of length 7 days with a step of 24 hours for BTC log-returns $R$ (top), trading volume $V$ (middle), and the number of transactions $N$ (bottom) on Binance, Bitget, Kraken, and KuCoin. The time coordinate of each point denotes the end of the corresponding rolling window.}
\label{fig::AprEntroknoBTC}
\end{figure}

\begin{figure}[ht!]
\centering
\includegraphics[width=0.99\textwidth]{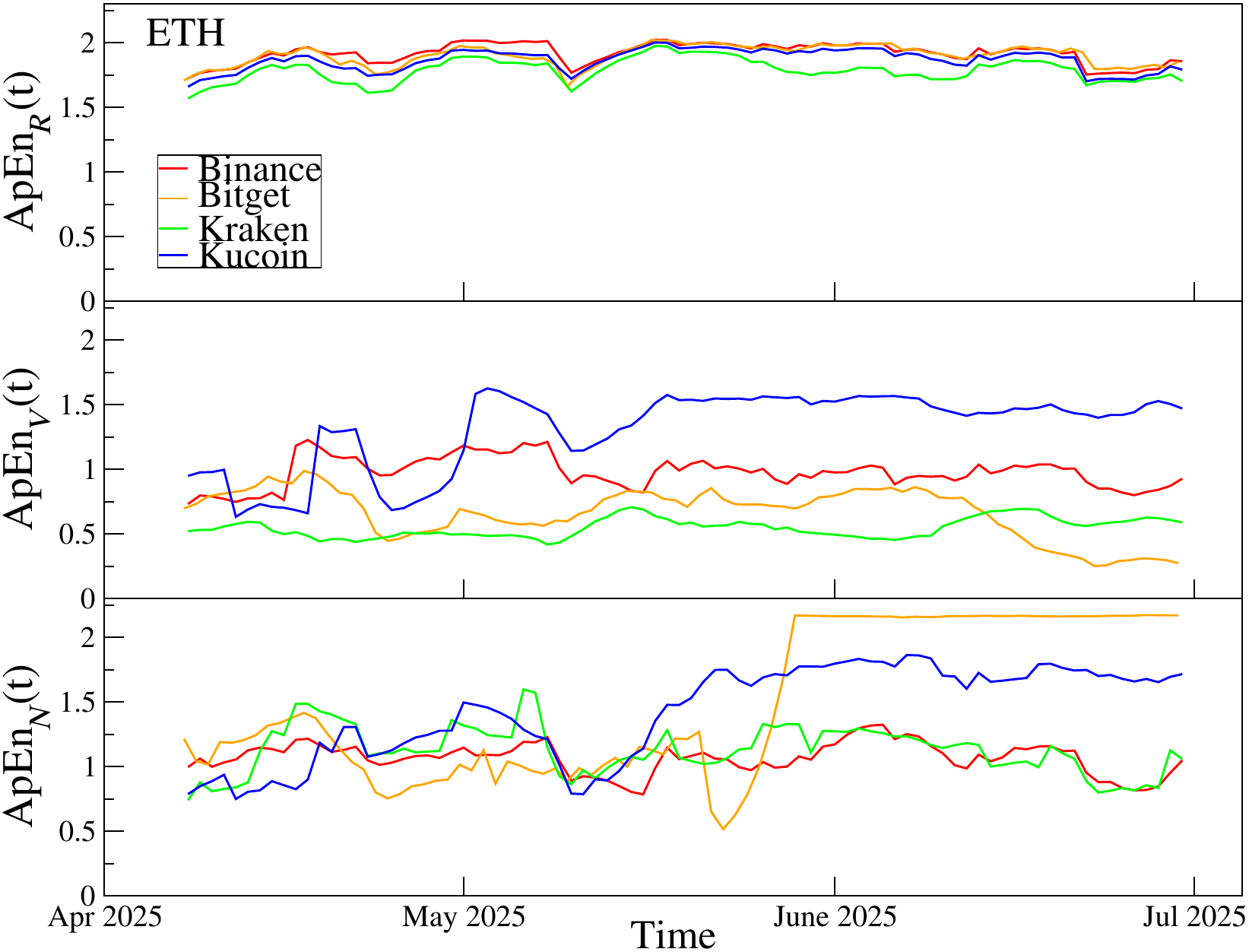}
\caption{The same as in Fig.~\ref{fig::AprEntroknoBTC}, but for ETH.}
\label{fig::AprEntroknoETH}
\end{figure}

\begin{figure}[ht!]
\centering
\includegraphics[width=0.99\textwidth]{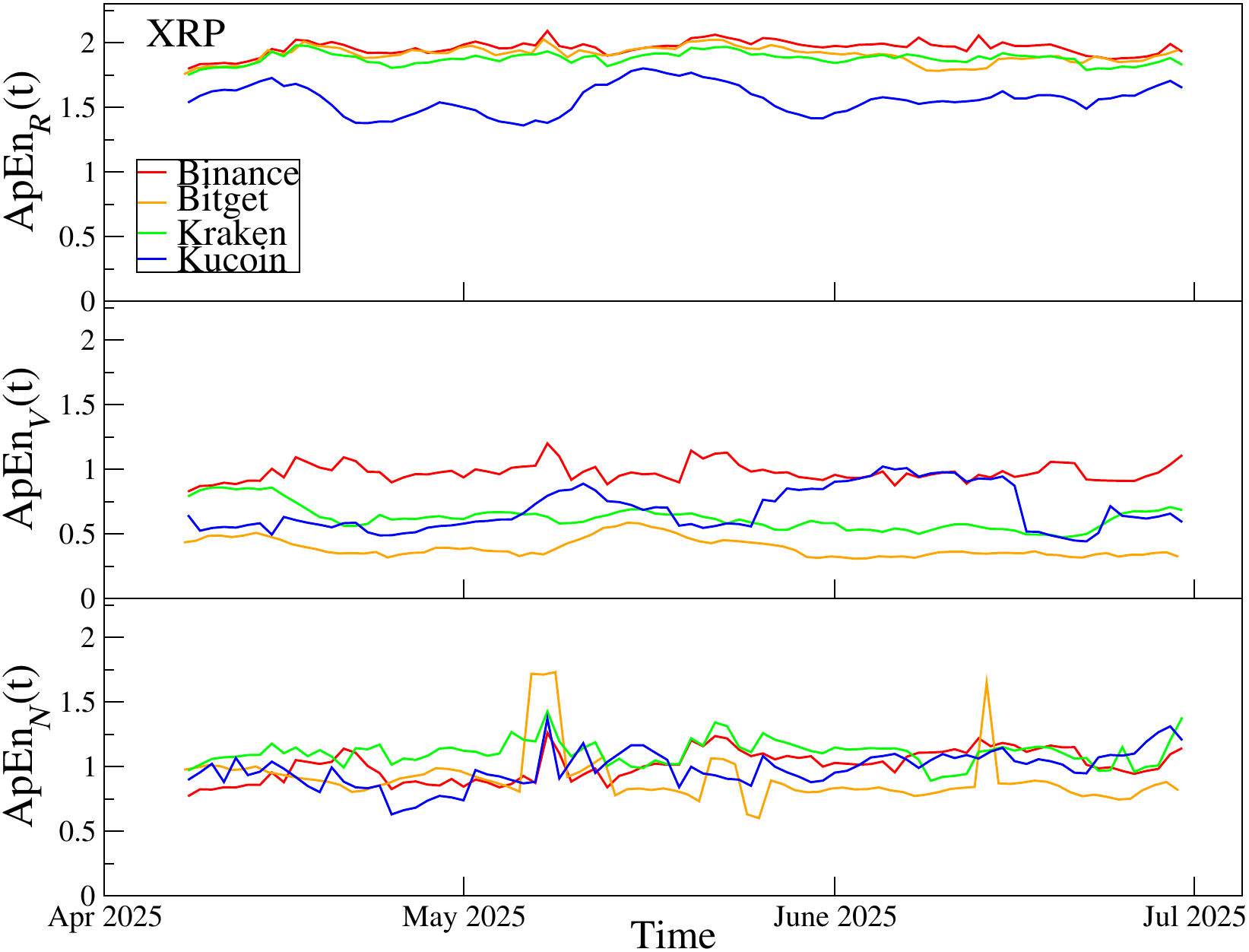}
\caption{The same as in Fig.~\ref{fig::AprEntroknoBTC}, but for XRP.}
\label{fig::AprEntroknoXRP}
\end{figure}

The most stable behaviour is observed for log-returns, shown in the top panels of Figs.~\ref{fig::AprEntroknoBTC},~\ref{fig::AprEntroknoETH}, and~\ref{fig::AprEntroknoXRP}. The ApEn values remain high across all three assets and exchanges, typically within a relatively narrow range. This indicates that return dynamics remain strongly irregular throughout the sample and do not undergo a major structural simplification. Among the exchanges, Binance is often characterised by the highest values of ApEn, whereas Kraken is frequently somewhat lower, especially for BTC and ETH, suggesting slightly more regular return dynamics. For XRP, the return-based ApEn is also high and stable across Binance, Bitget, and Kraken, which remain close to one another, while KuCoin stays somewhat lower during much of the period.

Much stronger exchange-specific differences emerge for the trading volume $V$ and the number of transactions $N$. For BTC, the ApEn of volume, shown in the middle panel of Fig.~\ref{fig::AprEntroknoBTC}, is clearly heterogeneous across exchanges. Binance and KuCoin remain at moderate levels, Kraken is lower and more stable, whereas Bitget shows the strongest temporal variation, with elevated values early in the sample followed by a marked reduction after mid-April and relatively subdued levels thereafter. The most striking pattern, however, appears in ApEn of the number of transactions, which is shown in the bottom panel of Fig.~\ref{fig::AprEntroknoBTC}. On Bitget, ApEn for $N$ increases sharply from the second half of May and then remains near its upper range for a prolonged period, clearly separating this exchange from Binance, Kraken, and KuCoin. This indicates that the transaction-count dynamics on Bitget become substantially more irregular and less repeatable during the same period in which the rolling-window cross-correlations weaken.

For ETH, the picture shown in Fig.~\ref{fig::AprEntroknoETH} is even clearer. Return ApEn again remains high and relatively stable across exchanges, whereas the largest differences are observed for $V$ and, especially, for $N$. In the middle panel of Fig.~\ref{fig::AprEntroknoETH}, KuCoin has the highest ApEn for volume over a large part of the sample, Binance remains at an intermediate level, Kraken is lower, and Bitget declines strongly toward the end of the period. This indicates a transition toward more regular volume dynamics on this exchange. In the bottom panel of Fig.~\ref{fig::AprEntroknoETH}, the dominant feature is again Bitget, for which the entropy of the number of transactions jumps abruptly around mid-May and then remains close to the upper boundary of the observed range. This regime shift is 
more pronounced than for BTC and coincides with the period in which the cross-correlations involving $N$ on Bitget fall close to zero. Hence, for ETH, the decrease in cross-correlation is accompanied by a simultaneous increase in transaction activity irregularity.

For XRP, the behaviour is clearly different, as shown in Fig.~\ref{fig::AprEntroknoXRP}. Although cross-exchange differences in ApEn are still visible, the patterns are much more stable, and there is no persistent Bitget-specific regime shift comparable to that observed for BTC and ETH. In particular, neither the volume ApEn nor the transaction-count ApEn on Bitget exhibits the same abrupt and long-lasting change. This is consistent with the rolling-window cross-correlation results, where XRP did not display the strong post-mid-May decrease observed for BTC and ETH on Bitget.

In the previous figures, the embedding dimension was set to $m=2$ and the lag was set to $\tau=1$. To test the robustness of the results with respect to the embedding dimension, additional calculations were performed for $m=3,4,5$, while keeping $\tau=1$ fixed. Since the probability of finding matching patterns decreases with increasing embedding dimension in finite samples, ApEn values can depend strongly on $m$. Therefore, the focus is placed on the qualitative time evolution rather than on the absolute values~\cite{Richman2000,Mesin2018}. 

\begin{figure}[ht!]
\centering
\includegraphics[width=0.49\textwidth]{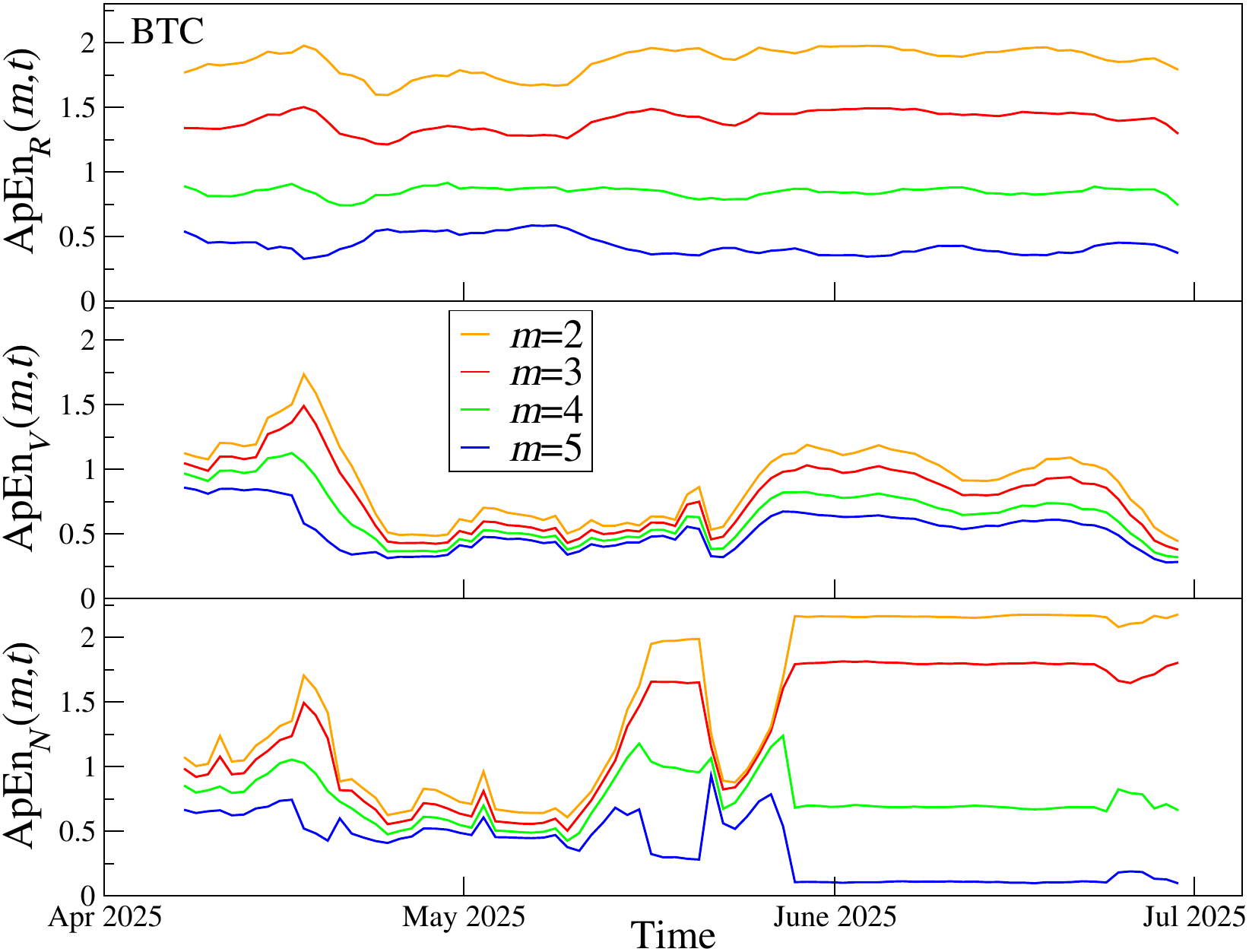}
\includegraphics[width=0.49\textwidth]{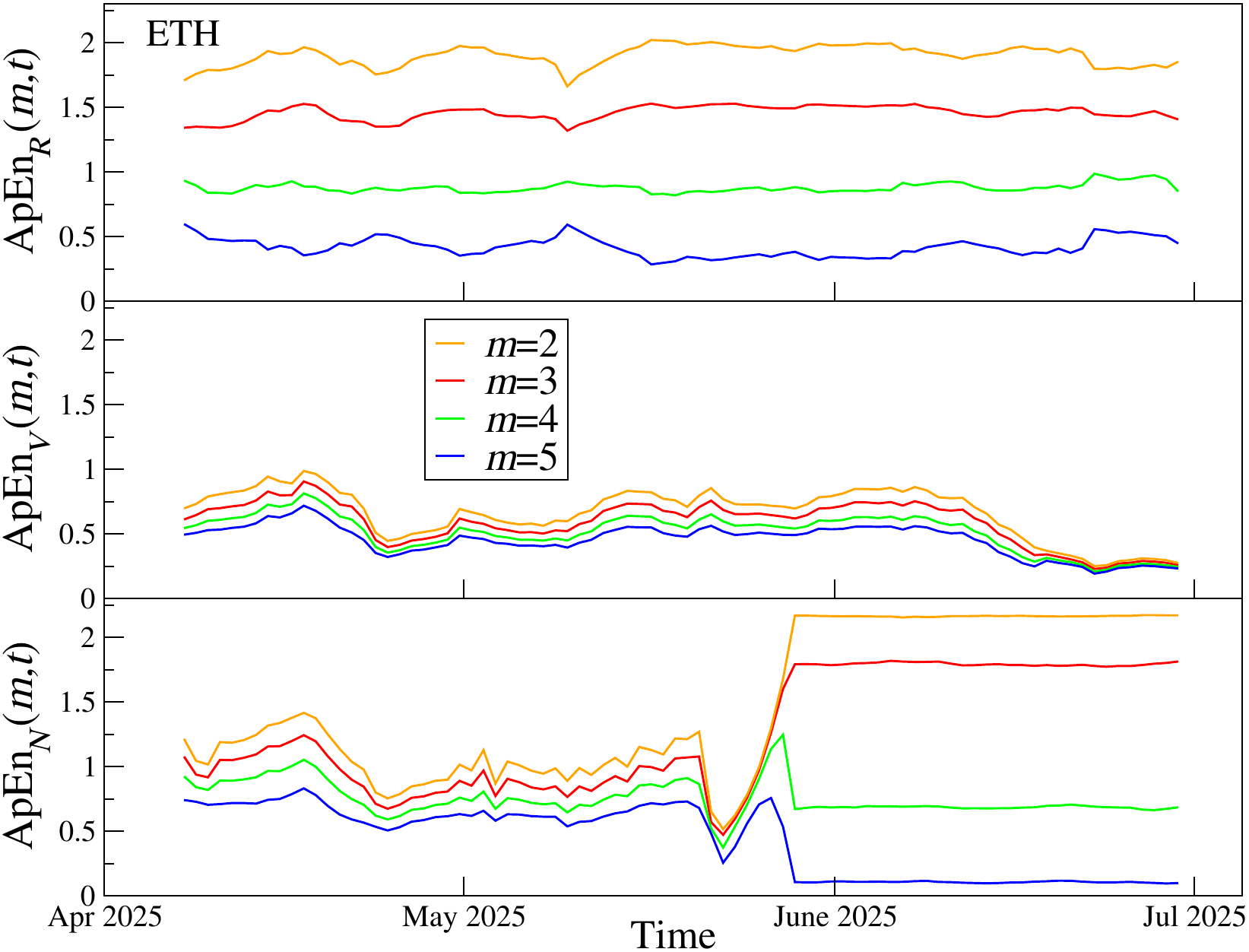}
\caption{Approximate entropy calculated in a rolling window of length 7 days with a step of 24 hours for BTC and ETH log-returns $R$ (top), trading volume $V$ (middle), and the number of transactions $N$ (bottom) on Bitget for different embedding dimensions $m=2,3,4,5$ and fixed $\tau=1$. The time coordinate of each point denotes the end of the corresponding rolling window.}
\label{fig::AprEntroknoDim}
\end{figure}

The results presented in Fig.~\ref{fig::AprEntroknoDim} show that the detected regime changes are not an artefact of a single parameter choice. For both BTC and ETH, the return-based entropy preserves the same qualitative ordering over time for all values of $m$, and the curves remain relatively stable, while higher dimensions produce lower ApEn values as expected. The volume-based entropy also shows a similar time profile across dimensions, with the main changes occurring in similar time intervals, although the absolute levels decrease with increasing $m$.

The strongest effect is again observed for the number of transactions. For both BTC and ETH on Bitget, a sharp structural change occurs around mid-May. For smaller embedding dimensions, $m=2$ and $m=3$, ApEn increases markedly and then remains elevated. This indicates that short transaction-count patterns become less repeatable and less predictable after the regime change. For larger dimensions, $m=4$ and $m=5$, the levels are much lower and in some cases decrease toward very small values. This difference reflects ApEn's dependence on the length of the compared patterns. For larger $m$, longer sequences are compared, and the number of effectively matched patterns in a finite rolling window decreases. Consequently, the numerical values of ApEn become more sensitive to the structure of the dominant activity regime and to finite-sample effects. Therefore, the opposite response observed for smaller and larger embedding dimensions should not be interpreted as a contradiction, but rather as evidence that the temporal organisation of $N$ changes differently for short and longer local patterns.
The rolling-window ApEn analysis shows that the post-mid-May weakening of cross-correlations on Bitget for BTC and ETH is accompanied by a simultaneous change in the temporal organisation of market activity, especially in the transaction-count series. This supports the interpretation of an exchange-specific and asset-dependent regime shift rather than a purely local fluctuation.

The timing of the decrease in cross-correlations and the increase in ApEn on Bitget coincides with the unusual increase in the number of transactions observed after mid-May 2025 in Figs.~\ref{fig::szeregiBTC} and~\ref{fig::szeregiETH}. Therefore, to better understand the nature of this process, the Bitget time series are divided into two subsamples: before and after the mid-May regime change. The properties of these two periods are analysed in the following section.

\begin{figure}[ht!]
\centering
\includegraphics[width=0.49\textwidth]{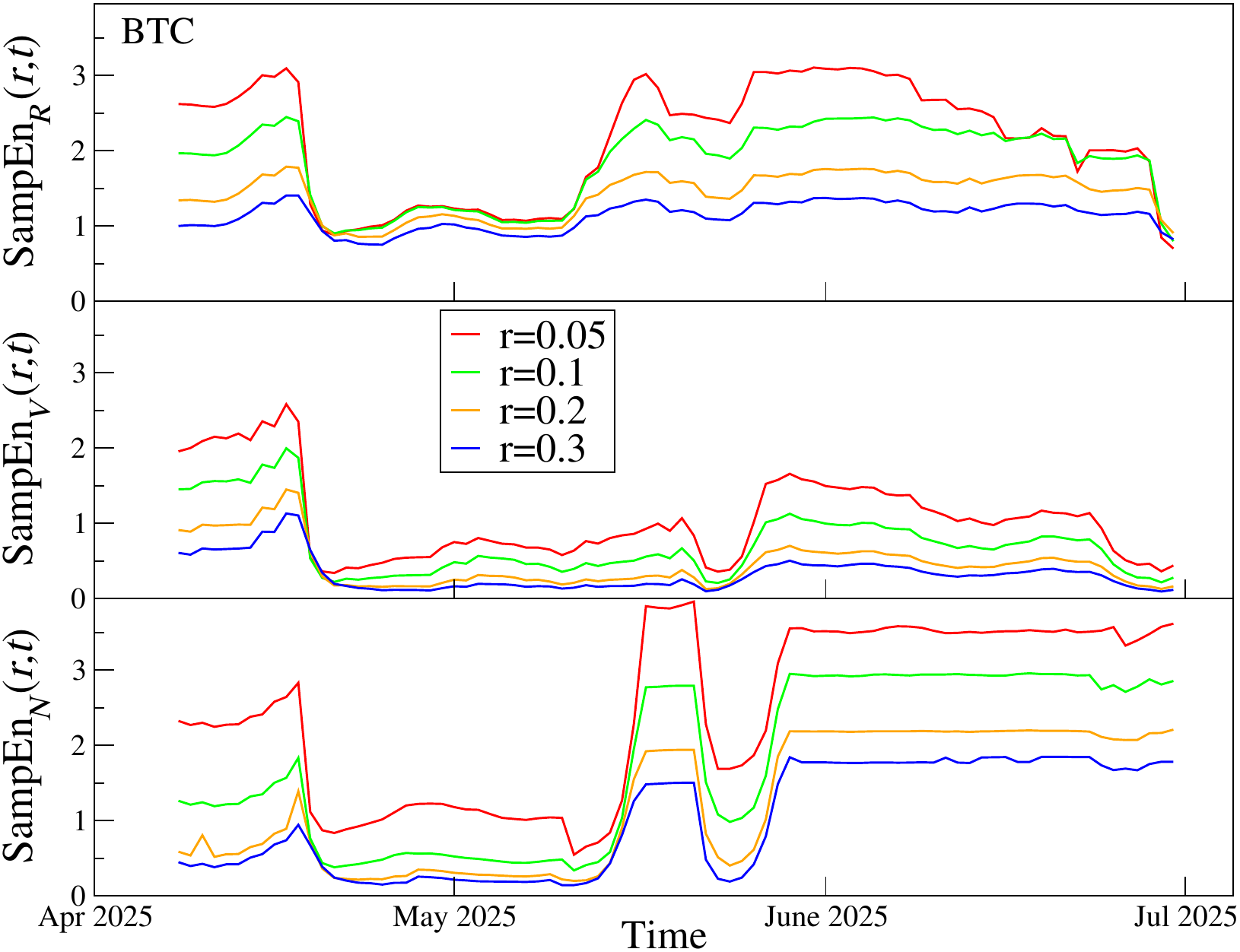}
\includegraphics[width=0.49\textwidth]{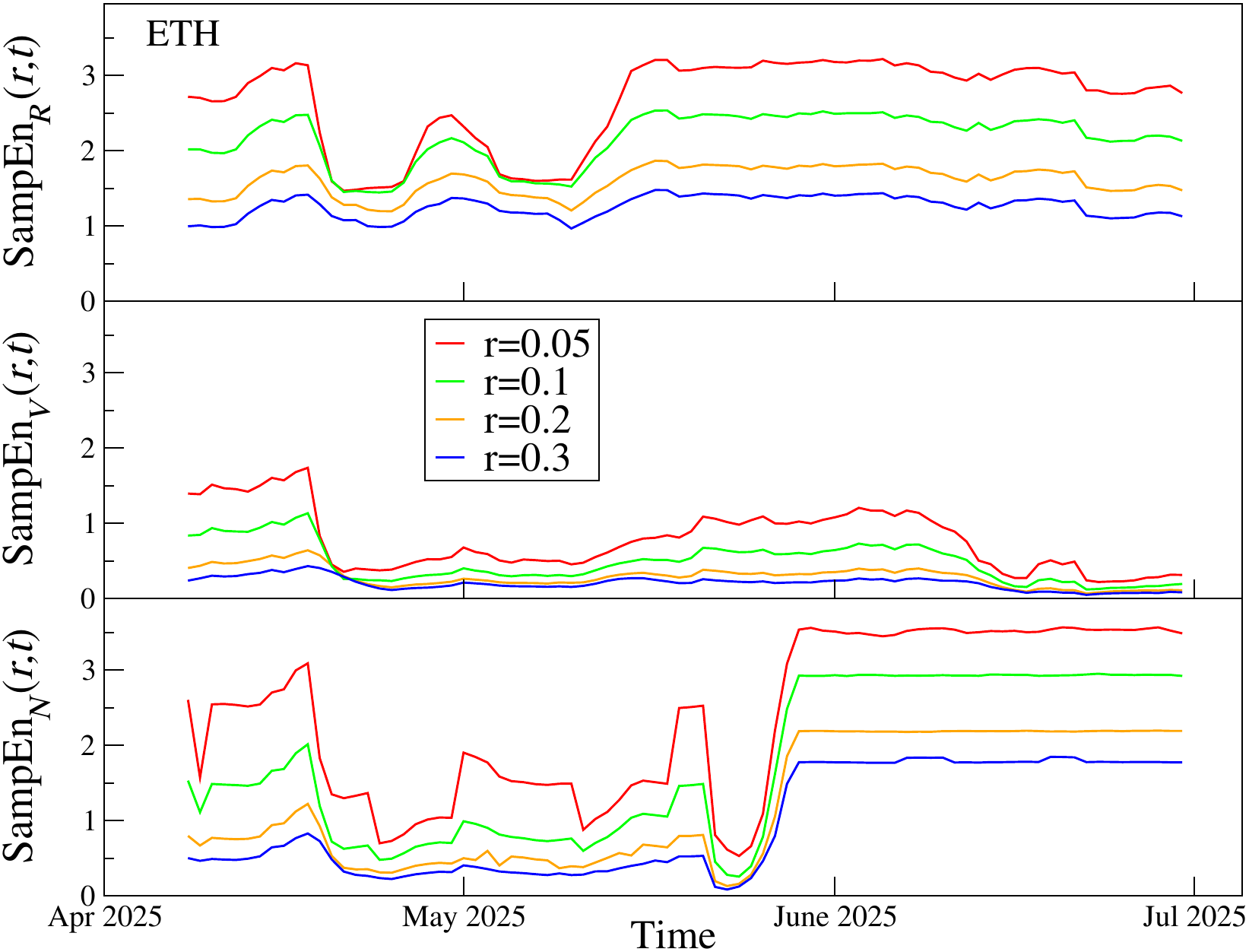}
\caption{Sample entropy calculated in a rolling window of length 7 days with a step of 24 hours for BTC and ETH log-returns $R$ (top), trading volume $V$ (middle), and the number of transactions $N$ (bottom) on Bitget for different tolerance values $r=0.05,0.1,0.2,0.3$ and fixed $m=2$. The time coordinate of each point denotes the end of the corresponding rolling window.}
\label{fig::SampEntrokno_r}
\end{figure}

\begin{figure}[ht!]
\centering
\includegraphics[width=0.49\textwidth]{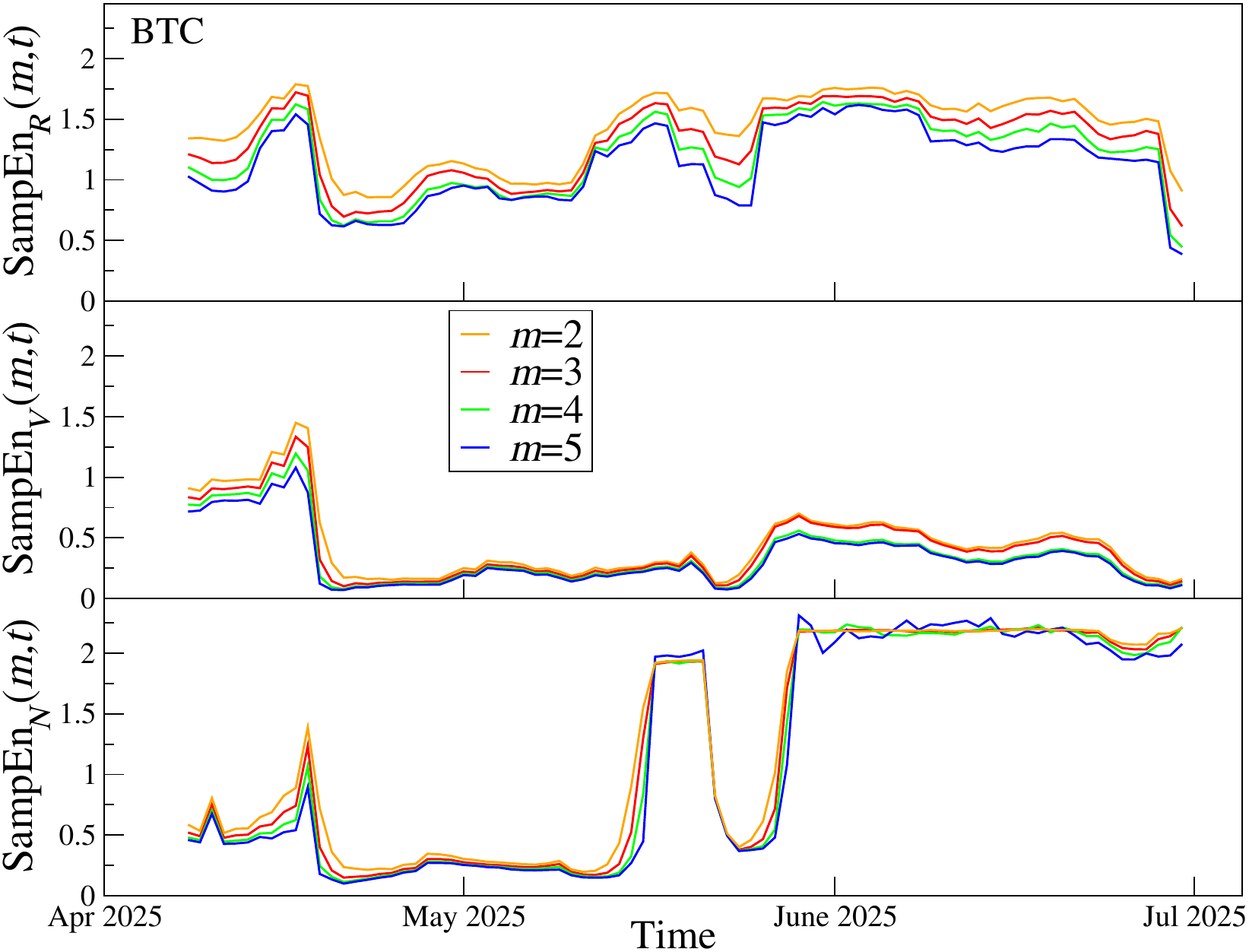}
\includegraphics[width=0.49\textwidth]{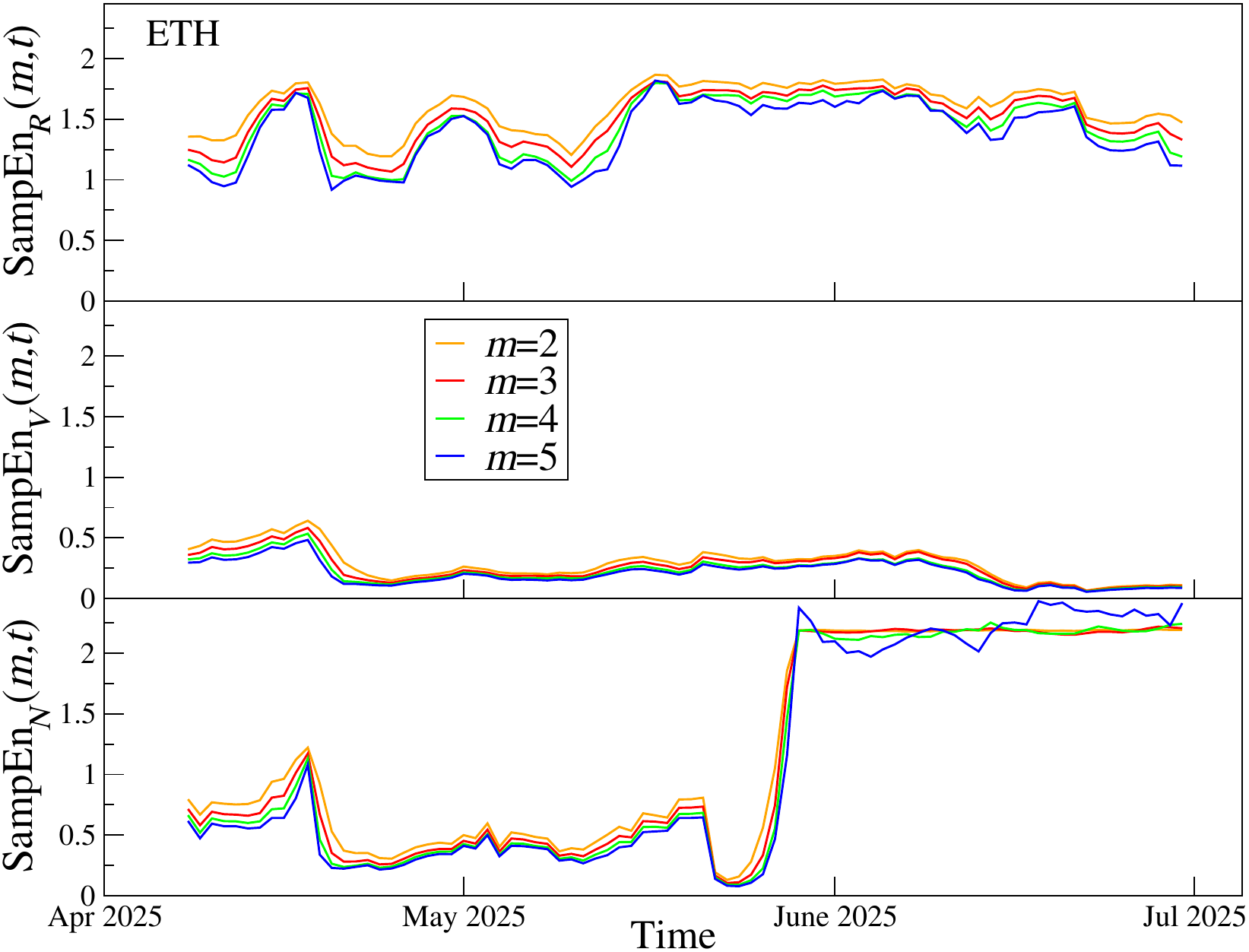}
\caption{Sample entropy calculated in a rolling window of length 7 days with a step of 24 hours for BTC and ETH log-returns $R$ (top), trading volume $V$ (middle), and the number of transactions $N$ (bottom) on Bitget for different embedding dimensions $m=2,3,4,5$ and fixed $r=0.2$. The time coordinate of each point denotes the end of the corresponding rolling window.}

\label{fig::SampEntrokno_dim}
\end{figure}

To assess whether the entropy-based conclusions are robust to the known limitations of approximate entropy, sample entropy (SampEn), defined in Section~\ref{Entr} by Eq.~\eqref{eq:sampen}, was also calculated. In contrast to ApEn, SampEn excludes self-matches and is therefore less biased, especially for finite samples~\cite{Richman2000}. The calculations were performed for Bitget BTC and ETH, where the strongest regime change had been detected. First, the tolerance parameter was varied, $r=0.05,0.1,0.2,0.3$, while keeping $m=2$ fixed. Second, the embedding dimension was varied, $m=2,3,4,5$, while keeping $r=0.2$ fixed. The results are shown in Figs.~\ref{fig::SampEntrokno_r} and~\ref{fig::SampEntrokno_dim}, respectively.

The SampEn results confirm the main conclusions obtained from ApEn. For log-returns and traded volume, the qualitative temporal evolution is very similar to that obtained from approximate entropy. The most important change is again observed for the number of transactions on Bitget. For both BTC and ETH, SampEn for $N$ increases sharply around mid-May and remains elevated afterwards, indicating a clear change in the temporal organisation of the transaction-count series. This confirms that the increase in irregularity of $N$ is not an artefact of self-matching in ApEn.

The sensitivity to the embedding dimension is also weaker for SampEn than for ApEn. In particular, the opposite behaviour observed for ApEn at smaller and larger embedding dimensions is no longer dominant: although the numerical level of SampEn changes with $m$, the timing and qualitative character of the regime shift remain stable. The sensitivity to the tolerance parameter $r$ mainly affects the absolute entropy level, but not the occurrence of the mid-May transition. Additional calculations using a shorter rolling window of 24 hours (not shown) produced more irregular and noisy curves, as expected due to the smaller number of observations in each window, but the increase in SampEn for the Bitget transaction-count series occurred at the same time. Thus, the SampEn analysis supports the robustness of the entropy-based evidence for a Bitget-specific regime shift in BTC and ETH transaction counts.

\section{Bitget regime-change analysis}

To complement the visual identification of regime changes in previous sections, a formal change-point detection procedure was applied to the transaction-count time series. Such methods are widely used to identify structural breaks in time series~\cite{Killick2012,Truong2020}. For each asset and exchange, the analysed series was defined as
\begin{equation}
    x_t = \log\left(1 + N_{\Delta t=1\mathrm{min}}(t)\right),
\end{equation}
where \(N_{\Delta t=1\mathrm{min}}(t)\) denotes the number of transactions observed in the \(t\)-th one-minute interval. The logarithmic transformation was used to reduce the influence of extreme transaction-count bursts and to focus on relative rather than absolute changes in trading intensity.

Change points were estimated using the MATLAB \texttt{findchangepts}~\cite{Matlabchpoint,Lavielle2005} procedure with the mean-shift statistic. In the baseline specification, the maximum number of change points was set to one in order to identify the dominant structural break. In addition, a minimum distance of 1440 observations was imposed, corresponding to one day of one-minute data. Thus, the procedure was designed to identify persistent changes in the average level of \(\log(1+N_t)\), rather than short-lived intraday fluctuations.

For BTC and ETH traded on Bitget, the dominant change point was detected on May 21, 2025. This date coincides with the visually observed transition in the transaction-count time series in Figs.~\ref{fig::szeregiBTC} and~\ref{fig::szeregiETH} and was therefore used to divide the Bitget sample into two subperiods: Bitget1 from April 1 to May 20, 2025 and Bitget2, from May 21 to June 30, 2025. By contrast, no analogous dominant regime shift was detected for XRP on Bitget. This supports the interpretation that the regime change is asset-specific and affects mainly BTC and ETH transaction-count dynamics on Bitget.

To examine the structure of the post-break increase in the number of transaction records, an additional within-minute decomposition of the transaction-count series was performed. The aim was to determine whether the increase in $N_t$ is mainly due to a denser concentration of records within already active seconds or whether it reflects a more continuous low-volume transaction process. Since Bitget timestamps are recorded with one-second resolution, identical timestamps cannot be interpreted as direct evidence of child-fill splitting. Instead, for each one-minute interval $t$, the number of transaction records was decomposed into the number of active seconds and the average number of records per active second.

Let $n_{t,j}$ denote the number of transaction records observed in second $j=1,\ldots,60$ of minute $t$. Then
\begin{equation}
N_t=\sum_{j=1}^{60} n_{t,j},
\end{equation}
is the total number of transaction records in minute (t). The number of active seconds is defined as
\begin{equation}
A_t=\sum_{j=1}^{60}\mathbf{1}(n_{t,j}>0),
\end{equation}
where $\mathbf{1}(\cdot)$ is the indicator function. Finally, the average number of transaction records per active second is defined as
\begin{equation}
M_t=\frac{N_t}{A_t},
\end{equation}
for $A_t>0$.

\begin{figure}[ht!]
\centering
\includegraphics[width=0.44\textwidth]{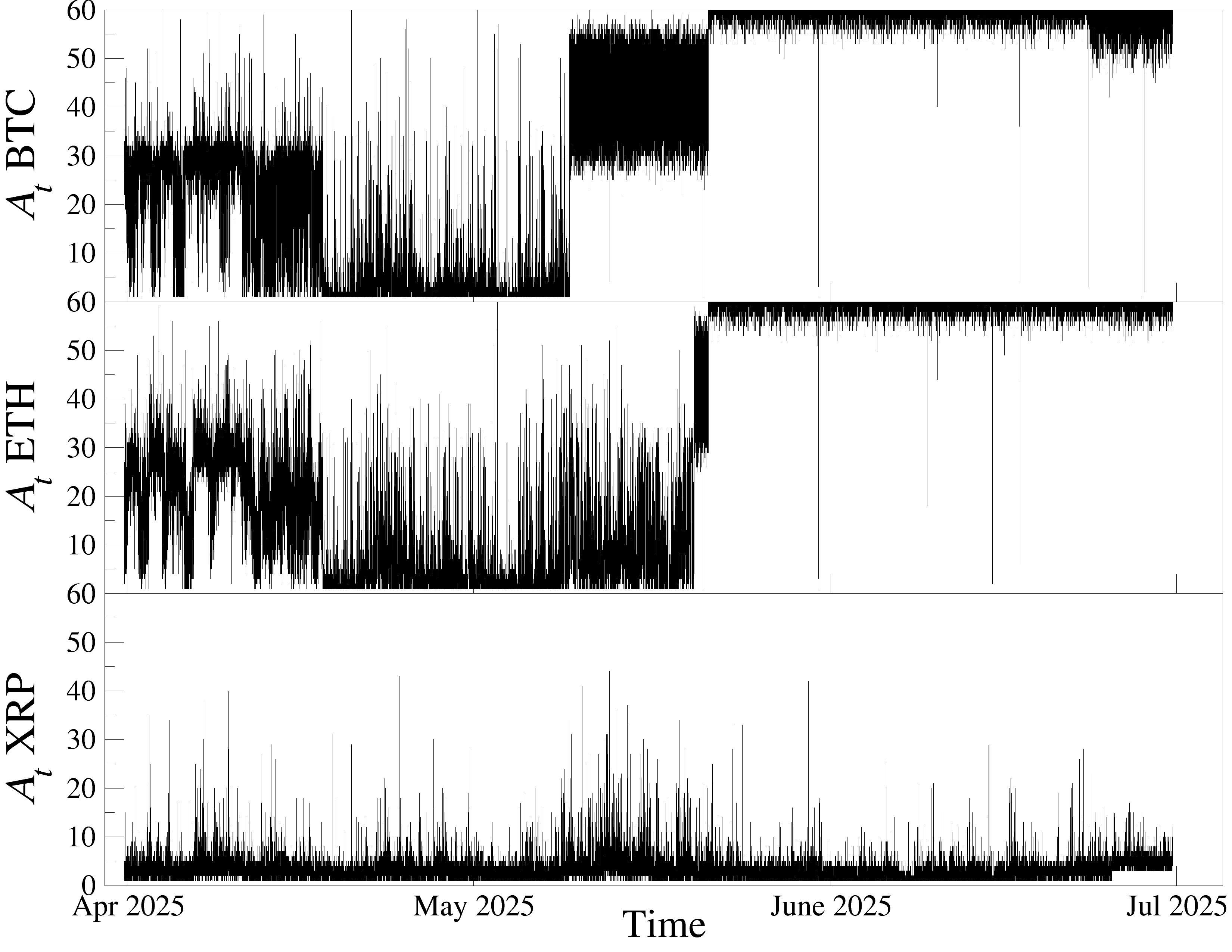}
\includegraphics[width=0.44\textwidth]{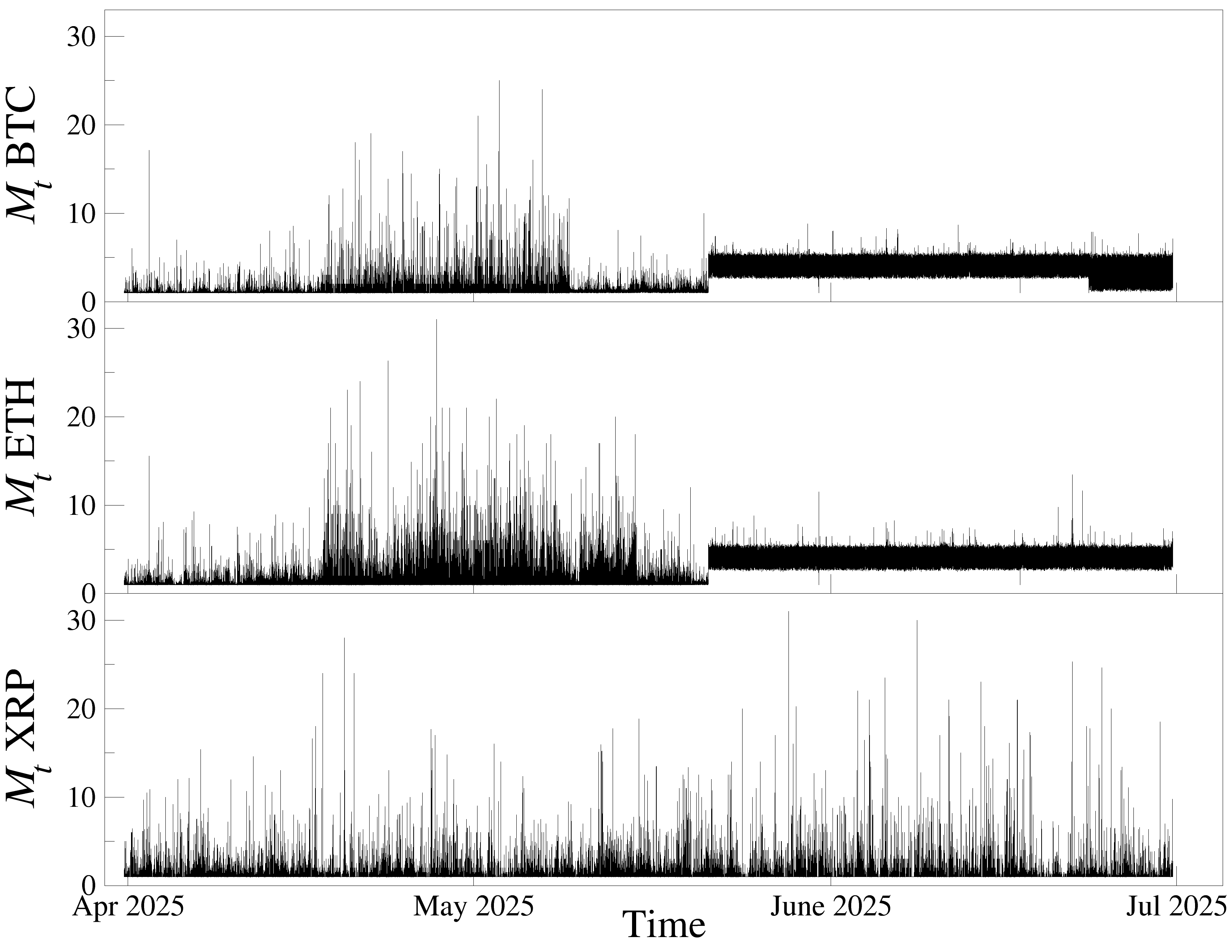}
\caption{Decomposition of the one-minute transaction count on Bitget. Left: number of active seconds per one-minute interval, $A_t$. Right: average number of transaction records per active second, $M_t=N_t/A_t$. Results are shown for BTC, ETH, and XRP.}
\label{fig::statBitget2}
\end{figure}

The results are shown in Fig.~\ref{fig::statBitget2}. For BTC, the average number of active seconds increased from approximately $12.0$ in Bitget1 to $55.5$ in Bitget2. At the same time, the average number of transaction records per active second increased from approximately $1.9$ to $3.3$. Thus, the increase in $N_t$ is not driven only by denser reporting of records within already active seconds. Rather, a large part of the effect comes from the fact that, after the break, trading records appear in almost every second of a minute. In multiplicative terms, the increase in $A_t$ is about $55.5/12.0 \simeq 4.6$, whereas the increase in $M_t$ is about $3.3/1.9 \simeq 1.7$. Therefore, even under the conservative aggregation in which all records with the same one-second timestamp are collapsed into a single event, the post-break activity remains more than four times higher than before the break. A similar qualitative pattern is observed for ETH, whereas XRP does not exhibit an analogous transition. This cross-asset difference makes a purely exchange-wide reporting-change explanation unlikely.

\begin{figure}[ht!]
\centering
\includegraphics[width=0.44\textwidth]{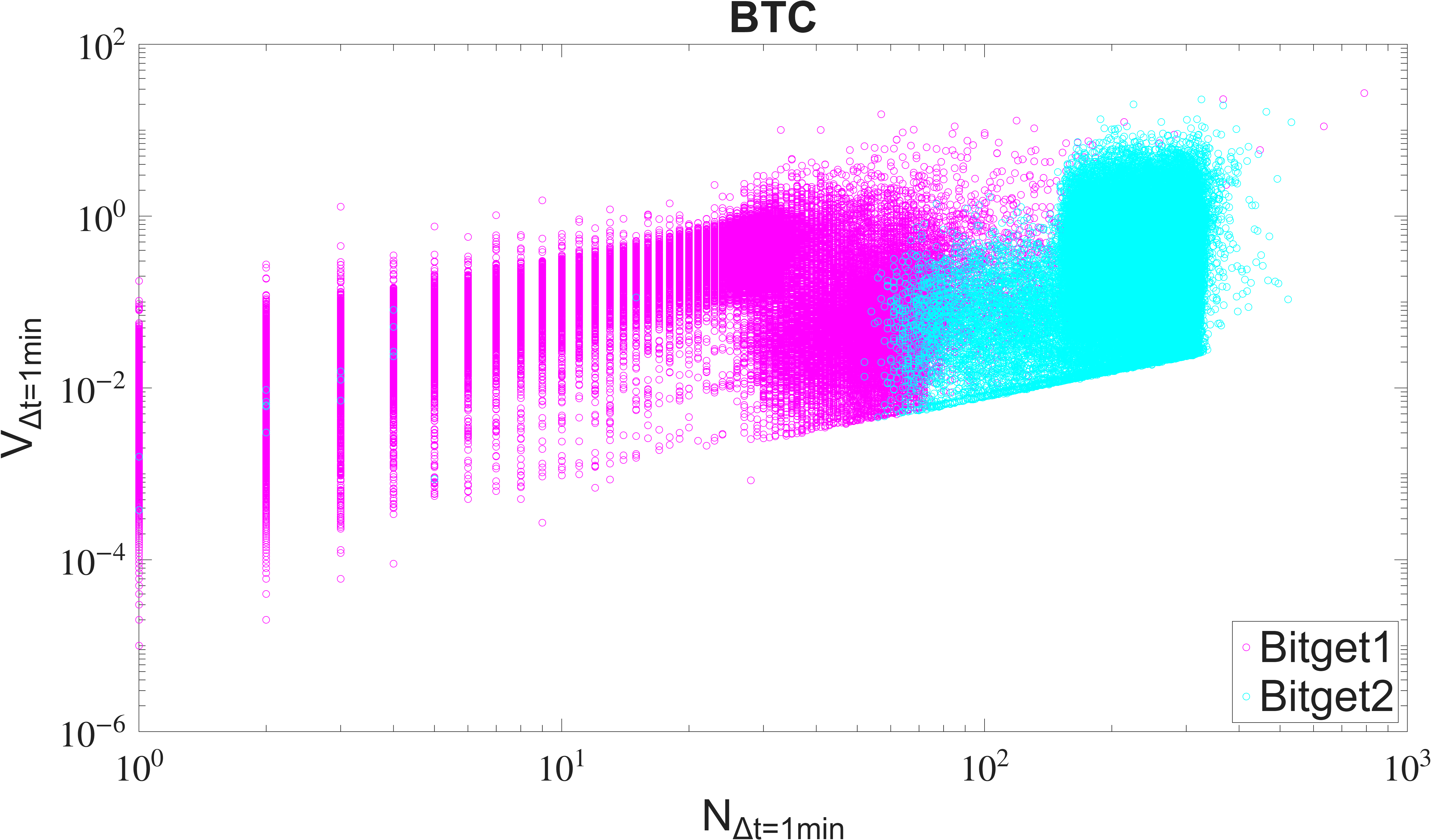}
\includegraphics[width=0.44\textwidth]{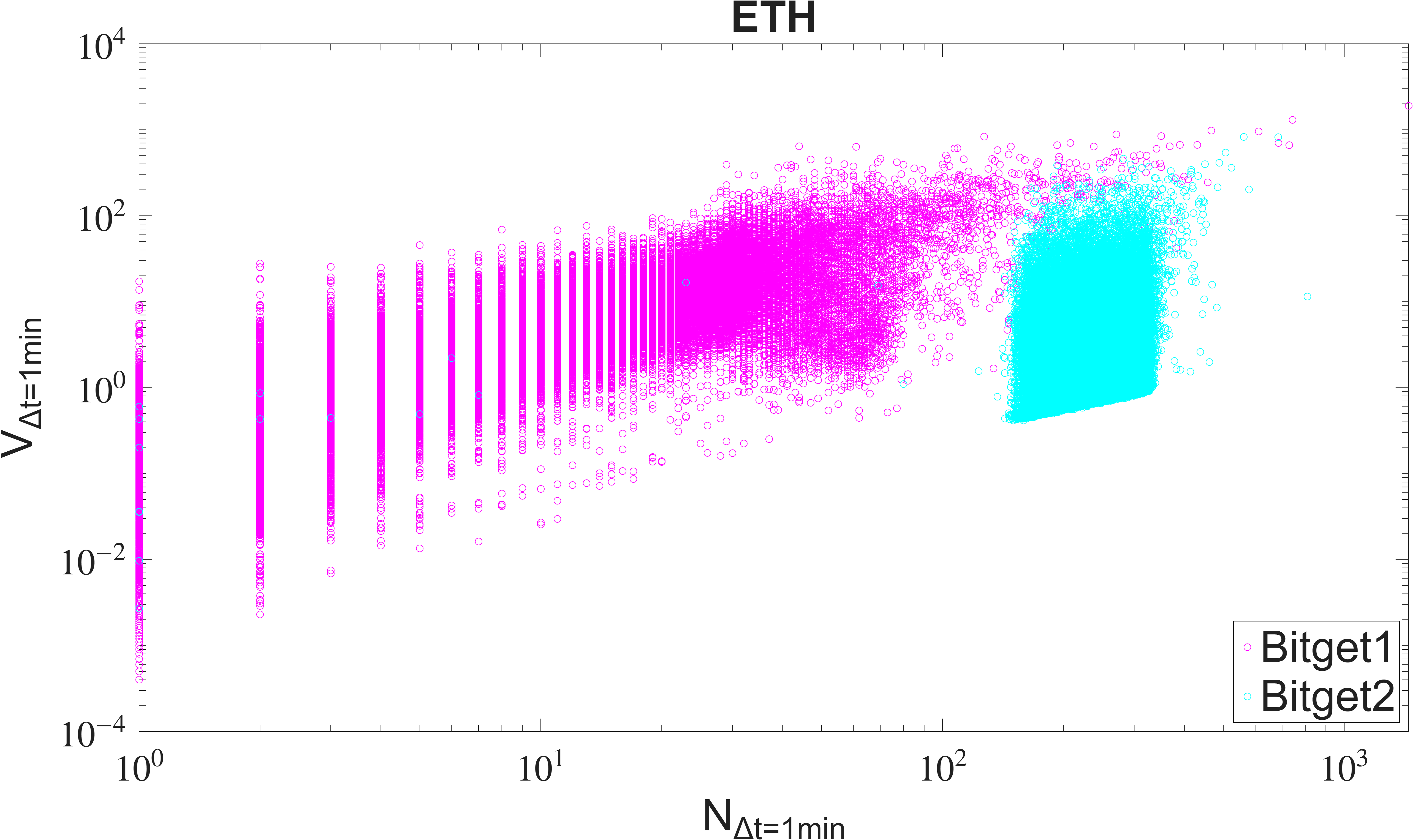}
\includegraphics[width=0.44\textwidth]{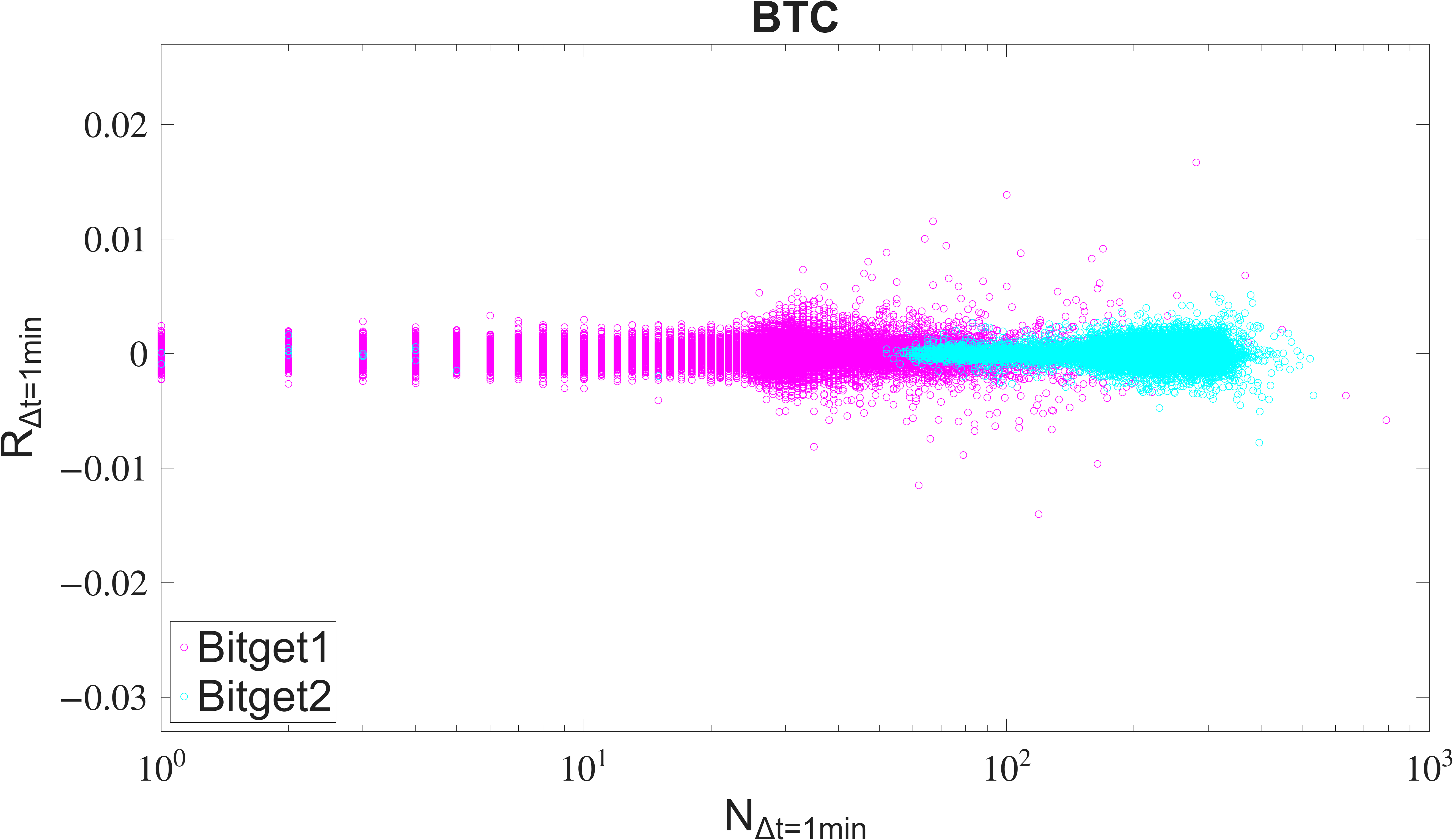}
\includegraphics[width=0.44\textwidth]{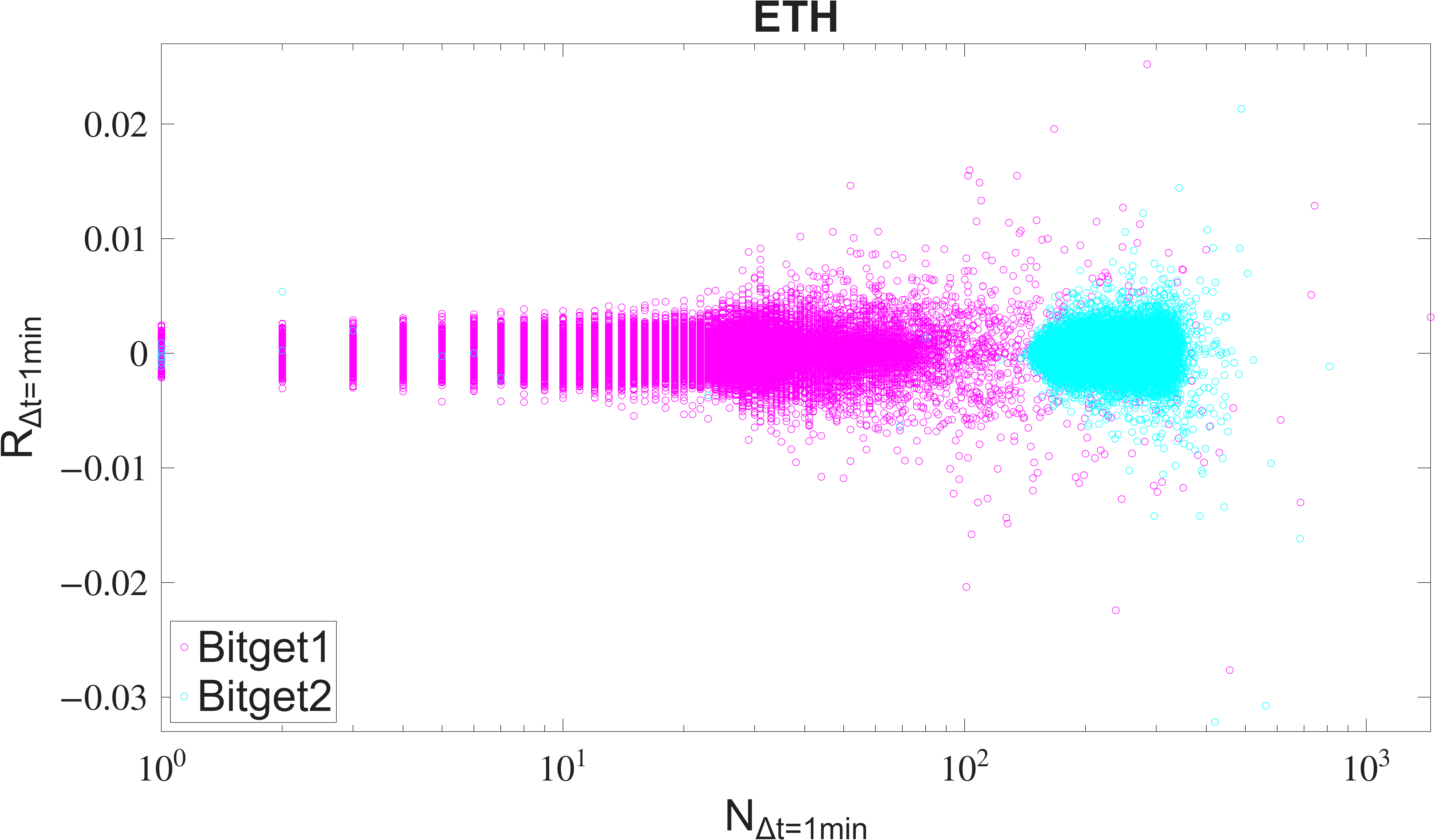}
\caption{Relations between the number of transactions $N_{\Delta t=1\mathrm{min}}$ and trading volume $V_{\Delta t=1\mathrm{min}}$ (top panels), and between log-returns $R_{\Delta t=1\mathrm{min}}$ and $N_{\Delta t=1\mathrm{min}}$ (bottom panels), on Bitget in the two periods of interest for BTC (left) and ETH (right).}
\label{fig::BTC_ETH_V_N}
\end{figure}

After splitting the sample into two periods, a clear difference in the relation between the number of transactions and trading volume is observed for Bitget, as shown in the upper panels of Fig.~\ref{fig::BTC_ETH_V_N}. The observations recorded in Bitget2 form a separate cluster, in which a higher number of transactions does not translate into higher traded volume, especially for ETH. This indicates that the second period is dominated by numerous low-volume transactions, which substantially increase $N_{\Delta t=1\mathrm{min}}(t)$ without a proportional increase in $V_{\Delta t=1\mathrm{min}}(t)$. The separation between the two periods is also visible in the $R$--$N$ relation, shown in the lower panels of Fig.~\ref{fig::BTC_ETH_V_N}. In Bitget2, a larger number of transactions is not associated with larger deviations in returns. This means that the increase in the number of transactions after the regime change does not translate into greater price variability.

\begin{figure}[ht!]
\centering
\includegraphics[width=0.99\textwidth]{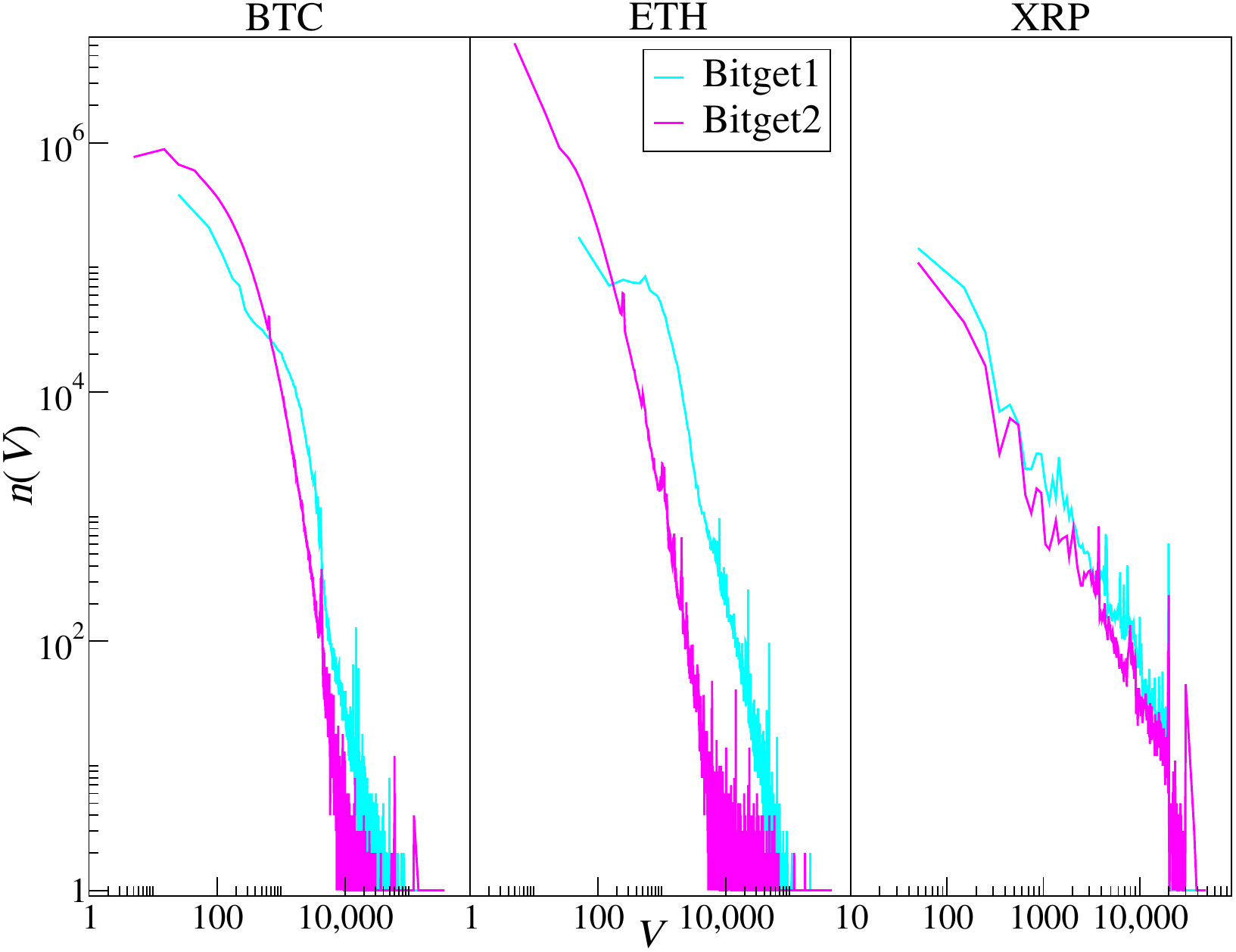}
\caption{Trade-size histograms for BTC, ETH, and XRP on Bitget in the two periods of interest. The vertical axis shows the number of trades $n(V)$ in each trade-size bin.}
\label{fig::tradesizeBitget.pdf}
\end{figure}

The trade-size distributions in Fig.~\ref{fig::tradesizeBitget.pdf} provide additional evidence that the post-break BTC and ETH regimes are dominated by smaller transaction records. This distributional diagnostic is consistent with standard trade-size-based approaches used to detect suspicious trading activity on centralised cryptocurrency exchanges~\cite{Chen2022,Cong2022CryptoWashTrading,Amiram2025}. For BTC and ETH, Bitget2 is shifted toward smaller trade sizes relative to Bitget1, whereas no comparable shift is observed for XRP. This is consistent with the time-series evidence showing that the elevated number of records in Bitget2 is generated by low-volume transactions.

\begin{figure}[ht!]
\centering
\includegraphics[width=0.44\textwidth]{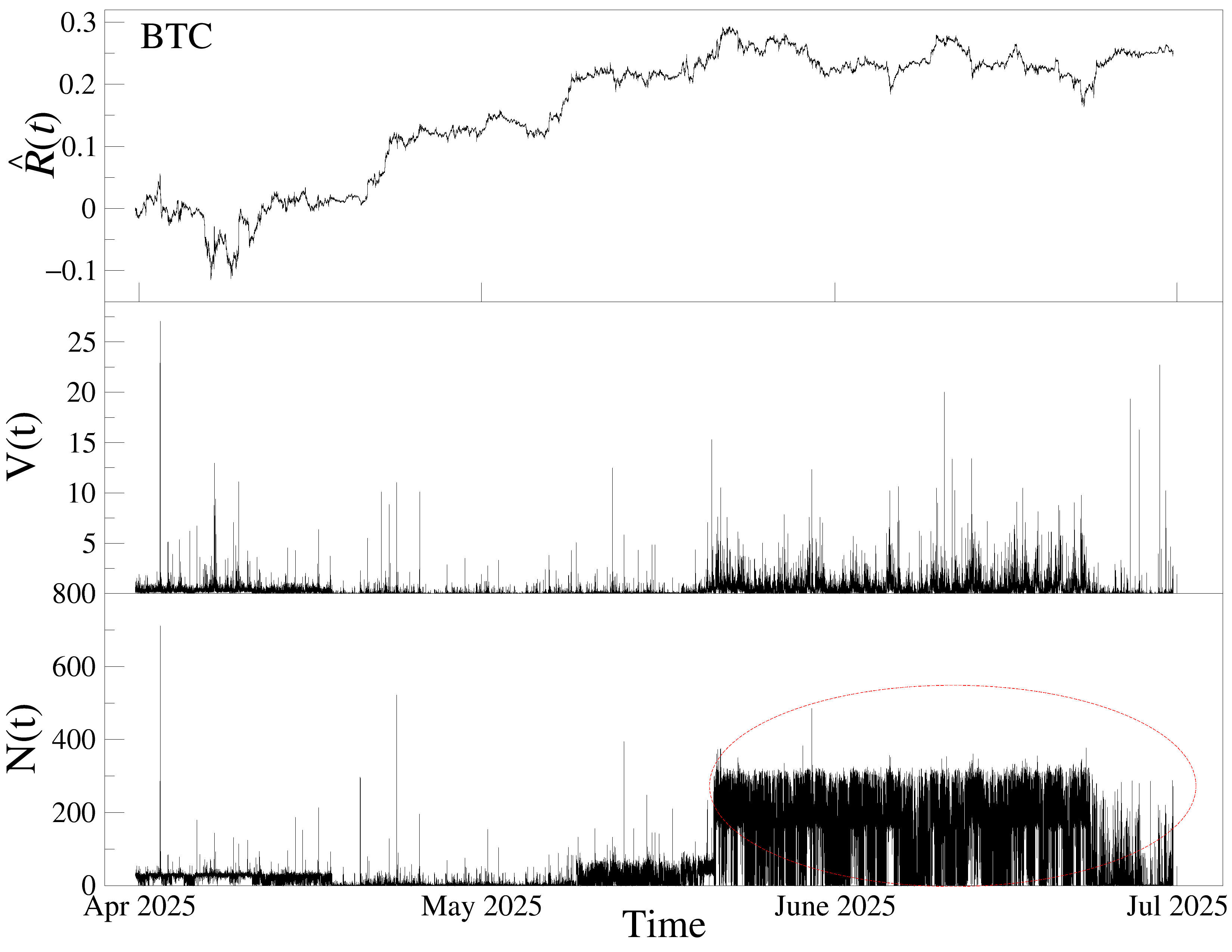}
\includegraphics[width=0.44\textwidth]{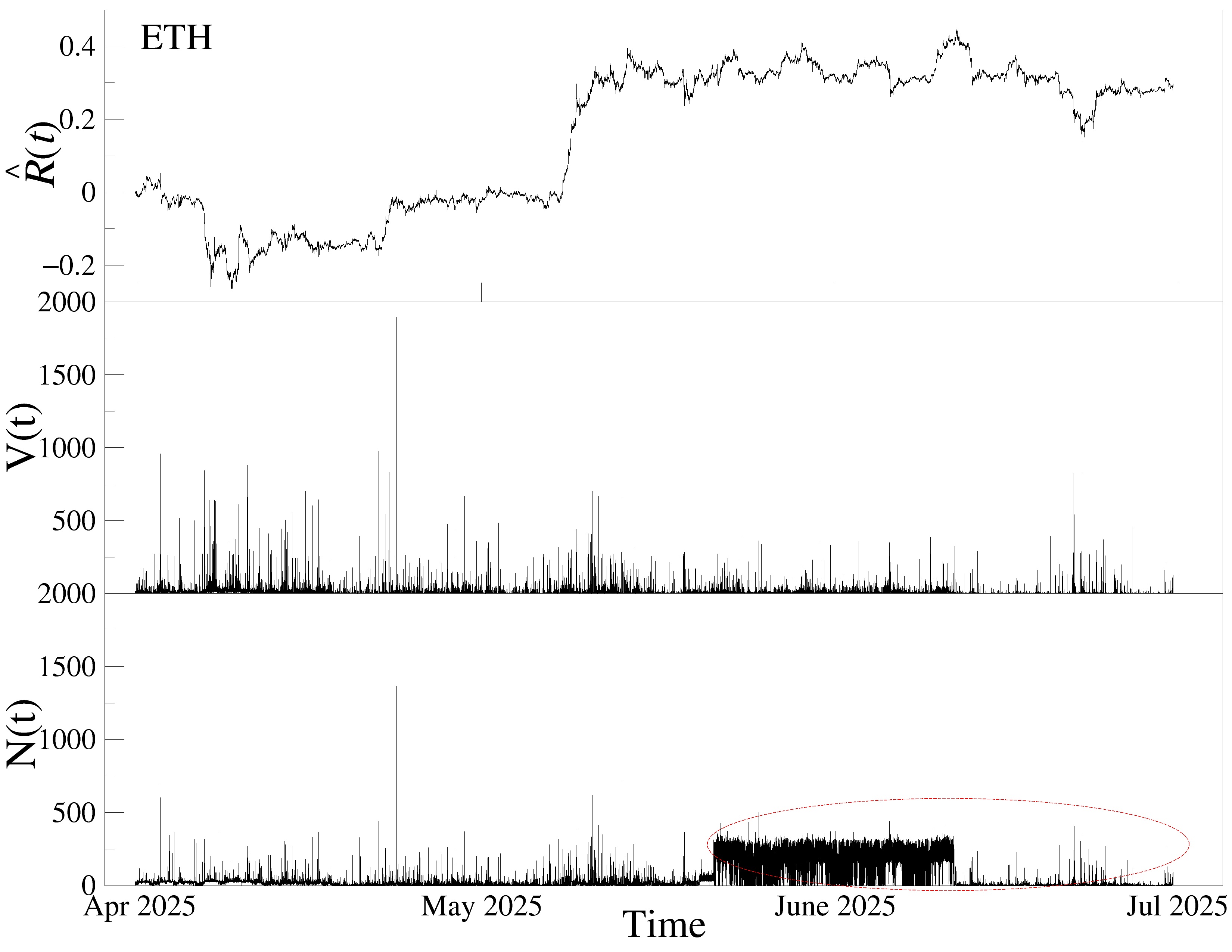}
\caption{Time series from Bitget after removing transactions with volumes below $0.001$ BTC (left) and $0.01$ ETH (right). The average volume per transaction is $0.0078$ BTC in Bitget1 and $0.0024$ BTC in Bitget2, and $0.6636$ ETH in Bitget1 and $0.0268$ ETH in Bitget2. The periods with increased transaction activity from Figs.~\ref{fig::szeregiBTC} and~\ref{fig::szeregiETH} are marked by a red dashed ellipse.}
\label{fig::szeregiBTC_ETHu}
\end{figure}

This interpretation is further supported by Fig.~\ref{fig::szeregiBTC_ETHu}. After removing transactions with the smallest volumes, below $0.001$ BTC and below $0.01$ ETH, periods without trading become visible. Such periods are not visible in the original full transaction-count series presented in the bottom panels of Figs.~\ref{fig::szeregiBTC} and~\ref{fig::szeregiETH}. This confirms that the elevated transaction count in Bitget2 is largely driven by very small trades. Since public data do not include account identifiers or order-level information, these results should be interpreted as evidence of an unusually low-volume transaction component rather than direct proof of wash trading.

\begin{figure}[ht!]
\centering
\includegraphics[width=0.99\textwidth]{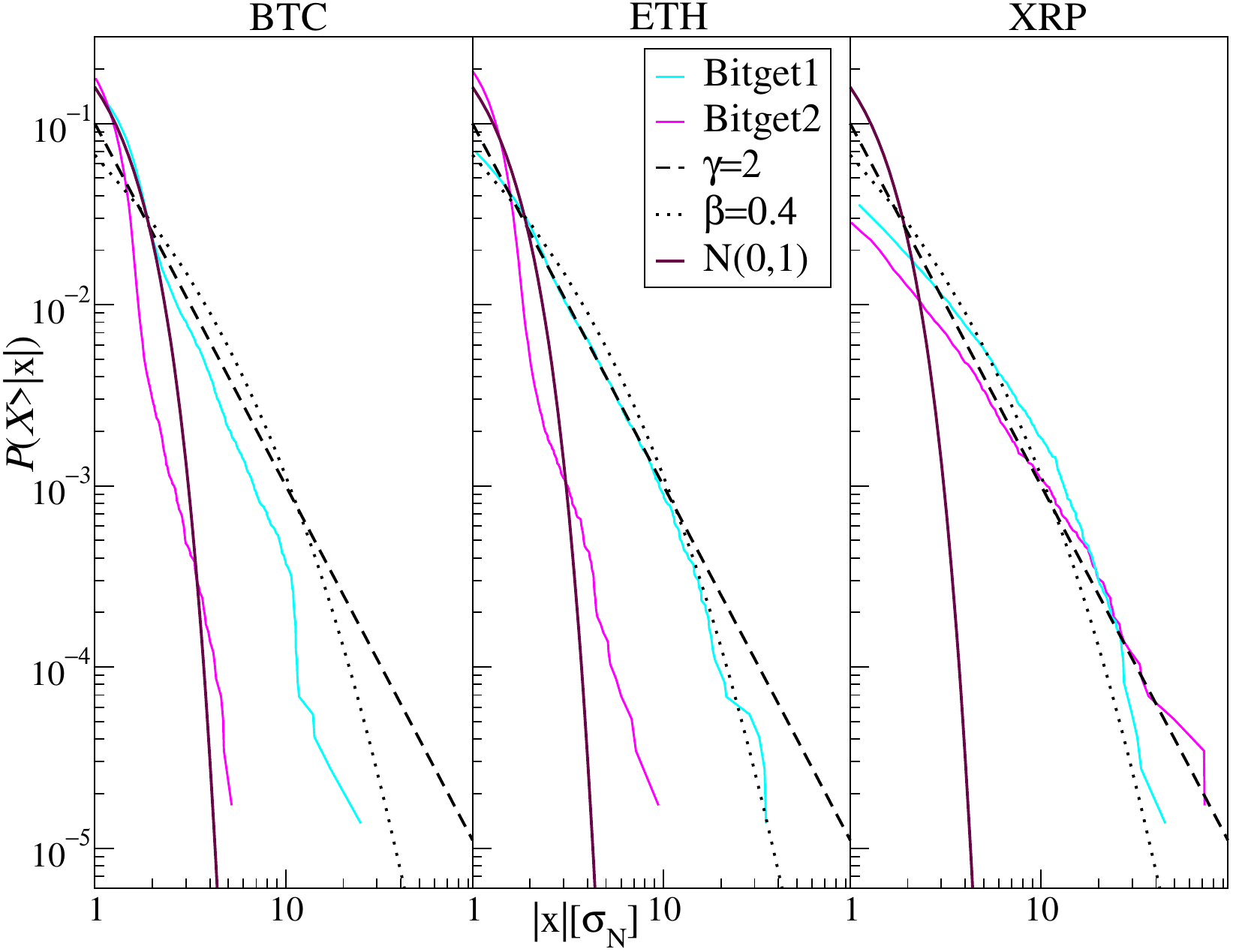}
\caption{Complementary cumulative distribution functions for the number of transactions $N$ for BTC, ETH, and XRP on Bitget in the two periods of interest, together with the Gaussian, power-law with $\gamma=2$, and stretched exponential with $\beta=0.4$ reference distributions.}
\label{fig::returnsNBitget.pdf}
\end{figure}

The difference between the two periods is also clearly visible in the CCDFs of the number of transactions shown in Fig.~\ref{fig::returnsNBitget.pdf}. For BTC and ETH, the distributions on Bitget2 are close to Gaussian. In contrast, the distribution in Bitget1 resembles those observed on the other exchanges in Fig.~\ref{fig::returnsN.pdf} and can be fitted with a stretched exponential distribution. For XRP, no comparable difference between Bitget1 and Bitget2 is observed.

\begin{figure}[ht!]
\centering
\includegraphics[width=0.99\textwidth]{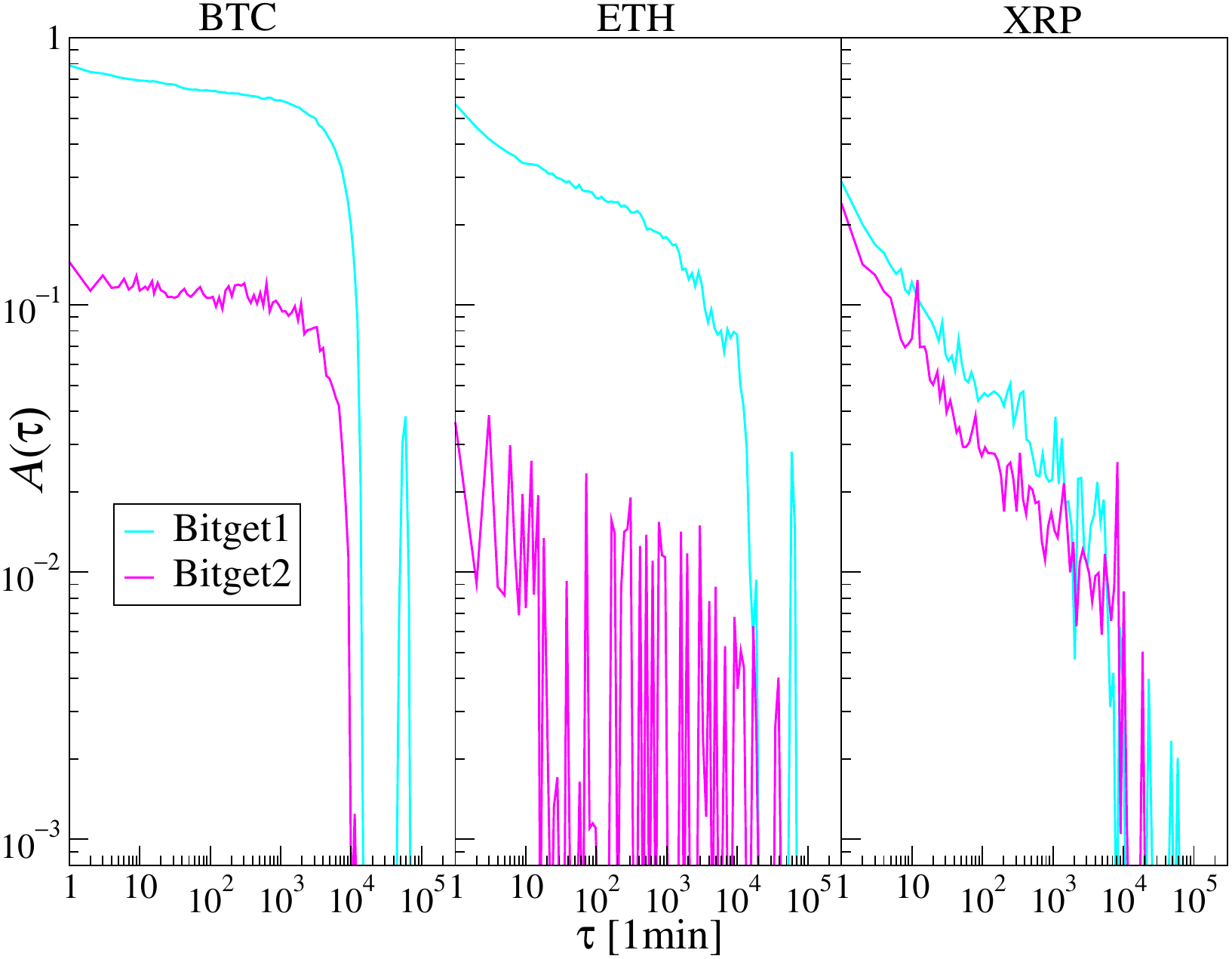}
\caption{Autocorrelation functions of the number of transactions $N$ for BTC, ETH, and XRP on Bitget in the two periods of interest.}
\label{fig::acfNBitget.pdf}
\end{figure}

A similar conclusion follows from ACFs of the number of transactions shown in Fig.~\ref{fig::acfNBitget.pdf}. In Bitget1, the autocorrelation of $N_{\Delta t=1\mathrm{min}}(t)$ behaves similarly to the autocorrelations observed on the other exchanges. In Bitget2, however, ACFs for BTC and ETH become much weaker. For ETH, they are close to the behaviour expected for an almost uncorrelated process. Again, XRP does not exhibit a comparable change between the two periods.

\begin{figure}[ht!]
\centering
\includegraphics[width=0.49\textwidth]{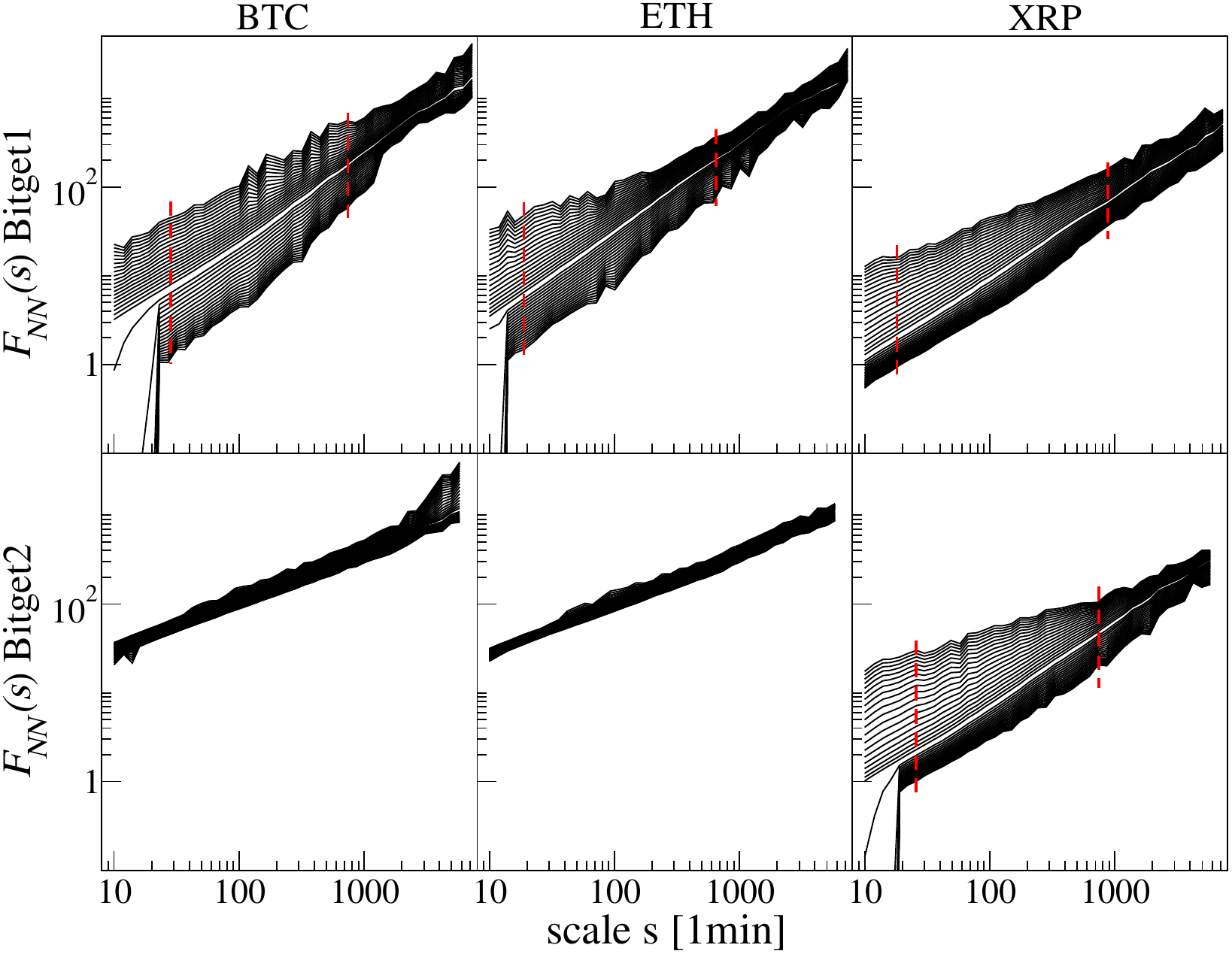}
\includegraphics[width=0.49\textwidth]{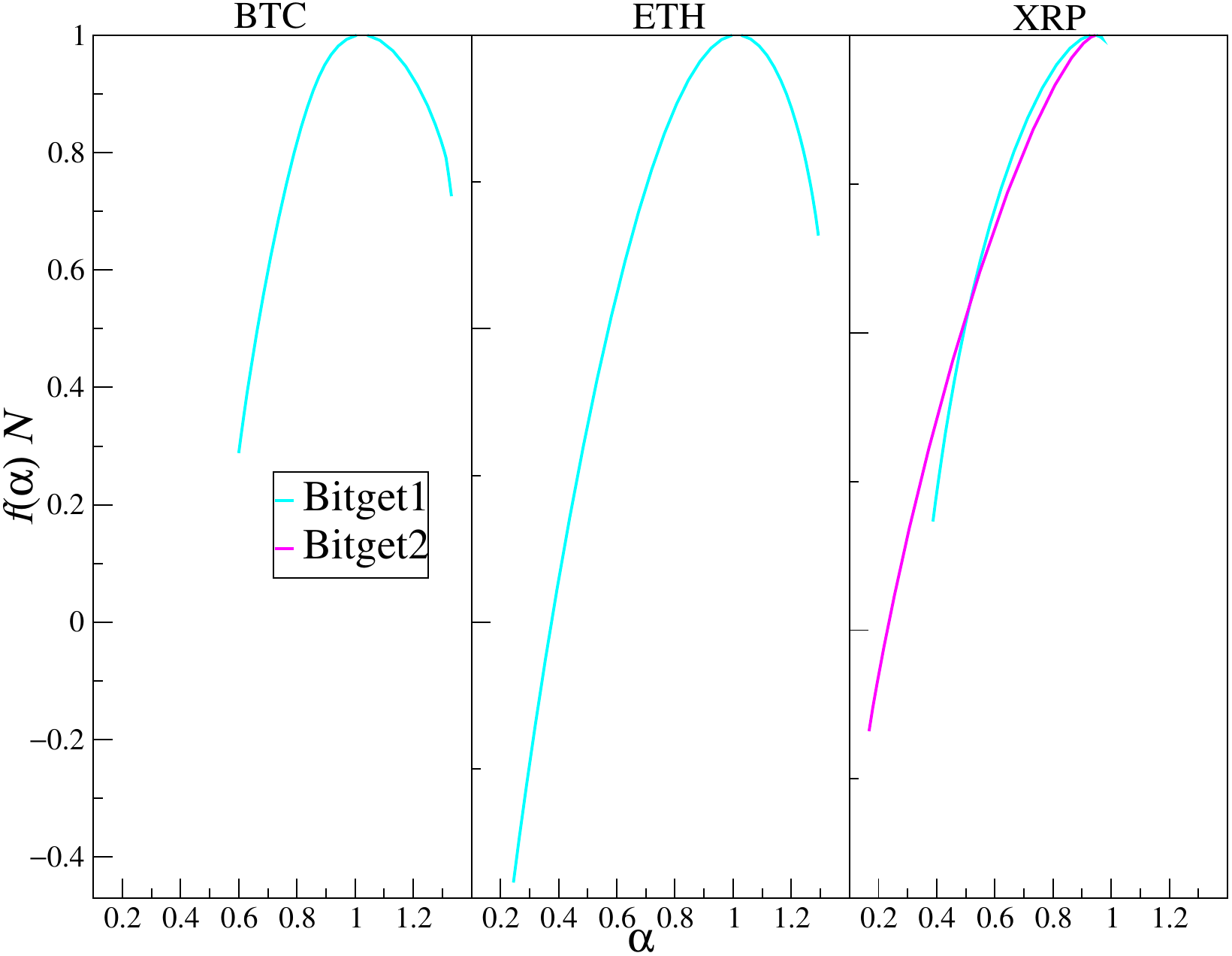}
\caption{Fluctuation functions $F_{NN}(s)$ (left panel) and the corresponding multifractal spectra (right panel) for the number of transactions $N$ for BTC, ETH, and XRP on Bitget in two periods. Dashed red lines indicate the scale range selected for determining the multifractal~spectra.}
\label{figBitget::FqNR}
\end{figure}

The hypothesis about the random-like nature of the transaction-count process in the second period on Bitget for BTC and ETH is supported by the fluctuation functions shown in the left panel of Fig.~\ref{figBitget::FqNR}. While in Bitget1 it is possible to determine the multifractal spectrum, the fluctuation functions in Bitget2 exhibit a monofractal character. The spectra presented in the right panel of Fig.~\ref{figBitget::FqNR} for Bitget1 are similar to those observed for the number of transactions on the other exchanges in Fig.~\ref{fig::spectra}. This indicates that the regime change reduces the multifractal organisation of the transaction-count series and makes the process more homogeneous across scales.

\begin{figure}[ht!]
\centering
\includegraphics[width=0.99\textwidth]{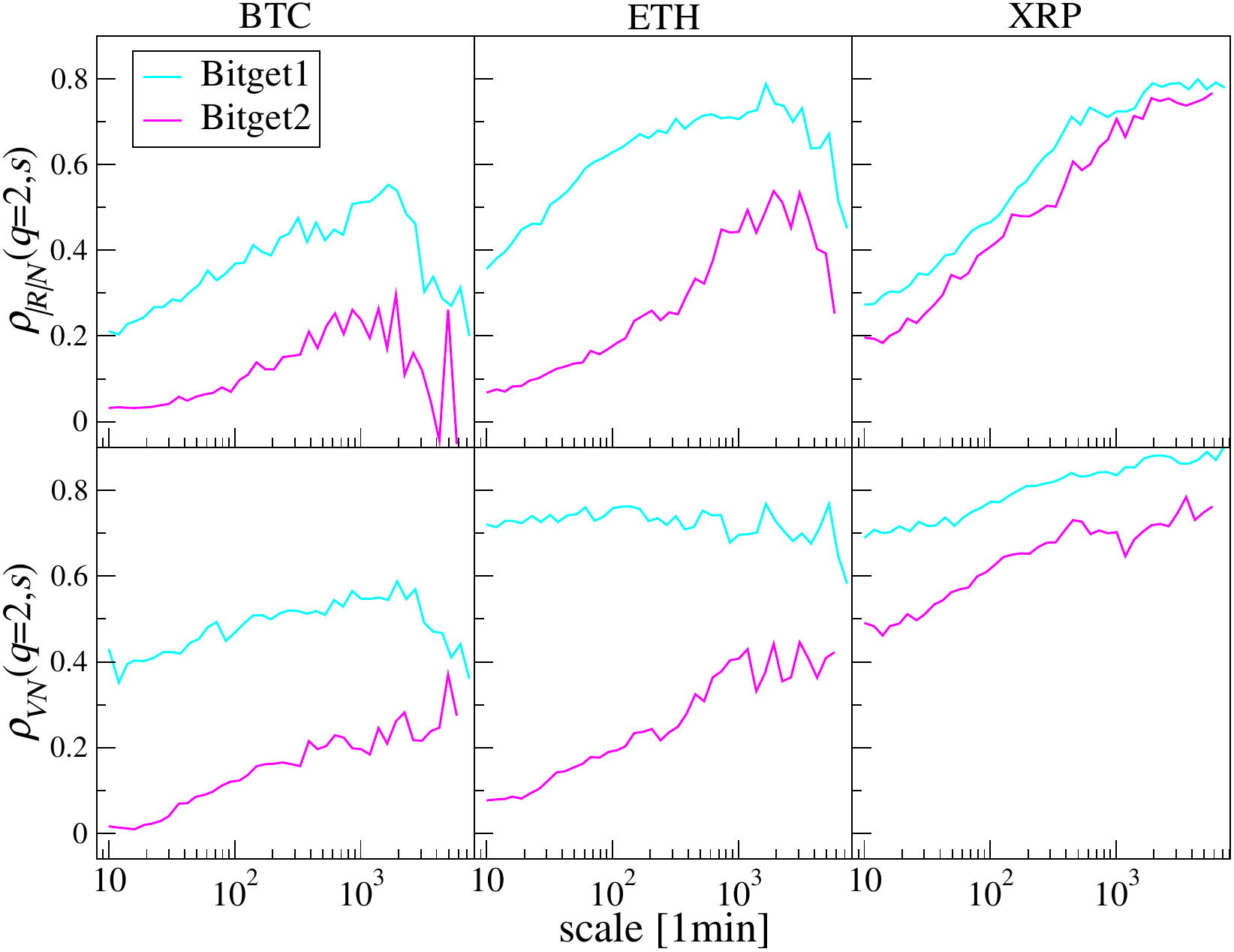}
\caption{Detrended cross-correlation coefficient $\rho(q=2,s)$ calculated between absolute log-returns $|R|$ and the number of transactions $N$ (top panels) and between trading volume $V$ and the number of transactions $N$ (bottom panels), for BTC, ETH, and XRP on Bitget in the two periods of interest.}
\label{fig::rhor.XYBitget}
\end{figure}

The observed differences in the $V$--$N$ and $|R|$--$N$ relations are confirmed by the detrended cross-correlation coefficients shown in Fig.~\ref{fig::rhor.XYBitget}. For BTC and ETH, the values of $\rho(q=2,s)$ in Bitget2 are substantially lower than in Bitget1. At the shortest scales, the coefficients are close to zero. Although they increase with $s$, the gap between Bitget1 and Bitget2 remains large over a broad range of scales. The effect is particularly strong for ETH in the $V$--$N$ relation, where the coefficient in Bitget2 remains much lower than in Bitget1. For XRP, the difference between the two periods is considerably weaker, which is consistent with the previous results.

The previous observations indicate that BTC and ETH trading on Bitget in the second period is characterised by a large number of weakly correlated, low-volume transactions that have little impact on either traded volume or price changes. This interpretation is supported by several observations: the nearly Gaussian distribution of $N_{\Delta t=1\mathrm{min}}(t)$, the weak autocorrelation of the transaction-count series, the monofractal form of the fluctuation functions for $N_{\Delta t=1\mathrm{min}}(t)$, and the weak cross-correlations between $N$ and both $V$ and $|R|$. Moreover, the change in the transaction-count characteristics for BTC and ETH occurs approximately at the same time as shown in Figs.~\ref{fig::szeregiBTC} and~\ref{fig::szeregiETH}. This raises the question of whether the trading characteristics of BTC and ETH became mutually synchronised during the anomalous period.

\begin{figure}[ht!]
\centering
\includegraphics[width=0.99\textwidth]{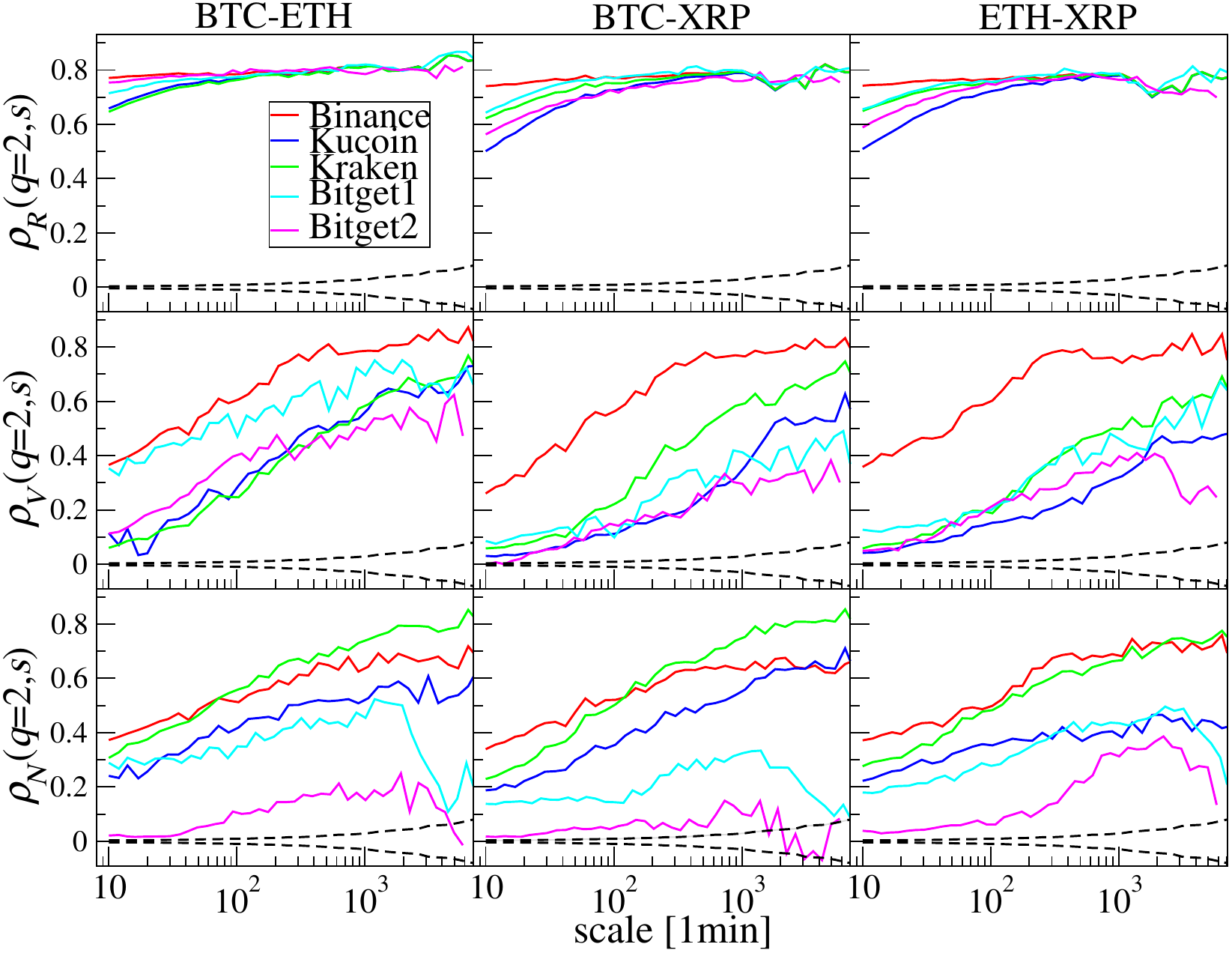}
\caption{Detrended cross-correlation coefficient $\rho(q=2,s)$ calculated between cryptocurrency pairs for  log-returns $R$ (top panels), trading volume $V$ (middle panels), and the number of transactions $N$ (bottom panels). The columns correspond to BTC--ETH, BTC--XRP, and ETH--XRP. Results are shown for Binance, Kraken, KuCoin, and Bitget split into Bitget1 and Bitget2. The dashed curves denote the scale-dependent $\pm 1$ standard-deviation surrogate interval of $\rho(q=2,s)$ obtained from 100 independently shuffled realisations preserving the distributions of the original time series.}
\label{fig::rhor.XY_BTCETHXRP_Bitget}
\end{figure}

This issue was examined by calculating $\rho(q=2,s)$ between pairs of cryptocurrencies: BTC--ETH, BTC--XRP, and ETH--XRP. The calculation was performed separately for log-returns $R$, trading volume $V$, and the number of transactions $N$. The results are presented in Fig.~\ref{fig::rhor.XY_BTCETHXRP_Bitget}.

For log-returns (top panels in Fig.~\ref{fig::rhor.XY_BTCETHXRP_Bitget}), strong correlations are observed between all cryptocurrency pairs and across all exchanges, with values typically in the range 0.7--0.8. This indicates that price dynamics remain strongly synchronised across assets and that the Bitget regime change does not affect the cross-asset return dependence. More pronounced differences are visible for trading volume (middle panels in Fig.~\ref{fig::rhor.XY_BTCETHXRP_Bitget}). The strongest volume correlations are observed on Binance, followed by Bitget1. In Bitget2, the volume correlations are slightly weaker, but they remain comparable to those observed on Kraken and KuCoin. Thus, the volume dynamics on Bitget do not show a complete breakdown of the cross-asset synchronisation.

The largest differences occur for the number of transactions (bottom panels in Fig.~\ref{fig::rhor.XY_BTCETHXRP_Bitget}). In Bitget2, the correlations between transaction-count series are the weakest, typically ranging from 0 to 0.2. Moreover, at longer scales the Bitget2 correlations remain close to, and in some cases fall within, the surrogate-based reference interval, unlike the correlations observed on Binance, Kraken, and KuCoin, which clearly increase with scale. In Bitget1, the correlations are approximately 0.2 higher and reach levels comparable to those observed on KuCoin. The strongest transaction-count correlations are observed on Binance and Kraken, where the $\rho(q=2,s)$ values start at about 0.3 at the shortest scales and increase to 0.7--0.8 at the longest scales. The weak cross-asset correlations in $N$ on Bitget2 further support the interpretation that the elevated number of transactions in this period is driven by a noise-like component rather than coordinated market-wide trading activity.

Overall, the subsample analysis shows that the Bitget anomaly is concentrated in the second period and mainly concerns the transaction-count process for BTC and ETH. The results indicate that, after May 21, 2025, the elevated number of transactions is largely driven by low-volume trades that are weakly connected with both traded volume and return fluctuations. This supports the interpretation of a noise-like component in the post-change transaction-count dynamics.

\section{Summary}

This study examined the statistical and temporal properties of 1-min cryptocurrency trading activity for BTC, ETH, and XRP on four major centralised exchanges: Binance, Bitget, Kraken, and KuCoin. The analysis covered log-returns, trading volume, and the number of transactions, with a particular emphasis on exchange-specific differences in their distributions, autocorrelations, multifractal properties, cross-correlations, and time-dependent structural changes. Unusual behaviour was therefore interpreted as a persistent departure from the empirical stylised organisation of market activity expected under regular trading conditions.

The first important result was that the price dynamics was highly consistent across exchanges. For a given cryptocurrency, the cumulative log-returns and the distributions of standardised absolute log-returns were very similar on all considered platforms. The return distributions exhibited heavy tails and were much broader than the Gaussian distribution, which was consistent with the stylised facts of financial markets. This similarity indicated that, at the 1-min resolution, exchange-specific discrepancies in price formation were largely suppressed, most likely due to arbitrage mechanisms. In contrast, trading volume and the number of transactions displayed much stronger exchange-specific features, reflecting differences in liquidity, order fragmentation, and trading activity.

Among the analysed exchanges, Binance and Kraken exhibited the most regular and internally consistent market structure. The relationships between trading volume, number of transactions, and return fluctuations on these exchanges were smooth and broadly consistent with the standard market-microstructure expectations: larger transaction counts were associated with larger traded volume, and both higher volume and higher transaction intensity were associated with stronger return fluctuations. KuCoin followed the same general structure, but the corresponding relationships were more dispersed, which was consistent with its lower liquidity and more intermittent trading activity.

The most pronounced anomaly was observed on Bitget and was concentrated in BTC and ETH. A clear regime change appeared after mid-May 2025, when the number of transactions per minute increased sharply and remained elevated for an extended period. This change was not observed in XRP, suggesting that the effect was not a general exchange-wide transformation but rather an asset-specific phenomenon. The anomaly was visible across several measures. In the second period, several properties of the transaction-count process for BTC and ETH changed: the distributions shifted toward Gaussian behaviour, the autocorrelations weakened substantially, the fluctuation functions exhibited monofractal scaling, and the approximate entropy revealed a marked change in short-term regularity. At the same time, the cross-correlations involving $N$ decreased markedly. 

The subsample analysis showed that the elevated number of transactions on Bitget after May 21, 2025, was largely driven by numerous low-volume trades. These trades did not produce a proportional increase in traded volume and had no impact on return fluctuations. This was supported by the separation of Bitget1 and Bitget2 in the $V$--$N$ and $|R|$--$N$ scatter plots as well as by the fact that, after removing the smallest transactions, intervals without trading became visible again. Thus, the apparent continuity of trading activity in the original transaction-count series was largely produced by very small trades.

An important observation was that the regime changes for BTC and ETH occurred approximately simultaneously, but the anomalous transaction-count processes were not strongly correlated with each other. Cross-asset detrended cross-correlations showed that return correlations remained high across all exchanges and all cryptocurrency pairs, and volume correlations on Bitget decreased only moderately in the second period. In contrast, the correlations between transaction-count series in Bitget2 were very weak, typically close to zero at short scales and much lower than on Binance, Kraken, KuCoin, and Bitget1. This indicated that, despite the temporal alignment of the BTC and ETH regime changes, the resulting transaction-count dynamics were not driven by a synchronised market-wide process.

Taken together, the results presented in this work suggest that the post-change transaction-count series on Bitget contain a substantial noise-like component. A plausible interpretation is that, in the second period, many low-volume BTC and ETH trades were weakly connected with genuine market activity, thereby increasing transaction counts without a proportional increase in traded volume or price impact. This pattern may reflect changes in order fragmentation, trading algorithms, liquidity provision, or other platform-specific mechanisms. It may also be consistent with artificial activity, including wash-trading-like behaviour. However, this interpretation should be treated with caution. The analysis was based on public transaction-level data and statistical properties of aggregated time series. Therefore, it could not unambiguously distinguish between these possible mechanisms or identify the underlying source of the observed behaviour. What can be concluded is that the transaction-count process for BTC and ETH on Bitget after May 21, 2025, is statistically different from its earlier behaviour and from the corresponding processes observed on Binance, Kraken, KuCoin, and XRP on Bitget.

The results demonstrate that the number of transactions can be a highly informative variable for identifying exchange-specific anomalies that are not visible from price dynamics alone. The anomaly observed on Bitget would be difficult to detect using standard price-based indicators, because the log-return characteristics and cross-asset return correlations remain broadly comparable to those observed on the other exchanges. However, when price- and activity-related characteristics are analysed jointly within the proposed complexity-based framework, a clear structural change in trading activity becomes visible. This suggests that complexity-based characteristics can be useful diagnostic tools for detecting unusual trading patterns in cryptocurrency markets. Future work can extend this approach by incorporating order-book data, trade directions, account-level information, and longer samples to more clearly distinguish between natural market-making activity and potentially artificial transaction generation.

\vspace{6pt} 

\authorcontributions{Conceptualization M.W. and J.Z.; methodology S.D., J.K. and M.W.; software M.W. and J.Z.; validation S.D., J.K. and M.W.; formal analysis M.W. and J.Z.; investigation S.D., J.K., M.W. and J.Z.; resources, M.W. and J.Z.; data curation, M.W. and J.Z.; writing---original draft preparation M.W. and J.Z.; writing---review and editing S.D., J.K., M.W. and J.Z.; visualization M.W. and J.Z.; supervision, S.D. and M.W.; project administration, S.D. and M.W.; funding acquisition M.W. All authors have read and agreed to the published version of the manuscript.}

\funding{This research received no external funding.}

\institutionalreview{Not applicable.}

\informedconsent{Not applicable.}

\dataavailability{The data are available through the public API of the exchanges.} 

\conflictsofinterest{The authors declare no conflicts of interest.} 

\reftitle{References}
\bibliography{refs}

\end{document}